\documentclass[%
 aip,
 amsmath,amssymb,
 reprint,%
]{revtex4-1}

\usepackage{graphicx}
\usepackage{dcolumn}
\usepackage{bm}

\usepackage[utf8]{inputenc}
\usepackage[T1]{fontenc}
\usepackage{mathptmx}
\usepackage{etoolbox}
\usepackage{upgreek}

\let\oldsqrt\sqrt
\def\sqrt{\mathpalette\DHLhksqrt}
\def\DHLhksqrt#1#2{%
\setbox0=\hbox{$#1\oldsqrt{#2\,}$}\dimen0=\ht0
\advance\dimen0-0.2\ht0
\setbox2=\hbox{\vrule height\ht0 depth -\dimen0}%
{\box0\lower0.4pt\box2}}

\makeatletter
\def\@email#1#2{%
 \endgroup
 \patchcmd{\titleblock@produce}
  {\frontmatter@RRAPformat}
  {\frontmatter@RRAPformat{\produce@RRAP{*#1\href{mailto:#2}{#2}}}\frontmatter@RRAPformat}
  {}{}
}%
\makeatother

\begin{document}

\preprint{AIP/123-QED}

\title[Cavity-Enhanced HHG for Time-Resolved Photoemission]{Cavity-Enhanced High-order Harmonic Generation for High-Performance Time-resolved Photoemission Experiments}
\author{Thomas K. Allison}%
\affiliation{Department of Physics and Astronomy, Stony Brook University, Stony Brook, NY 11794-3800}
\affiliation{Department of Chemistry, Stony Brook University, Stony Brook, NY 11794-3400}
\email{thomas.allison@stonybrook.edu}

\author{Alice Kunin}
\affiliation{Department of Chemistry, Princeton University, Princeton, NJ 08544}

\author{Gerd Sch\"onhense}
\affiliation{Johannes Gutenberg-Universit\"at, Institut f\"ur Physik, D-55099 Mainz, Germany}
\date{\today}

\begin{abstract}
Recent breakthroughs in high repetition-rate extreme ultraviolet (XUV) light sources and photoelectron analyzers have led to dramatic advances in the performance of time-resolved photoemission experiments. 
In this perspective article, we discuss the application of cavity-enhanced high-order harmonic generation (CE-HHG), with repetition rates exceeding 10 MHz, to photoemission experiments using advanced endstations incorporating time-of-flight photoelectron analyzers.
We discuss recent results, perspective on future areas for further technological improvement, and the wide array of science enabled by ultrafast XUV photoemission experiments, now in a qualitatively new regime.

\end{abstract}

\maketitle

\section{INTRODUCTION}\label{sec:introduction}

\begin{figure*}
  \includegraphics[width = 6 in]{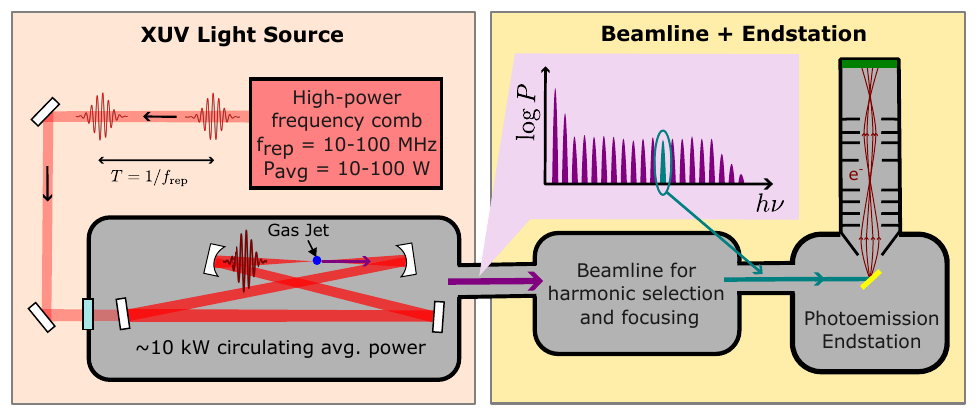}
  \caption{
    Sketch of layout for a CE-HHG-based photoemission apparatus. 
    The pulse train from high-power frequency comb laser is resonantly enhanced in a ``bow-tie'' style cavity with a small focus.
    A broad range of high-order harmonics is generated at $>$ 10 MHz repetition rate and outcoupled from the cavity; the figure is deliberately agnostic as to the outcoupling method, since several methods exist.\cite{Pupeza_NatPhot2021}
    For experiments with high spectral resolution, a single harmonic is selected with an XUV beamline and delivered to an endstation for photoelectron spectroscopy. 
  }
  \label{fig:overview}
\end{figure*}

Photoemission spectroscopy using extreme ultraviolet light (XUV, $h\nu = 10-100$ eV) is a powerful approach to study the electronic structure of matter, and thus it is not surprising that nearly every synchrotron light source has XUV beamlines dedicated to photoemission measurements using a sophisticated suite of electron analyzers and photo-emission electron microscopes (PEEM).
The power of photoemission lies in its simplicity of interpretation.
Conservation of energy gives direct access to electron binding energy spectra of atoms, molecules, liquids, and crystals and careful analysis of binding energy shifts provides detailed information regarding local chemical bonding environments.\cite{Hufner_book2003,Kolasinski_Book2012,Siegbahn_Book1967}
For angle-resolved photoemission (ARPES) from clean surfaces with crystalline order, conservation of momentum gives direct access to energy-momentum dispersion relations, formally described by the one-particle spectral function $A(\mathbf{k}, E)$, reducing to band structure in the absence of strong correlations.\cite{Hufner_book2003, Sobota_RMP2021}
ARPES measurements are widely regarded as the ``gold standard" for studying electronic structure, and thus there has been considerable investment in synchrotron ARPES facilities for ground-state photoemission measurements studying a wide variety of systems.

Time-resolved (i.e. pump/probe) photoemission measurements can track dynamics of excited states and unravel the complex coupling of electronic, nuclear, and spin degrees of freedom that occur in condensed-phase systems.
However, whether angle-resolved or not, time-resolved measurements are considerably more difficult. 
While there are many technical challenges associated with conducting pump/probe photoemission studies on condensed-phase samples, the required data rates present the most severe problem.
Adding a pump pulse to excite the sample adds four additional dimensions to the measurement space: pump-probe time delay, pump fluence, pump photon energy, and polarization. 
Ultrafast relaxation generally negates the utility of synchrotron radiation, which has $\gtrsim$100 MHz pulse repetition rates but intrinsically provides pulses with long pulse durations greater than 10 ps - too long to study short-lived excited states.
Most critically, photoemission signals from excited states are inherently much smaller than ground-state signals, in some cases by many orders of magnitude.
In particular it has been a long-standing problem to achieve time-resolved ARPES (tr-ARPES) studies covering the whole Brillouin zone in the limit of low excitation fluence, where one studies individual quasi-particle dynamics as opposed to a more collective response observed in high-fluence experiments,\cite{Basov_RMP2011} and the signal can be described as a low-order (imaginary $\chi^{(3)}$) perturbative response from a nonlinear optics perspective.\cite{Mukamel_book1995} 
This is especially true for spin-resolved measurements.\cite{Fanciulli_PhysRevRes2020,Plotzing_RSI2016}
Another major challenge in the field has been studying dynamics of solvated species at chemically relevant dilute concentrations in liquid microjet experiments.\cite{Suzuki_JChemPhys2019}

The critical parameters determining the data rate in a time-resolved photoemission experiment are the XUV/x-ray pulse repetition rate and the electron collection/detection efficiency. 
Recently, major progress has been made on both of these factors.
XUV light sources based on single-pass high-harmonic generation (HHG) of high-power femtosecond lasers at $\sim$1 MHz repetition rate\cite{Hadrich_JPhysB2016, Madeo_Science2020, Puppin_RSI2019, Keunecke_RSI2020} are now more-or-less commercially available and marketed towards the photoemission community.\cite{AFS_web,class5_web,KMLabs_web}
For photoelectron analysis and detection, the new technique of time-of-flight momentum microscopy (ToF k-mic) enables parallel detection of all photoelectron momenta and energies emitted from the sample,\cite{Chernov_Ultramicroscopy2015, Medjanik_NatMat2017} with a corresponding increase in data rate of several orders of magnitude over conventional hemispherical analyzers.
Over the past four years, this combination has enabled a variety of breakthrough measurements on excited states in the low-fluence limit (or close), a qualitatively new regime mostly inaccessible to time-resolved photoemission with XUV pulses until recently.
For example, in 2020 Madeo et al. reported the momentum-space imaging of excitons in 2D materials, including the resolution of dark excitons not observable with optical measurements.\cite{Madeo_Science2020}
Work on excitons in 2D material heterostructures, including the effects of moiré potentials, followed shortly thereafter.\cite{Schmitt_Nature2022, Karni_Nature2022} 
Paradigm-shifting photoemission measurements in molecular systems have also been reported.
For example, Wallauer et al. have reported time-resolved orbital imaging of large polycyclic aromatic hydrocarbons oriented on a Cu(001) surface\cite{Wallauer_Science2021} and Neef et al. measured diffuse momentum-space signatures attributed to singlet fission in crystalline pentacene.\cite{Neef_Nature2023}

However, data rate is \emph{still} the limiting factor in these experiments, and compromises must be made on the usable pump laser fluence or the range of parameter space that can be covered within reasonable accumulation times.
For example, for exciton studies in 2D materials, accumulation times can reach tens of hours per pump/probe delay point for studies at low excitation fluence.\cite{Man_SciAdv2021}
The data rate in these experiments is also not high enough to enable low-fluence time- and spin-resolved ARPES measurements, since the addition of spin filters in the electron analyzer decreases the attainable signal by 1-2 orders of magnitude.\cite{Kolbe_PRL2011,Schoenhense_Ultramicroscopy2017}

Since efficient HHG, especially for photon energies exceeding 30 eV, requires high pulse energies $\gtrsim$100 $\upmu$J, achieving HHG at high repetition rate requires a high average laser power for the driving laser.
Thus, the laser systems employed to achieve single-pass HHG at $\sim$1 MHz repetition rate and beyond rapidly become expensive and complicated.
For example, several efforts have implemented systems based on femtosecond lasers with several hundred Watts to kW average power from coherently combined fiber amplifiers, which require extensive feedback mechanisms, considerable engineering, and large expensive optics for the handling the high average power.\cite{Keunecke_RSI2020, NSFNeXUS}
The accessibility of these large laser systems remains limited, and despite major engineering investment, there are still challenges with reliability.

Cavity-enhanced HHG (CE-HHG) is an elegant alternative method for generating very high repetition rate ultrashort XUV pulses.\cite{Pupeza_NatPhot2021, Mills_JPhysB2012}  
In the femtosecond enhancement cavity (fsEC) approach, successive low-energy $\sim$1 $\upmu$J pulses from a frequency-comb laser are coherently added in an optical resonator.\cite{Gherman_OptExp2002, Jones_OptLett2002, Jones_OptLett2004}
Power enhancement factors of several hundred to several thousand can be obtained in this matter, and circulating average powers of up to 400 kW have been achieved while still preserving femtosecond pulse durations for the intracavity pulse.\cite{Carstens_OptLett2014}
These average power levels far exceed that available from any femtosecond laser system implemented or reasonably conceived.
High-harmonic generation can then be achieved in a gas jet at a focus of the fsEC, and coupled out with various means, providing ultrashort XUV light pulses in the photon energy range of 10-100 eV, at repetition rates from 10's to 100's of MHz rates.\cite{Carstens_OptLett2014}
 
Cavity-enhanced HHG has now been shown to work well as a stable and reliable light source and has been successfully applied to photoemission experiments by several research groups.\cite{Corder_StructDyn2018, Mills_RSI2019, Saule_NatComm2019}
The CE-HHG method has been previously reviewed by Pupeza et al.\cite{Pupeza_NatPhot2021} and Mills et al.,\cite{Mills_JPhysB2012} and the general problem of laser development for ARPES has recently been exhaustively reviewed by Na et al.\cite{Na_PhysicsReports2023}
In this perspective article we focus on the application of CE-HHG to particularly difficult problems in time-resolved photoemission, surveying the past work and also discussing future directions. 
We discuss how the high repetition rates and broad photon energy ranges provided by CE-HHG, coupled with continuously advancing photoelectron analyzers, can routinely enable time-resolved photoemission experiments that are either heroic or impossible with conventional methods.
The perspective article is organized as follows.
In section \ref{sec:review}, we briefly survey some results already obtained by several research groups applying CE-HHG to time-resolved photoemission.
In section \ref{sec:ideal} we discuss parameters for an ``ideal'' light source for pump/probe photoemission experiments.
In section \ref{sec:HHGphysics} we discuss some of the basic physics of high harmonic generation that impact design decisions of driving lasers and HHG-based beamlines for photoemission setups in general, and in section \ref{sec:CE-HHGpractice} we discuss practical considerations that specifically drive the design of CE-HHG sources for time-resolved photoemission.
In section \ref{sec:analyzers} we discuss recent developments in high-performance photoelectron analyzers and their coupling with CE-HHG light sources, and in section \ref{sec:outlook} we provide an outlook of future experiments possible at CE-HHG-based photoemission beamlines.

\section{Examples of Several CE-HHG-based Photoemission Experiments}\label{sec:review}

\begin{figure}
  \includegraphics[width = 0.9\columnwidth]{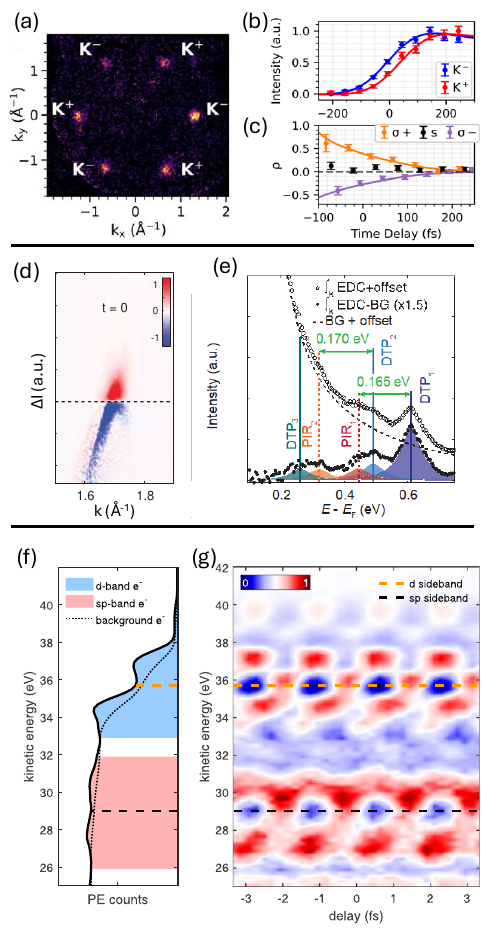}
  \caption{
  Some highlights of CE-HHG-based photoemission experiments. 
  (a) 2D momentum maps of excitons in monolayer WS$_2$ after excitation with 2.4 eV light.
  (b) Time-resolved signals for the K$^{-}$ and K$^{+}$ valleys after excitation with $\sigma^{-}$ polarized light.
  Rapid valley depolarization due to the Coulomb exchange interaction leads to the K$^{+}$ valley signal rising only 50 fs after the K$^{-}$ valley signal.
  (c) Valley polarization vs. pump/probe delay for $\sigma^{-}$, $\sigma^{+}$, and linearly-polarized ($s$) excitation.
  (d) An energy-momentum difference map (pump on - pump off) for graphite excited with 1.2 eV pump pulses at a fluence of 18 $\upmu$J/cm$^2$.
  (e) Peaks corresponding to the direct pump excitation energy and satellite peaks provide detailed information of electron-phonon coupling in graphite.
  (f) Valence band EDC from W(110) upon illumination with several harmonics.\cite{Heinrich_NatComm2021}
  The regions highlighted in red and blue correspond to signals from the sp-bands and d-bands, respectively, separated by 6.4 eV.
  (g) RABBITT signal showing sideband modulation vs. XUV/IR delay.
  More detailed analysis of the RABBITT signal gives a 39 $\pm$ 18 as delay between d-band photoemission and sp-band photoemission.
  (a) through (c) are adapted with permission from Kunin et al.\cite{Kunin_PRL2023}
  Copyright 2023, American Physical Society.
  (d) and (e) are adapted with permission from Na et al.\cite{Na_Science2019}
  Copyright 2019, American Association for the Advancement of Science.
  (f) and (g) are adapted from Heinrich et al.\cite{Heinrich_NatComm2021}
  with the author's permission under the Creative Commons CC BY license.
  }
  \label{fig:PreviousResults} 
\end{figure}

Figure \ref{fig:PreviousResults} shows three examples of high-performance time-resolved photoemission results obtained by three different research groups using CE-HHG, each of which has represented a technical landmark in terms of resolution, sensitivity, or acquisition speed.

The top panels, figures \ref{fig:PreviousResults}(a)-(c), show tr-ARPES measurements of valley polarization dynamics in monolayer WS$_2$ measured by Kunin et al.\cite{Kunin_PRL2023} using ToF k-mic and XUV pulses at 61 MHz repetition rate.
The monolayer WS$_2$ sample used was a 10 $\upmu$m exfoliated flake, supported by an insulating hexagonal boron nitride layer to prevent exciton quenching.
These measurements imposed three critical challenges met by the high performance of the beamline: (1) the small sample size,
(2) the need to perform the experiment at very low excitation fluences less than 5 $\upmu$J/cm$^2$ to keep the exciton density well below the Mott density, at which point interactions of the photoexcited carriers lead to unbound electron-hole pairs,\cite{Wang_RMP2018}
and (3) the need to vary many parameters to understand the effects of sample temperature, XUV photon energy, and pump polarization on the observed signals.  

Figure \ref{fig:PreviousResults}(a) shows a raw momentum-space image of the full Brillouin zone, acquired in parallel using ToF k-mic.
Figure \ref{fig:PreviousResults}(b) shows the signals in the inequivalent K$^{-}$ and K$^{+}$ valleys upon excitation with $\sigma^{-}$ circularly polarized light, which preferentially populates the K$^{-}$ valleys via the chiral optical selection rules of the transition-metal dichalcogenide monolayers.\cite{Wang_RMP2018}
Figure \ref{fig:PreviousResults}(c) shows the valley polarization $\rho(t)=\left\{\left[I_{K^{+}}(t)-I_{K^{-}}(t)\right] /\left[I_{K^{+}}(t)+I_{K^{-}}(t)\right]\right\}$ as a function of pump-probe delay for $\sigma^{-}$ excitation as in panel (b), and also the control experiments with $\sigma^{+}$ and $s$-polarized excitation.
The high signal-to-noise ratio of the measurement enabled the resolution of a small 50 fs shift between the signals from the K$^{-}$ and $K^{+}$ valleys, much smaller than the 200 fs instrument response function.
Further detailed analysis of the valley-resolved energy and momentum distributions identified the Coulomb exchange interaction as the dominant mechanism for valley depolarization.

The middle panels of figure \ref{fig:PreviousResults}, figures \ref{fig:PreviousResults}(d) and \ref{fig:PreviousResults}(e), show tr-ARPES measurements in bulk graphite obtained by Na et al.\cite{Na_Science2019} using a hemispherical analyzer and XUV pulses at 60 MHz repetition rate.
Here, a combination of high energy resolution (22 meV) and high sensitivity (excitation fluence only 18 $\upmu$J/cm$^2$) enabled the resolution of discrete peaks in a non-thermal excited electron distribution, and extraction of electron-phonon coupling strengths.
Figure \ref{fig:PreviousResults}(d) shows an energy momentum difference map, i.e. E vs. k for pump on minus pump off, along the $\bar{\Gamma} - \bar{\text{K}}$ direction, when pump and probe maximally overlap.
The valence bands are depleted (blue) as electrons are promoted to the conduction bands (red).
Figure \ref{fig:PreviousResults}(e) shows the energy distribution curve (EDC) after integration over momentum.
Subtraction of a smooth background signal (dashed line) reveals a series of peaks corresponding to direct excitation to the conduction bands and replica peaks due to electron-phonon coupling.
From analysis of the amplitudes of the replica peaks and their dependence on pump/probe delay, Na et al. determined values for the electron-phonon coupling matrix elements for several optical phonon modes.

The lower panels of figure \ref{fig:PreviousResults}, figure \ref{fig:PreviousResults}(f) and \ref{fig:PreviousResults}(g), show  results of attosecond photoemission experiments on W(110) conducted by Heinrich et al.\cite{Heinrich_NatComm2021} using a CE-HHG source with 18.4 MHz repetition rate and time-of-flight analyzer with $\pm$7 deg. angular acceptance.
In this experiment, an attosecond pulse train consisting of several high harmonics around 40 eV are simultaneously used for photoemission in the presence of an infrared laser field synchronized with interferometric precision.
  The infrared laser field produces sidebands on the photoemission signal due to the laser-assisted photoelectric effect,\cite{Miaja-Avila_PRL2006,Saathoff_PRA2008} and the interference of sidebands of adjacent harmonics contains information on attosecond time delays in the photoemission process.
Relative time delays are extracted using the Reconstruction of Attosecond Beating by Interference of Two-photon Transitions (RABBITT) method, originally developed for gas-phase attosecond experiments\cite{Paul_Science2001} and later applied to surface photoemission.\cite{Locher_Optica2015,Luchhini_PRL2015,Tao_Science2016}

Figure \ref{fig:PreviousResults}(f) shows the EDC for valence electrons near $\Gamma$ of W(110), with the momentum-space range selected by the narrow angular acceptance of the analyzer.
Regions of the spectrum corresponding to the d-bands and sp-bands of W(110) are highlighted in blue and red, respectively.
Figure \ref{fig:PreviousResults}(g) shows the delay-dependent RABBITT signal in the presence of the IR field.
Analysis of the RABBITT signals allowed the authors to determine that photoemission from the d-band is delayed by 39 $\pm$ 18 attoseconds relative to the sp-band.
The experiment required (relatively) high resolution for an attosecond experiment and also high count rates to achieve the high signal-to-noise ratio required to observe small modulations of the spectrum.\cite{Heinrich_NatComm2021, Saule_NatComm2019}
The high data rate enabled the authors to repeat the experiments for several different incident harmonic spectra and thus study the impact of systematic effects such as spectral phase of the XUV mirrors.
Another unique feature of CE-HHG exploited in this work was continuous tuning of the photon energy using the effect of repeated plasma blueshifting of the fundamental intracavity pulse.\cite{Allison_PRL2011}

\section{What is the ideal ultrafast XUV light source for photoemission?}\label{sec:ideal}

In this section, we discuss several design considerations for the ``ideal'' time-resolved photoemission beamline in general. 
What the ``ideal'' setup is depends to some extent on the scientific goals, but many design goals are common to all experiments.
We return to how CE-HHG can meet all the demands in the itemized points below in section \ref{sec:CE-HHGpractice}.
\subsection{Photon energy and tunability}
The parallel momentum of an electron emitted from a surface at angle $\theta$ with kinetic energy $E$ is
\begin{align}
  k_{\parallel}^{(\text{vacuum})} & = \sqrt{\frac{2 m E}{\hbar^2}} \sin \theta \nonumber \\ 
                                  &= 0.512 \; \text{\AA}^{-1} \sqrt{E\; \text{[eV]}} \; \sin \theta,
\end{align}
where $m$ is the mass of the electron and $\hbar$ is Planck's constant.
Thus, to cover the full Brillouin zone of crystalline materials, which may extend to ${\sim}2$ \AA$^{-1}$, with reasonable emission angles less than 90 degrees, one needs at least ${\sim}20$ eV photon energies.\cite{Na_PhysicsReports2023}
A similar momentum extent is required in molecular photoemission in order to perform molecular orbital tomography.\cite{Luftner_JElecSpec2014,Puschnig_Science2009,Wallauer_Science2021}
Furthermore, for performing experiments on weakly excited samples, with small signals above the Fermi level, it is critical to obtain a large suppression of spurious photons with energy above that selected by the beamline for use in the experiments, as higher-energy photons will produce large background signals. 

While it is easier to develop an HHG beamline that generates sufficient flux with sufficient isolation of adjacent harmonics at one fixed photon energy, it has long been appreciated by the ground-state photoemission community that it is generally important to tune the XUV photon energy in photoemission experiments.
A commonly used expression for the photocurrent $I$ of electrons detected at a particular momentum $\mathbf{k}$ with energy $E$ is \cite{Damascelli_RMP2003,Hufner_book2003,Sobota_RMP2021}
\begin{equation}
    I(\mathbf{k},E) \propto\left|M_{f i}^{\mathbf{k}}\right|^2 \times f(E) \times A(\mathbf{k}, E) \;,
\end{equation}
where $M_{f i}^{\mathbf{k}}$ 
is the one-electron dipole matrix element describing interaction with the electromagnetic field, $f(E)$  describes the state occupation, 
and $A(\mathbf{k}, E)$ is the single-electron removal spectral function.
The photon energy affects the photoemission amplitude via two factors: 
(1) The final state energy determines the 3D subset of the 4D-spectral function recorded since $\mathbf{k}$ and $E$ are constrained by energy conservation, and in general the photon energy must be scanned to determine the dispersion of bands in the momentum direction perpendicular to the surface, $k_z$.
(2) The photoemission matrix element $M_{f i}^{\mathbf{k}}$ depends on the final state energy.
Thus, all synchrotron beamlines for photoemission use tunable monochromators and/or variable-gap undulators to scan the photon energy and discern both effects.

\begin{figure}
  \includegraphics[width = 0.8\columnwidth]{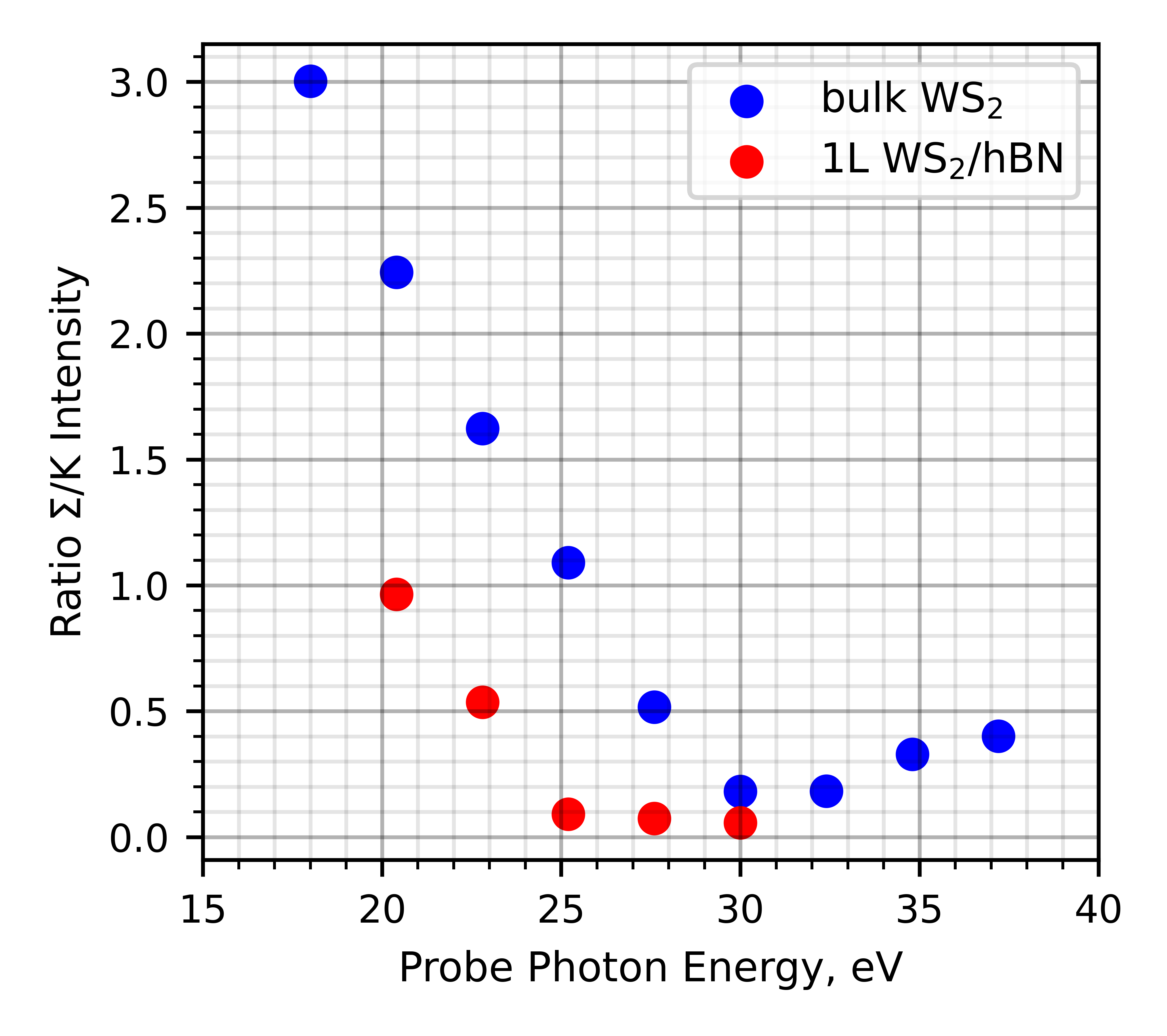}
  \caption{
    Matrix element effects in pump/probe photoemission.
    Ratio of excited-state signal at the $\Sigma$ valley of the Brillouin zone to that at the K valley in bulk (blue) and monolayer (red) WS$_2$, recorded after excitation with 517 nm light, as a function of the probe photon energy.
    Even in monolayer WS$_2$, a strong dependence on the photon energy is observed due to matrix element effects.
    Reproduced with permission from Kunin et al.\cite{Kunin_PRL2023}.
    Copyright 2023, by the American Physical Society.
    }
    \label{fig:KSigma}
\end{figure}

Of course, tuning the photon energy is also important for understanding excited-state signals, and arguably moreso since theoretical calculations one might use to correct for matrix element effects or calibrate the inner potential\cite{Hufner_book2003} are in general less reliable for excited states.
Figure \ref{fig:KSigma}, from Kunin et al.,\cite{Kunin_PRL2023} shows a striking example of matrix-element effects in excited-state photoemission.
Plotted is the ratio of the largest signals observed above the valence band maximum recorded from the $\Sigma$ valleys and the K valleys of WS$_2$, after 517 nm excitation, as a function of the XUV photon energy for both bulk and monolayer samples. 
For the bulk WS$_2$ sample the signals are due to conduction band electrons and for the monolayer they are from bound excitons, appearing below the conduction band.
In both cases, there is a strong photon energy dependence to the signals, even though the excitation conditions for each data point are 
the same.
Performing the monolayer experiment at only one unlucky photon energy, say 25 eV, one could falsely conclude that dark K-$\Sigma$ excitons\cite{Wang_RMP2018} are not present in this system.

\subsection{Energy and time resolution}

In all-optical ultrafast measurements such as transient absorption spectroscopy, the conversion of ultrafast light pulses into an electrical signal happens after the light pulses interact with the sample.
The instrumental energy- and time-resolutions of the experiment are then decoupled since analysis of spectral features happens separately (e.g. in a spectrograph) from the ultrafast experiment at the sample.\cite{Pollard_JCP1990}
So while samples with very fast dynamics will intrinsically exhibit broad spectral features and systems with slower dynamics may have narrower features, the experimenter need not decide the optimum resolutions for conducting the experiment before doing it, and there is in principle no spectral resolution penalty for doing the experiment with an arbitrarily short pulse. 
However, in photoemission, the ultrafast light pulses are converted to the electrical signal (i.e. photoelectrons) at the sample as part of the experiment, and thus the attainable resolution in photoemission measurements is fundamentally limited by a Fourier relation on the probe pulse bandwidth:
\begin{equation}\label{eqn:transform_limit}
  \delta E \delta t \ge \hbar \times 4\ln 2 = 1.83 \mbox{ eV-fs} \;,
\end{equation}
where $\delta t$ and $\delta E$ are the FWHM resolutions in time and energy, respectively.

In practice very few time-resolved photoemission experiments using XUV light have actually realized the transform limit of equation (\ref{eqn:transform_limit}) for a variety of technical reasons.
For one thing, the temporal resolution of a pump/probe experiment can also be impacted by the duration of pump pulse, and this is not included in equation (\ref{eqn:transform_limit}).
Also, producing nearly transform-limited femtosecond (or attosecond) XUV pulses requires care in the details of the HHG source and beamline.\cite{Wang_NatCom2015}
The energy resolution is also easily broadened by detector resolution, voltage fluctuations, sample quality, or stray magnetic fields unless great care is taken.

However, the most pernicious resolution problem for time-resolved photoemission experiments on solids and liquids has been the so-called vacuum space charge effect, in which the Coulomb repulsion of photoelectrons emitted from the sample nearly simultaneously during the ultrashort pulse causes shifts, broadening, and other distortions of the photoelectron energy and momentum distributions. 
Given the critical limitations imposed on time-resolved photoemission experiments by space charge, this problem has been extensively studied\cite{Hellmann_PRB2009, Plotzing_RSI2016, Frietsch_RSI2013, Mathias_Collection2010, Graf_JApplPhys2010, Passlack_JApplPhys2006, Zhou_JElecSpec2005,Rotenberg_JSyncRad2014} and recently revisited in the context of PEEM-based photoelectron analyzers.\cite{Schoenhense_NewJPhys2018,Schoenhense_RSI2021,Maklar_RSI2020,Chernov_UP2022}
For photoemission with XUV pulses, shifts and broadening of the photoelectron spectrum scale linearly with the number of photoelectrons emitted from the sample per pulse, $N_{\text{pulse}}$, and $N_{\text{pulse}}$ should typically be below 100-1000 for achieving resolution of $\sim$100 meV or less, depending on the type of photoelectron analyzer used.\cite{Maklar_RSI2020}
The limitations imposed by space charge have been one of the primary motivators for developing HHG sources at higher repetition rates, and particularly for the application of CE-HHG to time-resolved photoemission.
The reasoning is simple: if the rate of pulses impinging on the sample is increased, then the number of photoelectrons per pulse can be reduced while still maintaining a healthy sample current.


\subsection{Repetition rate}\label{sec:ideal:reprate}

The ideal repetition rate for a time-resolved photoemission experiment represents a balance between three competing interests.

First, the desire to mitigate the effects of space charge advocates for the largest possible repetition rate, such that the fewest number of photoelectrons per pulse are emitted for the same average sample current.
Synchrotron photoemission beamlines, for example, operate with very high repetition rates exceeding 100 MHz, such that only a few electrons per pulse may be emitted from the sample.\cite{Rotenberg_JSyncRad2014, Zhou_JElecSpec2005}

Second, for pump/probe experiments the decay of the excited state and or sample heating may limit the pump repetition rate. 
Ideally, to capture the intrinsic dynamics of the sample, the time between subsequent pump excitations should exceed the sample relaxation time such that each pump/probe experiment is performed on an equilibrium sample, although in some cases a background of long-lived excited states may be tolerable.
The limit this places on the repetition rate is strongly sample dependent.
For example, triplet excitons in organic semiconductor systems can have very long lifetimes exceeding 1 $\upmu$s,\cite{Stadtmuller_NatComm2019} restricting repetition rates to ${\sim}1$ MHz and below.\cite{Bennecke_NatComm2024}
For experiments in the high-fluence regime ($> 100$ $\upmu$J/cm$^2$), sample heating and/or damage also may limit the repetition rate to substantially less than 1 MHz.\cite{Na_PhysicsReports2023} 
However, for many experiments higher repetition rates can be used. 
For example, metallic or graphene/graphite-based sample relaxation rates are very fast and the sample thermal conductivity is also large, such that repetition rates may exceed 100 MHz, in principle, and many time-resolved 2-photon photoemission (2PPE) experiments using lower energy photons have been done on metal surfaces using ${\sim}100$ MHz pulse trains derived from Ti:Sapphire oscillators.\cite{Bovensiepen_Book2010,Hofer_Science1997,Petek_ProgSurfSci1997}
Excitons in transition metal dichalcogenide (TMD) monolayer and heterostructure systems also generally have sub-10 ns lifetimes due to rapid radiative recombination.\cite{Wang_RMP2018,Regan_NatureReviews2022}
Charge-transfer states between molecules and surfaces can also be very short lived and have been routinely studied using ${\sim}$100 MHz repetition rates in 2PPE experiments.\cite{Petek_Science2000, Onda_Science2005,Stahler_ChemSocRev2008}

Third is the limitations of time-of-flight analyzers.
In the time-of-flight method, photoelectron energies are mapped to a ToF window $1/f_{\text{rep}}$ wide using an electron detector with finite resolution $\delta\text{ToF}$.
If a total electron energy range $\Delta E$ is recorded in parallel, this then sets a resolution limit
\begin{equation}\label{eqn:dToF}
    \delta E >  \Delta E \times f_{\text{rep}} \times \delta\text{ToF}.
\end{equation}
For example, with $\delta$ToF = 200 ps, $f_{\text{rep}}$ = 100 MHz, and $\Delta E$ = 5 eV, this gives a resolution limit of 100 meV.
Limitations of the detector in terms of total count rate may place further constraints on the repetition rate or detected energy band.

The limitations on pump repetition rate are intrinsic to the sample physics under study, and are thus fundamentally limited. 
Space charge is somewhat dependent on the details of the analyzer,\cite{Maklar_RSI2020,Chernov_UP2022,Schoenhense_NewJPhys2018, Schoenhense_RSI2021} but in the end is mostly limited by fundamental aspects of the light-matter interaction at the sample surface intrinsic to the experiment.
In contrast, there are large gains still to be made in detector technology, as we discuss further in sections \ref{sec:analyzers} and \ref{sec:outlook}.

\subsection{Stability}\label{sec:ideal:stability}

Even with high repetition rates, low-fluence time-resolved photoemission experiments will intrinsically involve long acquisition times, especially if many parameters such as sample temperature, pump fluence, pump polarization, etc., are systematically varied.
For example, for pump/probe experiments on excitons in small exfoliated TMD systems, even with full $2\pi$ collection of the electrons using ToF k-mic, for exciton densities well below the Mott transition, signal levels on the order of ${\sim}$1 detected electron for every $10^6$ pump/probe pulse pairs are observed.\cite{Kunin_PRL2023,Madeo_Science2020}
So even with a very high pulse repetition rate of 100 MHz, there is only a count rate of ${\sim}100$ Hz.
If this signal is parsed even coarsely into 300 momentum/energy bins (e.g. $30 \times 10$), and 10 pump/probe delays are taken, it still requires ${\sim}3,000$ seconds, or roughly 1 hour to achieve a signal-to-noise ratio limited by Poisson statistics of only 10:1 in the pump/probe scan.
Varying other parameters implies at least multi-day measurement campaigns even at 100 MHz repetition rate, and thus source stability and repeatability on at least this time scale is critical.
Indeed, for lower repetition rate systems, it becomes clear that measurement campaigns can easily stretch into months of continuous operation, putting a premium on the stability of the system.

\subsection{Pump Pulses}\label{sec:ideal:pump}

Ideally, the pump pulse energy should be readily tunable over the range of excitations of interest to fully tease out the sample physics of interest.
This is particularly true for low-fluence experiments that aim to study particular quasi-particles addressed specifically by tuning the pump excitation energy,\cite{Schmitt_Nature2022, Bange_SciAdv2024}
but it can be true for high-fluence experiments as well.
For example, for studying Floquet physics or strong phonon excitation, tunable mid-IR or far-IR pump pulses are desired.\cite{McIver_NatPhys2020, Mahmood_NatPhys2016}

Fortunately, the pump pulse energies required are not large.
Particularly for experiments at high repetition rate, the maximum pump fluence an experiment might use at the sample is on the order of $\sim$1 mJ/cm$^2$.
For a 100 $\upmu$m spot size on the sample, this corresponds to pulse energies of only 100 nJ.
For the low-fluence experiments recently made possible with high-performance time-resolved photoemission beamlines, fluences may be as small as $\sim$1 $\upmu$J/cm$^2$, requiring only $\sim$1 nJ pump pulses.
In addition to restrictions imposed by the sample physics under study, space charge due to multi-photon photoemission of the pump can also limit the applied fluence.\cite{Ultstrup_JElecSpec2015, Oloff_JApplPhys2016, Borgwardt_Thesis2016}

In principle, using ToF detectors, the pump repetition rate can be decoupled from the XUV pulse repetition rate, since each electron can be measured individually and correlated with the incident optical pulse train.
For example, if pump pulses are incident at $f_{\text{rep}}/4$, time-stamping the electrons enables simultaneous recording of signals with femtosecond pump/probe delay $\tau$, and nanosecond pump/probe delays $\tau + 1/f_{\text{rep}}$, $\tau + 2/f_{\text{rep}}$, and $\tau + 3/f_{\text{rep}}$.
This is important for CE-HHG-based beamlines since it is not easy to adjust the XUV pulse repetition rate, but the pump can be adjusted separately.\cite{Wahl_Thesis2024}

\section{Some Fundamentals of High-order Harmonic Generation}\label{sec:HHGphysics}

In this section we provide a very brief survey of high-order harmonic generation physics most relevant to HHG source design for photoemission.
For a more comprehensive discussion of many technical and fundamental aspects of HHG, the reader is referred to the book by Chang.\cite{Chang_Book2011}

Gas-phase HHG is a nonlinear optical process that takes place in intense laser fields with electric field strengths on the order of $\mathcal{E} \sim 3$ V/\AA, or intensities of approximately $I \sim 10^{14}$ W/cm$^2$.
At these field strengths, the perturbation theory approach to describing light-matter interaction, for example as would be appropriate for studying more conventional nonlinear optical phenomena such as frequency doubling, sum/difference frequency generation, or four-wave mixing,\cite{Boyd:2003, Agrawal_NonlinearFiberOpticsBook} breaks down and must be replaced by non-perturbative approaches.
High-harmonic generation is always accompanied by strong-field ionization of the gaseous medium, and the two phenomena are closely related.\cite{Krause_PRL1992,Schafer_PRL1993}
Quantitative calculation of HHG spectra is very difficult and in general should involve numerical simulation of both the intensity-dependent high-frequency atomic dipole coupled with Maxwell's equations for propagation of both the fundamental and harmonic fields. 
However, much can be understood using the ``simple-man's model''\cite{Krause_PRL1992,Schafer_PRL1993,Corkum:1993} or its semiclassical extension\cite{Lewenstein:1994} for the single-atom dipole, and one-dimensional phase-matching considerations for field propagation.\cite{Constant_PRL1999, Durfee_PRL1999, Allison_Thesis2010, Heyl_JPhysB2012}
In the simple-man's model, high harmonics are emitted by field-ionized electrons driven back to their parent ion at high kinetic energy.
The maximum energy of returning trajectories gives the cutoff law for the maximum emitted photon energy via
\begin{equation}\label{eqn:cutoff}
  \hbar \omega_c = 3.17 U_p + I_p \;,
\end{equation}
where $U_p = e^2 \mathcal{E}^2/4m\omega^2$ is the ponderomotive potential and $I_p$ is the ionization potential of the gas-phase target atoms or molecules.
Equation (\ref{eqn:cutoff}) implies that the cutoff energy scales simply as the focused peak intensity, since $U_p \propto I$, however the choice of target gas also has a strong impact on the cutoff since the neutral atoms must survive the rising edge of the driver pulse without being ionized in order to emit harmonics at the driver peak intensity.
In practice the desired cutoff energy determines the target gas which in turn determines the minimum required intensity via the $\mathcal{E}$-dependence of ionization.\cite{Chang_Book2011}
A series of many odd harmonics of the driving frequency $\omega_0$ can be emitted with comparable efficiency, but the spectrum of harmonics is also affected by the macroscopic propagation of the fundamental and harmonics through the gas medium.

Phase mismatch between harmonics in different parts of the gas medium, characterized in a 1D (on-axis) model by a wave-vector mismatch between the fundamental ($k_0$) and the $q^{th}$ harmonic ($k_q$) via $\Delta k = q k_0 - k_q$, can be broken down into four contributions: dispersion of the neutral gas, dispersion of the free-electron plasma created upon strong-field ionization, the intensity-dependent dipole phase, and the Gouy phase of the focused driver beam.\cite{Allison_Thesis2010}
The first two factors are determined by the gas medium and its ionization dynamics, and the last two are determined by the focusing geometry of the driver pulses and gas jet.
Since the number of XUV photons per pulse that can be used at the sample in photoemission experiments is drastically limited by space charge, HHG-based ARPES beamlines typically aim to operate at the highest repetition rate possible given the available laser power, and this implies low pulse energies and tight focusing for the driver.
The dominant source of phase mismatch is then due to the Gouy phase, with the effective $\Delta k$ given by\cite{Allison_Thesis2010}
\begin{equation}\label{eqn:Gouy}
  \Delta k_{\text{Gouy}} = -\frac{q}{z_R} = -\frac{q \lambda_0}{\pi w_0^2} \;,
\end{equation}
where $\lambda_0$ is the wavelength of the driver, $w_0$ is the $1/e^2$ radius of the focus, and $z_R$ is the Rayleigh range.
Efficient HHG occurs when the coherence length $L_c = \pi/|\Delta k|$ is longer than the medium length. 
In practice, it is hard to make the gas medium length much smaller than 100 $\upmu$m,\cite{Hammond_OptExp2011} which implies $w_0 > 20$ $\upmu$m for $\lambda_0 = 1030$ nm and phase matching at 40 eV.
For a 200 fs driving pulse duration, typical of Yb-based lasers, achieving a peak intensity of $10^{14}$ W/cm$^2$ with $w_0 > 20$ $\upmu$m then requires pulse energies greater than 100 $\upmu$J.
This scenario corresponds to a cutoff energy for HHG in argon of $\hbar \omega_c = 47$ eV.
Achieving higher photon energies, for example by driving HHG in neon, requires significantly higher pulse energies in order to achieve the necessary cutoff energy, ionization of Ne,\cite{Chang_Book2011} and a larger focus to maintain reasonable phase matching at the higher harmonic order.  

For generating lower photon energies, somewhat tighter focusing and correspondingly smaller pulse energies can be used by driving with a shorter wavelength, due to the fact that both terms in the numerator of equation (\ref{eqn:Gouy}) are reduced.
Many single-pass HHG systems for photoemission beamlines have taken this approach, driving the HHG process with a tightly focused second-harmonic beam, e.g. 400 nm\cite{Wallauer_Science2021, Puppin_RSI2019} or 515 nm.\cite{Keunecke_RSI2020,Madeo_Science2020}
In these systems, (still) poor phase matching due to the very tight focusing geometry is partially compensated by the $\lambda$-dependence of the single-atom HHG dipole,\cite{DiChiara_IEEE2012, Brabec_RMP2000} and simply the fact that one can reach ${\sim}20$ eV with a lower-order harmonic of the driving laser.
Driving with the second harmonic of the fundamental also facilitates isolating a specific harmonic to use for photoemission, since the wider harmonic spacing makes it easier to select a single harmonic with a combination of multilayer mirrors and filters.
However, since the ponderomotive energy of equation (\ref{eqn:cutoff}) scales as $\lambda_0^2$, it is difficult to push this approach to higher photon energies, and HHG beamlines driven with the second harmonic of the fundamental laser are usually designed to achieve enough flux in a fixed harmonic (i.e., not tunable) between 20 and 25 eV, with a sharp cutoff in the emission for higher orders.

Pulse-energy requirements can also be loosened by using shorter driver pulses, but this can lead to broader bandwidth in the harmonics, which degrades the energy resolution at a minimum via the transform limit of equation (\ref{eqn:transform_limit}), and also larger broadening can occur via stronger time-dependent plasma blue shifting of the driver during the HHG process and ionization gating.\cite{Wang_NatCom2015,Hadrich_JPhysB2016}

\section{Cavity-enhanced HHG in Practice}\label{sec:CE-HHGpractice}

Cavity-enhanced HHG was first demonstrated using Ti:Sapphire oscillators in 2005 by research groups at JILA in Boulder CO, and the Max Planck Institute for Quantum Optics (MPQ) in Garching, Germany.\cite{Gohle_Nature2005,Jones_PRL2005}
The primary motivation for this work was the generation of phase-coherent frequency combs in the XUV, where no other light sources with long coherence times suitable for precision spectroscopy exist.
Initially, these results did not garner much interest outside the precision frequency metrology community, since the XUV powers achieved in this early work were very low, on the order of 100 pW/harmonic or less generated at the cavity focus.

The year 2011 was a breakthrough period for the CE-HHG method. 
Several groups reported dramatic improvements in the HHG yield, to $\gtrsim 100$ $\upmu$W/harmonic, by using higher-power frequency comb driving lasers based on either Yb:fiber\cite{Ruehl_OptLett2010} or injection-locked Ti:Sapphire.\cite{Paul_OptLett2008} 
A higher-power frequency comb to excite the fsEC enables the use of a lower fsEC finesse $\mathcal{F}$ and also a larger cavity focus size.
A larger focus size is beneficial for the reasons discussed in section \ref{sec:HHGphysics}.
Lower cavity finesse makes the system overall easier to work with and more reliable, and also is important for reducing the impact of intracavity nonlinear phase shifts imparted to the driver pulse in the HHG process, first described in detail in two 2011 papers by Allison et al.\cite{Allison_PRL2011} and Carlson et al.\cite{Carlson_OptLett2011}
This power scaling work led to the demonstration of XUV frequency comb spectroscopy by the JILA group,\cite{Cingoz_Nature2012} which was also the first use of CE-HHG as a light source for an experiment, instead of CE-HHG being an experiment in itself.

Substantial further development has continued since 2011. 
Pupeza and coworkers at MPQ demonstrated the generation and efficient out-coupling of higher photon energies using nonlinear compression of the driving laser and the pierced-mirror output coupling method.\cite{Pupeza_NatPhot2013, Pupeza_PRL2014}
The MPQ team has also done substantial work on achieving shorter intra-fsEC pulse durations and cavity-mirror optimization.\cite{Carstens_Optica2016,Hogner_NJP2017,Holzberger_OptLett2015,Lilienfein_NatPhot2019,Lilienfein_OptLett2017}
Speeding up the gas flow to refresh the gas medium faster and reduce accumulated plasma, the JILA team reported generated HHG powers of > 1 mW/harmonic for the 11th harmonic of 1070 nm  in Xe (12.8 eV), but only transiently.\cite{Porat_NatPhot2018}
The JILA team has also explored non-collinear HHG for generating and outcoupling low-order harmonics.\cite{Zhang_PRL2020}

Now, nearly 20 years since the first demonstration, CE-HHG is a mature technology with well-understood tradeoffs impacting system design, which we discuss below in regards to the parameters listed in section \ref{sec:ideal}.
\subsection{Photon Energy Range}

Whereas single-pass HHG systems using high pulse energies have demonstrated usable harmonic fluxes out to $\sim$500 eV,\cite{Buades_ApplPhysRev2021,Saito_Optica2019,Summers_UltrafastScience2023,Teichmann_NatComm2016}
CE-HHG systems have so far only achieved out-coupled photon energies up to 140 eV.\cite{Saule_NatComm2019}
This so far limits CE-HHG-based beamlines for time-resolved photoemission to probing valence electrons or shallow inner-valence levels.\cite{Kunin_PRL2023,Bakalis_NanoLett2024,Na_Science2019,Na_PRB2020,Heinrich_NatComm2021}

\begin{figure}[t]
  \includegraphics[width = \columnwidth]{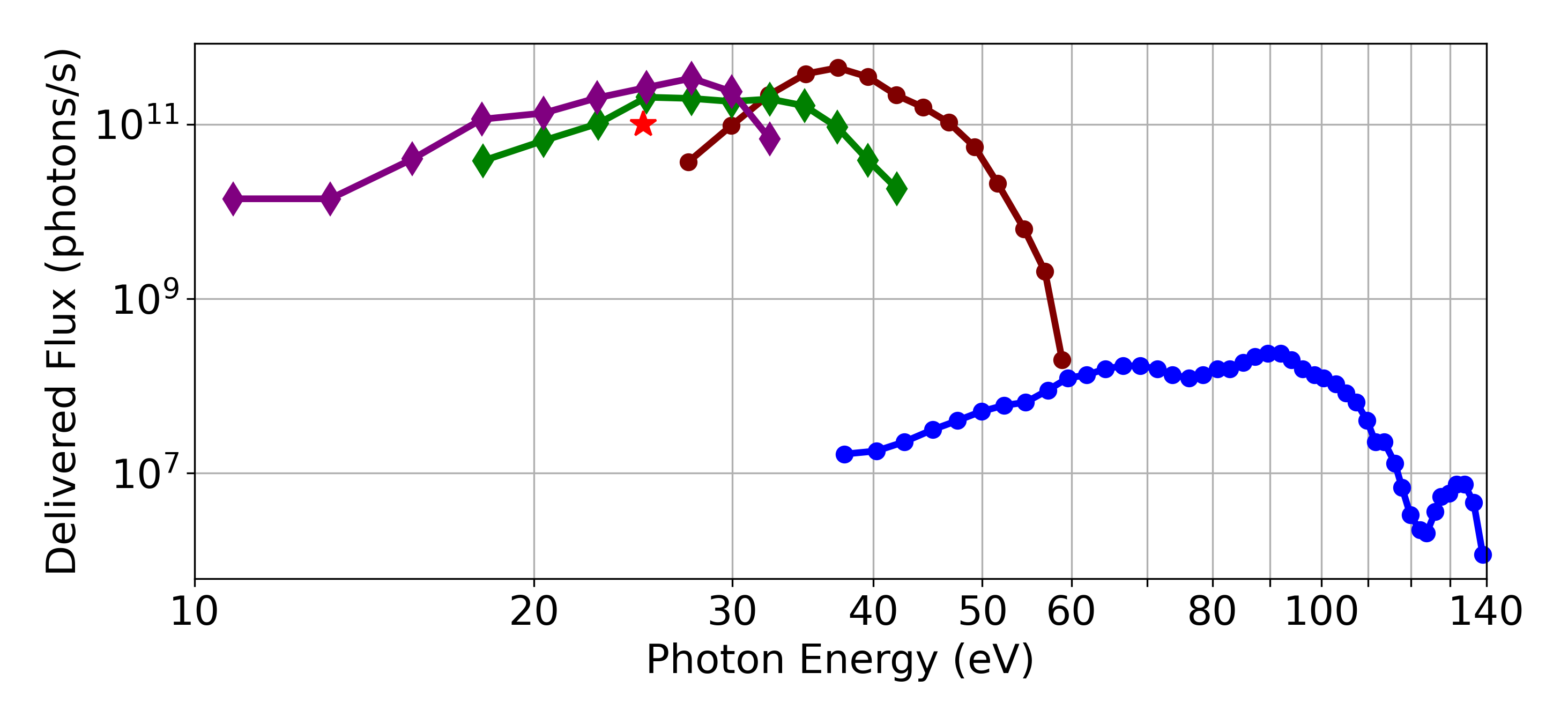}
  \caption{
    Reported fluxes for CE-HHG-based photoemission beamlines. 
    Purple Diamonds and green diamonds are fluxes delivered to the sample for CE-HHG in krypton and argon, respectively, using Brewster plate output coupling, reported by Corder et al. \cite{Corder_StructDyn2018} 
    Red star is the value reported by Mills et al. \cite{Mills_JPhysB2012}.
    Maroon and blue circles are the results reported by Saule et al. \cite{Saule_NatComm2019} for the flux coupled out of the cavity for argon and neon HHG, respectively, multiplied by 0.2 for a realistic XUV monochromator throughput.\cite{Frassetto_OptExp2011}
    In practice, Saule et al. achieved a lower beamline throughput for attosecond photoemission experiments.\cite{Saule_NatComm2019,Heinrich_NatComm2021}
  }
  \label{fig:fluxcompare}     
\end{figure}

The range of photon energies attainable via CE-HHG is determined by the intracavity pulse duration and peak intensity, phase matching, and output-coupling.
Figure \ref{fig:fluxcompare} shows spectra reported by CE-HHG-based photoemission beamlines from several research groups.
The values shown are fluxes \emph{at the sample}, i.e. after the losses of out-coupling and harmonic isolation.
Results from three different out-coupling methods are shown in figure \ref{fig:fluxcompare}.

The red star in figure \ref{fig:fluxcompare} was reported by Mills et al.\cite{Mills_RSI2019} using the grating output coupler method, in which a shallow diffraction grating is etched into the top surface of the cavity's dielectric mirror coatings.\cite{Yost_OptLett2008} 
This out-coupling method has the advantage of simultaneously coupling the harmonics out of the cavity and separating them spatially with a single optic.
However, it has the disadvantages of introducing significant pulse-front tilt on the outcoupled harmonics, which can cause an enlarged spot size for the harmonics on the sample.\cite{Mills_RSI2019} 
The large angular dispersion of the grating output coupler also presents challenges for alignment, since the $0^{th}$ order of the grating is not coupled out of the cavity and thus no fundamental or low-order harmonics are collinear with the XUV light, and designing a beamline that can collect all the harmonics and tune which one is delivered to the sample dynamically is very challenging.
Despite these challenges, Mills et al. have demonstrated an extremely stable light source with this scheme, and achieved $10^{11}$ photons per second at $h\nu = 25$ eV delivered to the sample in a $200 \times 400$ $\upmu$m$^2$ spot.

The diamonds in figure \ref{fig:fluxcompare} are from Corder et al.,\cite{Corder_StructDyn2018} where harmonics were coupled out of the cavity using a sapphire wafer at Brewster's angle.
This method has the advantage that all harmonics are coupled out collinearly, and residual fundamental or low-order harmonic light reflected from the Brewster plate can be used as an alignment guide for the rest of the beamline.
It has the disadvantages that the Brewster plate adds dispersion, the photon energy is limited by the reflectivity of the sapphire to less than $\sim$50 eV,\cite{Zhao_Thesis2019} and the nonlinear response of the Brewster plate can also distort the intracavity pulse, limiting the power enhancement.\cite{Corder_SPIE2018}
Using a pulse-preserving monochromator,\cite{Frassetto_OptExp2011} this system has achieved ${\sim}10^{11}$ photons per second delivered to the sample in $24 \times 16$ $\upmu$m$^2$ spot,\cite{Kunin_PRL2023} tunable over the range of 10-40 eV.

The circles in figure \ref{fig:fluxcompare} are from Saule et al.,\cite{Saule_NatComm2019} where harmonics were coupled out of the cavity using the pierced-mirror scheme, where the harmonics are coupled out of a small hole in the cavity mirror immediately following the HHG medium.
This output coupling method has the advantages of high output coupling efficiency for higher harmonics with lower divergence and collinear fundamental and low-order harmonics for alignment.
However, it has the disadvantages of substantial loss for the fundamental (limiting the attainable cavity finesse and power enhancement), poor outcoupling for lower-order harmonics less than $\sim$40 eV, and difficult manufacturing.
The high delivered photon energies and collinear and dispersion-less nature of the output coupling was critical for Saule et al. to demonstrate high-performance RABBITT measurements and resolve attosecond time delays in surface photoemission.\cite{Saule_NatComm2019,Heinrich_NatComm2021}

\subsection{Energy and time resolution}

Cavity-enhancement of frequency combs involves the simultaneous resonance of many comb teeth with many cavity resonances, typically order $10^4-10^5$.
While the frequency comb teeth are rigorously evenly spaced by $f_{\text{rep}}$,\cite
{Udem_OptLett1999} the mode spacing of the enhancement cavity varies due to intracavity group delay dispersion (GDD).
This sets a practical limit on the bandwidth that can be simultaneously resonantly enhanced, given by\cite{Silfies_OptLett2020}
\begin{equation}\label{eqn:dw}
  \Delta \omega < \sqrt{ \frac{8 \pi}{\mathcal{F} \phi_2}} \;,
\end{equation}
where $\Delta \omega$ is the FWHM intracavity bandwidth and $\phi_2$ is the cavity's net GDD.
Equation (\ref{eqn:dw}) implies that to achieve a 100 fs circulating intracavity pulse with $\mathcal{F} = 500$, one must limit the intracavity GDD to less than 60 fs$^2$, which is quite manageable even using simple mirrors with a ``quarter-wave stack" design, and fsEC operation with intracavity pulse durations in the range of 40-200 fs has been implemented in CE-HHG systems for photoemission. 
However, to achieve a few-cycle 10 fs circulating pulse, suitable for driving isolated attosecond pulse generation,\cite{Brabec_RMP2000} would require a very small GDD of less than 0.6 fs$^2$, which is currently not feasible due to limitations in the precision with which mirror coatings can be made. 

As mentioned in section \ref{sec:HHGphysics}, it is very easy in single-pass HHG systems to generate spectrally broad harmonics, much broader than the transform limit, due to dynamical plasma phase shifts on the fundamental beam causing loss of phase matching (ionization gating) and also via plasma blueshifting of the fundamental.\cite{Wang_NatCom2015}
However, as discussed by Corder et al.,\cite{Corder_StructDyn2018} limitations on the dynamic plasma phase shifts in CE-HHG limit this effect naturally, since (singe-pass) plasma phase shifts on the fundamental are limited in CE-HHG to order $\pi/\mathcal{F}$, whereas loss of phase matching for the $q^{th}$ harmonic occurs for phase shifts on the order of $\pi/q$, typically an order of magnitude larger.
CE-HHG systems then inherently are forced into the regime where distortions on the fundamental pulse during HHG are small, and the harmonics are closer to transform limited than is often attained in single-pass systems.

\subsection{Repetition Rate}

\begin{figure}[t]
  \includegraphics{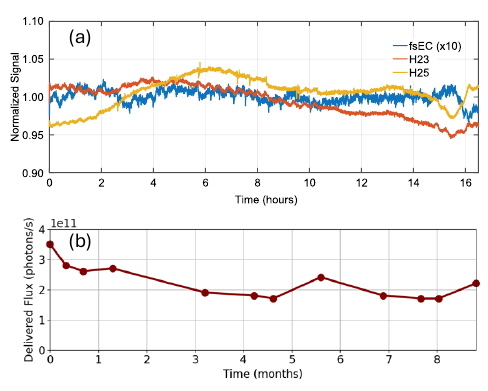}
  \caption{
  Short and long-term stability of working CE-HHG sources. 
  (a) Fluctuations less than $\sim$5\% over 16 hours of continuous operation reported by Mills et al.\cite{Mills_RSI2019}
  Reproduced with permission.
  (b) Flux delivered to the endstation over an 8 month period at Stony Brook with no venting or alignment of the CE-HHG source or XUV beamline.\cite{Zhao_Thesis2019}
  }
\label{fig:stab}
\end{figure}

CE-HHG has so far been reported at repetition rates between 250 MHz\cite{Carstens_Optica2016} and 10 MHz.\cite{Ozawa_OptExp2015,Ozawa_PRL2008}
Unlike single-pass HHG, in many ways it is actually easier to implement CE-HHG at higher repetition rates.
This is because the fsEC cavity length is given by $c/f_{\text{rep}}$ and the fsEC linewidth is given by $\Delta \nu = f_{\text{rep}}/\mathcal{F}$, such that as the $f_{\text{rep}}$ gets lower the fsEC gets longer and has narrower linewidth.
For example, a repetition rate of 10 MHz and a finesse of 500 gives an unwieldy cavity length of 30 m and a linewidth of only 20 kHz.
A 30 m cavity must either occupy a large space\cite{Ozawa_OptExp2015} or incorporate many folding mirrors, leading to more difficulties managing the net cavity dispersion and loss.\cite{Ozawa_PRL2008} 
Coupling the comb to the cavity with 20 kHz linewidth also requires a narrow-linewidth frequency comb and special care in the laser stabilization.\cite{Ozawa_OptExp2015, Saule_NatComm2019, Ozawa_PRL2008}
Generating frequency combs at lower repetition rates also presents additional challenges, in general.\cite{Canella_Optica2024}

However, although it is in some ways more challenging to achieve CE-HHG at lower repetition rates, several studies have indicated that lower repetition rates can achieve a higher HHG conversion efficiency due to reduced re-use of the gas medium.\cite{Porat_NatPhot2018,Saule_APLPhot2018}
The precise mechanism for this effect remains unclear.
Porat et al.\cite{Porat_NatPhot2018} implicated plasma-induced loss of phase matching for the reduced efficiency, but the plasma phase shifts they observed on the fundamental are still much smaller than the relevant scale of $\pi/q$.
Saule et al. have postulated that the reduced efficiency is instead due to depleted neutral atom density at the cavity focus and a reduced generation volume causing a larger divergence of the generated XUV light.
In practice, similar average XUV photon fluxes have been achieved by CE-HHG systems operating at very different repetition rates, with higher conversion efficiencies achieved at lower repetition rates, but larger circulating average powers achieved at higher repetition rates.

\begin{figure*}[t!]
  \includegraphics[width = 0.8\textwidth]{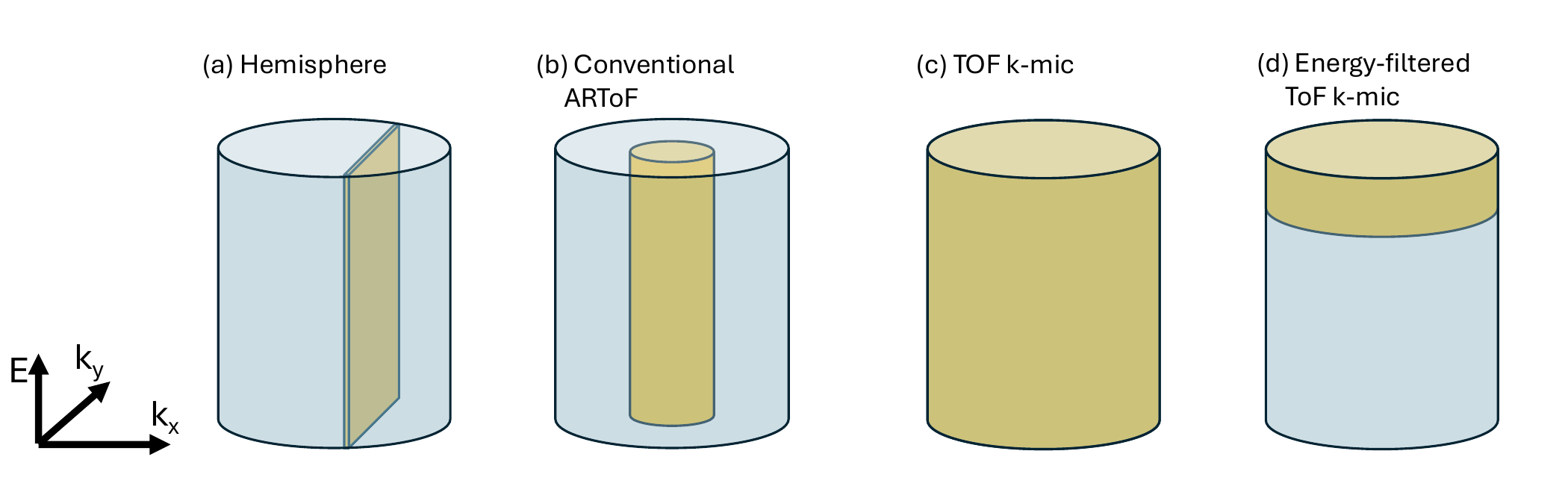}
  \caption{
    Illustration of collected signals for different types of photoelectron analyzers for ARPES.
    (a) A hemispherical analyzer collects only a small slice (beige) of the full 3D $(k_x, k_y, E)$ distribution of photoelectrons emitted from the sample.
    (b) A conventional angle-resolving time-of-flight analyzer, with no electric field between the sample and the first electron lens, collects all energies in parallel but usually has narrow angular acceptance.
    (c) The ToF k-mic collects the full $2\pi$ emission of the electrons from the sample, resolving their energy and momentum in parallel.
    (d) For time-resolved photoemission experiments using ToF k-mic, it is undesirable to attenuate the signal to avoid detector saturation and it is thus advantageous to filter the electron distribution that makes it to the detector to a narrow energy distribution. 
    }
  \label{fig:cubeslice}
\end{figure*}

\subsection{Stability}

Several groups now have demonstrated medium-term and long-term stability for CE-HHG.
Figure \ref{fig:stab}a) shows HHG fluxes for two harmonics recorded continuously for 16 hours by Mills et al.\cite{Mills_RSI2019}
Continuous frequency locking between the comb and cavity is maintained over this period with power drift of only 0.5\% for the fundamental and $\sim$5\% for the harmonics.
Zhao\cite{Zhao_Thesis2019} documented the stability of the Stony Brook system over an 8-month period, with no alignment of the in-vacuum optics of the cavity or beamline, and observed consistent flux of over $10^{11}$ photons/second delivered to experiments over this extended period, as shown in figure \ref{fig:stab}.
In another extended measurement campaign, we have now run the Stony Brook beamline for more than 2 years without venting the cavity or beamline vacuum systems or aligning any in-vacuum optics.

\subsection{Tunable Pump Pulses}\label{sec:CE-HHGpractice:pump}

One of the largest challenges in applying CE-HHG to time-resolved photoemission experiments has been the implementation of tunable pump pulses for exciting specific initial states in the sample.
Similar to HHG, the challenge lies in performing nonlinear-optical frequency conversion at high repetition rate starting with low fundamental pulse energies less than 1 $\upmu$J.
Indeed, this has been a challenge to the frequency comb community as a whole, and has thus been an area of sustained intense research over the past decade.\cite{Schliesser_NatPhot2012, Kobayashi_JOpt2015, Timmers_Optica2018, Seidel_SciAdv2018, Chen_ApplPhysB2019, Lesko_NatPhot2021, Rutledge_JOSAB2021, Catanese_OptLett2020,Wu_NatPhot2024}

Considering the pulse energy requirements discussed in section \ref{sec:ideal:pump}, for repetition rates of $\sim$100 MHz, average pump powers of 100 mW to 1 W are required for tunable pump sources, even for low-fluence experiments, which is at the state-of-the art in tunable frequency comb development.
Thus, so far CE-HHG-based photoemission experiments with tunable pump pulses have not been reported, but it is anticipated that this will be achieved soon.
The Allison group has recently developed an optical parametric amplifier (OPA) setup with arbitrary (0-$f_{\text{rep}}$) repetition rate and sufficient power.\cite{Wahl_Thesis2024}
For fixed repetition rate, optical parametric oscillators (OPOs) have also been demonstrated in this power range\cite{Chen_ApplPhysB2019, Steinle_OptLett2016, Ghotbi_OptLett2006} and there are commercially available options.

\section{Coupling CE-HHG with Photoelectron Analyzers}\label{sec:analyzers}

As discussed in the preceding sections, CE-HHG-based beamlines can deliver ${\sim}10^{11}$ photons/second to an endstation with repetition rate of $\sim$100 MHz.
Assuming a total quantum efficiency, including secondary electrons, of 10\%,\cite{Hufner_book2003} this corresponds to a sample current of ${\sim}1$ nA, comparable to that used in high-resolution synchrotron photoemission experiments,\cite{Corder_SPIE2018,Rotenberg_JSyncRad2014}
while operating with only ${\sim}100$ photoelectrons per pulse.
For the highest-performance experiments looking at very small pump-induced changes in the energy-momentum distributions, this large photocurrent should be collected and parsed with maximum efficiency.

The traditional workhorse of photoemission experiments has been the hemispherical analyzer, which records a narrow slice of the full 3D $(k_x, k_y, E)$ distribution emitted from the sample, as illustrated in figure \ref{fig:cubeslice}(a).
Hemispherical analyzers can have very high energy resolution even better than 1 meV, can work with pulsed or continuous light sources, and detector saturation is rarely a concern due to the strong filtering of the electrons outside the analyzer's pass band.
However, in order to record the full distribution, either the sample must be rotated or the analyzer must be rotated (or some combination of both) and the experiment repeated many times.

Angle-resolving time-of-flight (ToF) detectors can record a range of momenta and energy in parallel, and this can dramatically improve the acquisition speed.\cite{Sie_NatComm2019}
However, conventional ToF detectors collect only a limited angular range of the photoemitted electrons, as illustrated in figure \ref{fig:cubeslice}(b), and thus repeating measurements for different sample orientations is still generally necessary to record the full distribution.

In the ToF k-mic method, the front end of the analyzer is a photoemission electron microscope (PEEM), where the sample itself forms part of a cathode lens.\cite{Bruche_KolloidZ1942,Recknagl_ZPhys1943}
Electrons are emitted into a field of typically several kV/mm, and this strong electric field ensures that all photoelectrons emitted in the full $2\pi$ solid angle are pulled into the microscope column, even for electron kinetic energies $\gtrsim$100 eV.
Subsequent imaging of the back focal plane of the cathode lens to the detector gives direct momentum-space images on the detector.
Recording data with a 2D + time-resolving detector, such as a delay-line detector, gives the the momentum coordinates and the electron time-of-flight, which can then be calibrated to give the full $(k_x, k_y, E)$ distribution of photoemitted electrons in parallel, as illustrated in figure \ref{fig:cubeslice}(c).
The PEEM front end of the analyzer can also be used to select a micron-scale region of interest on the sample to record data from, enabling micro-ARPES measurements without needing a tightly focused XUV beam.
As described in the introduction, in just a few years this new analyzer technology has led to dramatic advances in what is possible in data-rate-limited time-resolved photoemission experiments, as well as for challenging ground-state ARPES measurements, such as hard x-ray ARPES (HAXPES).\cite{Medjanik_JSynchRad2019, Elmers_ACSNano2020,Schoenhense_JVST2022}

Integrating the ToF k-mic with the high repetition rate of CE-HHG presents several challenges.
The first challenge was discussed in section \ref{sec:ideal:reprate}: finite ToF resolution of the detector can degrade the energy resolution of the experiment via equation (\ref{eqn:dToF}).
A related problem is saturation of the detector.
A uniformly exposed microchannel plate can handle at most ${\sim}10^{7}$ events per second, due to the finite recharge time of the large-capacitance micro-channels, much smaller than the ${\sim}10^{10}$ photoelectrons per second that can be produced cleanly at the sample using a CE-HHG-based beamline.
Given the small signals involved in low-fluence time-resolved photoemission experiments (for example as discussed in section \ref{sec:ideal:stability}), attenuating the light source to avoid saturation is undesirable.

In the absence of a non-MCP-based time and position sensitive detector technology with sufficient resolution, the solution to both these problems is to filter the photoelectron distribution, such that a narrower range $\Delta E$ is detected, as illustrated in figure \ref{fig:cubeslice}(d).
While this does reduce the information recorded in the experiment, for most pump/probe experiments the region of interest is in a narrow range of energies at and above the Fermi level, so the impact is minimal.
Limiting $\Delta E$ both improves the ToF-limited energy resolution via equation (\ref{eqn:dToF}) and reduces the number of electrons reaching the detector, mitigating saturation.

\begin{figure}
  \includegraphics[width = 3.0 in]{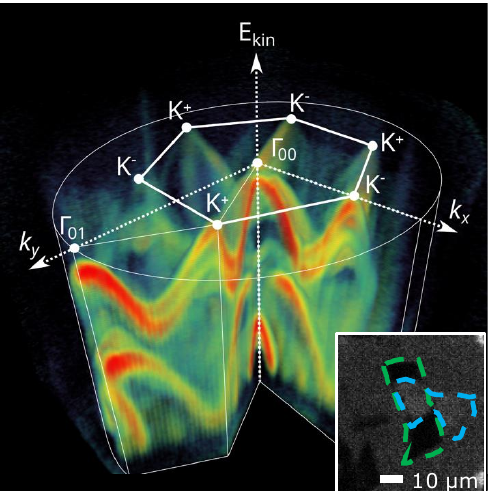}
  \caption{
  High-speed 3D data acquisition with a 61 MHz CE-HHG source and a ToF momentum microscope. 
  Occupied bands of a small monolayer WS$_2$ sample, recorded with $h\nu = 27.6$ eV. 
  Inset: Real-space PEEM image of the sample with monolayer WS$_2$ regions and hBN support outlined in blue and green, respectively.
  A field aperture in the momentum microscope selects only electrons from the $\sim$10 $\upmu$m region where the WS$_2$ and hBN flakes overlap.
  Despite the very small monolayer sample, the full 3D $(k_x, k_y, E)$ band structure is recorded in $\sim$ 1 minute.
  }
  \label{fig:3DWS2}
\end{figure}

\begin{figure}
  \includegraphics{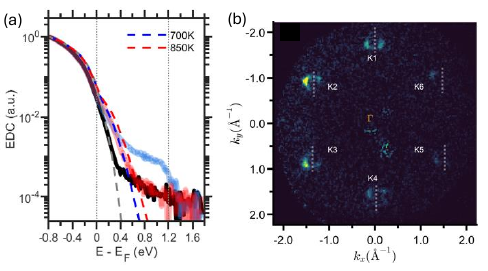}
  \caption{Example of tr-ARPES with high dynamic range using CE-HHG and ToF k-mic: Non-themal pseudospin distributions in graphene. 
  (a) Energy distribution curves produced upon exciting graphene with 517 nm ($h\nu = 2.4$ eV) light with a fluence of 45 $\upmu$J/cm$^2$, adapted from Bakalis et al.\cite{Bakalis_NanoLett2024}
  The blue curve corresponds to maximal pump/probe overlap ($\Delta t=0$), the red curve is for $\Delta t = 500$ fs, and the black curve is for the unexcited sample ($\Delta t = -1$ ps)
  Superimposed are EDCs assuming thermalized electrons, calculated via the graphene density of states, the photoemission matrix elements from tight-binding theory, and a Fermi-Dirac distribution at 700 K (blue dashed) and 850 K (red dashed).\cite{Bakalis_NanoLett2024}
  (b) Momentum-space distributions for electrons between 1.05 and 1.21 eV above the Fermi level for $\Delta t = 0$ and $s$-polarized pump excitation.
  A strong pseudospin polarization is observed, with corresponding nodes in the momentum-space distribution along the light polarization direction.
  Adapted with permission from reference \citenum{Bakalis_NanoLett2024}.
  Copyright 2024, American Chemical Society.
  }
  \label{fig:graphene}
\end{figure}

There are several methods one can use to filter the energy distribution.
At Stony Brook, we have implemented a simple method using a retarding field between two grids placed right before the detector.
This simple method achieves good suppression of low-energy electrons, but does degrade the energy and momentum resolution somewhat, especially close to the cutoff energy.\cite{Chernov_UP2022,Bakalis_Thesis2024}
Figure \ref{fig:3DWS2} shows an example of ARPES data taken from a small monolayer WS$_2$ flake, using the high-pass filter grids to pass the highest 4 eV photoelectrons to the detector.
With no attenuation of the XUV flux incident on the sample, the full 3D band structure of the monolayer is recorded with high quality in under 1 minute for a few-micron-sized spot selection.
Figure \ref{fig:graphene} from Bakalis et al.\cite{Bakalis_NanoLett2024} shows another example of using CE-HHG and ToF k-mic to achieve high dynamic range in time-resolved ARPES measurements.
In this example, the high-pass filter is set to reject most of the valence band signal, such that the detector sensitivity is preserved to observe a small nonthermal electron distribution in monolayer graphene created by optical excitation at low fluence.
Achieving ${\sim}10^4$ dynamic range in the EDC as shown in figure \ref{fig:graphene}(a) also requires excellent suppression of adjacent harmonics by the XUV beamline, in this case achieved using a pulse-preserving monochromator.\cite{Frassetto_OptExp2011}

\begin{figure}
  \includegraphics[width = \columnwidth]{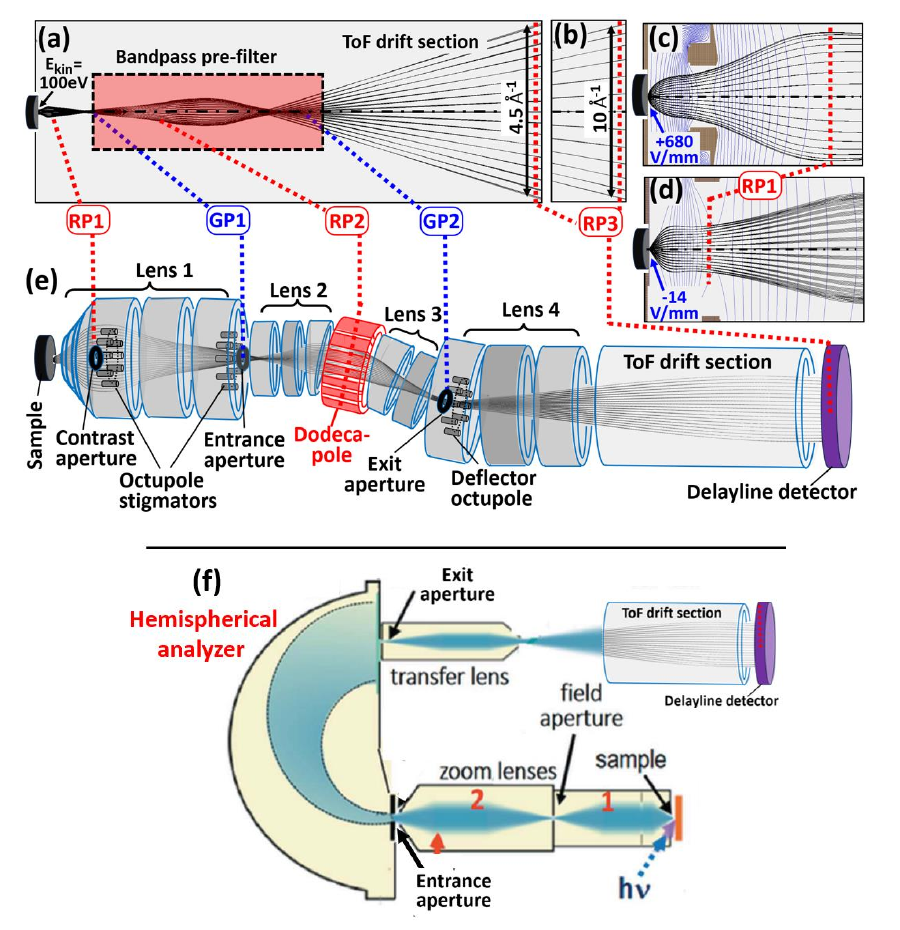}
  \caption{Next-generation momentum microscopes with integrated bandpass prefilter and front lens with space charge suppression option. (a) Ray tracing for a kinetic energy of 100 eV in repeller mode; reciprocal and Gaussian planes are denoted as RP1-3 and GP1-2, respectively. The red shaded area indicates the region where a bandpass filters can be inserted. Details of the trajectories for the gap-lens mode in the final k-image (b), near the sample (c), and for the repeller mode (d). (b,c) Polar angle intervals 0-90$^{\circ}$; (a,d) 0-27$^{\circ}$. (e) Lens system with superimposed trajectory calculation for an instrument with integrated dodecapole bandpass filter. (f) Scheme of a hemispherical analyzer as a bandpass filter followed by a ToF analyzer.}
  \label{fig:NextGen}
\end{figure}

More sophisticated high-pass and band-pass filtering techniques can give higher performance.
One obvious approach is to implement a dispersive element as a bandpass filter in the microscope column. 
Here we discuss three implementations of this basic idea. 
First, high-pass filtering can be done using the first lens of the k-mic.
Figure \ref{fig:NextGen}(a) shows electron trajectories in the so-called repeller mode of the front lens for an initial kinetic energy of 100 eV. 
A retarding field (here -14 V/mm) redirects all slow electrons back to the surface within a few 100 \textmu m from the sample surface. 
This elimination of the slow electron background provides a way to reduce the bandwidth of the electrons entering the ToF section. However, the cost of this mode is a reduction in the observed polar angle range to 0 - 27$^{\circ}$ (k-field diameter 4.5 \AA$^{-1}$). 
In the so-called gap lens mode shown in figure \ref{fig:NextGen}(b), with an accelerating field of 680 V/mm at the sample the full range of 0 - 90$^{\circ}$ (diameter 10 \AA$^{-1}$) is imaged. 
The trajectories in the region close to the sample and the first k-image in the reciprocal plane RP1 are shown in figures \ref{fig:NextGen}(c) and \ref{fig:NextGen}(d) for the gap lens and repeller modes, respectively.  

Second, band-pass filtering can be done using the microscope column itself or additional filters.
Figure \ref{fig:NextGen}(e) shows the layout of a momentum microscope with an integrated compact band-pass filter consisting of an electrostatic dodecapole with asymmetric spacing of the 12 electrodes. 
The system consists of 4 lens groups, 3 piezomotor adjustable aperture arrays (contrast, entrance, and exit apertures), and 3 octupole deflectors/stigmators. 
The core element is the dodecapole framed by two lens groups, lenses 2 and 3, and the entrance and exit apertures. 
Lens 1 produces a 1st Gaussian (real space) image in the entrance aperture, lens 2 a k-image in the center of the dodecapole and lens 3 a 2nd Gaussian image in the exit aperture. 
In the first prototype\cite{Tkach_JSynchRad2024} the deflection angle of the dodecapole is only $4^{\circ}$. The radial axis is stretched, so the angle appears larger in the figure. 
The octupole behind the exit aperture deflects the beam by $-4^{\circ}$, parallel to the initial optical axis. Finally, lens 4 focuses the final k-image on the detector via the ToF drift section. This compact instrument is in permanent use at the hard X-ray beamline P22 at PETRA III (DESY, Hamburg, Germany).

Figure \ref{fig:NextGen}(f) shows the third and most sophisticated way of energy band pre-selection, a large hemispherical dispersive analyzer (radius 225 mm) followed by a ToF analyzer. The latter divides the selected band into a stack of momentum patterns with different kinetic energies. This hybrid approach has been achieved using laser radiation at 80 MHz pulse rate\cite{Schoenhense_JSynchRad2021} and soft X-ray synchrotron radiation at 500 MHz pulse rate at the Diamond Light Source (Didcot, UK),\cite{Schmitt_arXiv2024} and is well suited for tr-ARPES measurements using CE-HHG. 

Similar to the development of cryogenic electron microscopy of biological samples, new detector paradigms can change the current landscape significantly,\cite{Cichocka_JAppCryst2018,Faruqi_QuartRevBiophys2011,Milne_FEBS2013} and offer the next revolution in high-sensitivity photoemission measurements.
Specifically, detectors based on ``direct electron detection," in which electrons impinge directly onto a pixelated silicon detector, offer nearly 100\% quantum efficiency.
Pixelated silicon detectors with sub-ns time resolution are being developed for x-ray free electron laser experiments\cite{Markovic_SLACPub2017} and high-energy physics,\cite{Giacomini_NIMA2019,Markvoic_IEEE2018} and these architectures can be adapted to direct electron detection.
It is clear from these efforts that a detector with sub-30 ps resolution capable of handling 1 GHz count rates is attainable, and this would lead to an improvement of high-performance photoemission experiments by several orders of magnitude, allowing for the use of the full photon flux generated by CE-HHG-based beamlines.
This advancement will also likely be of key importance for facilitating high data rate spin-, time- and angle-resolved photoemission studies.


\section{Conclusions and Outlook}\label{sec:outlook}

Leveraging the technical developments in the last 20 years since its introduction, cavity-enhanced HHG has now emerged as a powerful, stable, and versatile light source for demanding time-resolved photoemission experiments. 
The $\sim$ 100 MHz repetition rate directly addresses the persistent challenge of mitigating or eliminating vacuum space charge effects, and the attainable photon flux and range of XUV photon energies achievable with long-term stability provide the key elements necessary for state-of-the-art time-resolved photoemission experiments, especially when coupled with advanced ToF k-mic-based detection schemes.
An emerging related technology is intra-oscillator HHG, where HHG is performed inside the cavity of a thin-disk laser oscillator with high circulating average power.\cite{Labaye_OptLett2017}
Recently, this approach has shown substantial improvements in flux and out-coupled photon energies,\cite{Drs_OptExp2024} and may soon also be applied to high-performance photoemission experiments.

From our perspective, the remaining areas for further technical development in the application of CE-HHG to time-resolved photoemission lie in the advancement of: 
(1) nonlinear optical frequency conversion schemes for generating widely-tunable pump photon energies with 100s of mW of average power at $\sim$100 MHz as well as the ability to select lower pump photon repetition rates;
(2) circumventing the limitations imposed by MCP-based electron detection using new detector paradigms to eliminate the need for energy filtering the photoelectron signal;
(3) extending the attainable XUV photon energy markedly beyond 100 eV to gain access to atomic core levels;
(4) integrating CE-HHG with spin-resolving momentum microscopes to achieve spin- and time-resolved ARPES experiments with low fluence excitation.
In section \ref{sec:CE-HHGpractice:pump}, we surveyed the latest technical breakthroughs regarding (1), namely frequency combs with tunable pump wavelength and tunable pump repetition rate, and their application to CE-HHG-based photoemission experiments is on the near horizon.\cite{Wahl_Thesis2024}
Item (2) was discussed in section \ref{sec:analyzers} in the context of energy filtering.
Below we discuss frontiers (3) and (4).

As identified in section \ref{sec:CE-HHGpractice}, CE-HHG systems have now demonstrated outcoupled photon energies in excess of 100 eV,\cite{Pupeza_NatPhot2013, Carstens_Optica2016, Saule_NatComm2019} 
though it is desirable to extend this limit to higher photon energies to probe transient chemical shifts of atomic core levels.
The cutoff for the maximum attainable photon energy from the HHG process, equation (\ref{eqn:cutoff}), extends to higher emitted energies for longer wavelength driving lasers.
The high-power frequency comb systems driving the CE-HHG beamlines producing the results described in section \ref{sec:review} were all based on Yb-doped fiber amplification stages operating at 1030 nm to achieve scaling to the necessary high powers.
Recently, thulium (Tm)-doped fiber frequency combs near 2 \textmu m wavelength have also demonstrated robust operation at high average powers,\cite{Gaida_OptLett2018_2}
presenting a promising way forward for future CE-HHG beamlines driven by 2 $\upmu$m lasers.
This can in principle extend the attainable HHG cutoff into the soft x-ray, allowing access to core-level transitions such as the carbon K-edge.
However phase matching the generation of such high photon energies in a CE-HHG system, which typically use tight focusing, will likely be challenging.

As discussed in sections \ref{sec:introduction} and \ref{sec:analyzers}, the ToF k-mic method has quickly been adopted by many tr-ARPES efforts and has also been applied to challenging hard x-ray ARPES measurements.
An additional area where ToF k-mic can provide large improvements in photoemission measurements is in spin-resolved ARPES.
The low figure-of-merit (FoM) of classical single-channel spin detectors (FoM about $10^{-2}$) is prohibitive for spin-resolved band mapping of many materials of current interest, 
and the required acquisition time precludes spin mapping of chemically reactive surfaces, even in ultra-high vacuum. 
In the 2010's, this motivated the development of parallel imaging ARPES analyzers and imaging spin filters in Halle and Mainz, Germany,\cite{Kolbe_PRL2011,Tusche_APL2011} and their integration with momentum microscopy.
The advantage over conventional spin-resolved experiments is that spin-filtered k-mic directly observes a large field-of-view in momentum space in parallel, without any scanning or sample rotation. 
A benchmark example is the spin-texture measurement of the Au(111) Shockley state.\cite{Tusche_Ultramicroscopy2015}
In particular, the ToF k-mic approach is advantageous for spin-resolved experiments because a whole energy band is recorded in parallel, which greatly increases the recording efficiency. 
Suga and Tusche performed a quantitative comparison of different types of spin detectors (Fig. 28 in reference \citenum{Suga_JElecSpec2015}), where the 3D recording ToF k-mic solution outperforms the current single-channel detectors by 4-5 orders of magnitude.
Spin-resolving ToF k-mic has now been used in a number of different experiments, including band-structure tuning of Heusler compounds,\cite{Chernov_PRB2021} and spin mapping of surface and bulk Rashba states in $\alpha$-GeTe(111) films\cite{Elmers_PRB2016} and the Tamm state of Re(0001).\cite{Elmers_PhysRevResearch2020}
From a scientific point of view, the overarching long-term goal of this work has been to realize the so-called \emph{complete experiment}, a concept borrowed from atomic photoemission.\cite{Kessler_CommentsAMO1981} 

To date, tr-ARPES measurements with spin resolution using ToF k-mic have not been reported, likely due to the fact that despite the gains of parallel detection, the reflectivity of imaging spin filters is still low, and the signal levels for spin-resolved measurements are reduced by between 1 and 2 orders of magnitude compared to spin-integrated measurements.
In particular, this implies that for low-fluence experiments, such as TMD exciton imaging as discussed in section \ref{sec:ideal:stability}, spin-resolved tr-ARPES measurements are still prohibitively difficult using single-pass HHG systems, but feasible with CE-HHG beamlines operating with 1-2 orders of magnitude higher repetition rate.
Clean momentum-space imaging of excited quasi-particle states across the full Brillouin zone, including the spin texture, would represent the complete experiment for excited states, and a paradigm shift for ultrafast spectroscopy in general.

Although the primary scientific focus of high-performance photoemission measurements enabled by CE-HHG thus far has been the study of dynamics in perfect inorganic crystals,
there are also significant advantages for future studies in complex chemical systems and environments.
The high repetition rate, and thus the lower number of photons per pulse, is advantageous for studies of more fragile crystalline organic semiconductors that are highly susceptible to degradation and damage under intense laser irradiation.
Moreover, the high sensitivity afforded by photoemission at these repetition rates can enable significant advances in liquid-phase time-resolved photoemission studies of weak solute signals in dilute solutions, where the signal from dissolved solutes present in flowing liquid microjets is generally 4-6 orders of magnitude smaller than that of the solvent.\cite{Suzuki_JChemPhys2019}
Coupled with the recent development of the zero-field operation mode of the ToF k-mic that enables efficient electron collection while maintaining zero electric field at the surface of the sample, this could in principle also enable advanced angle-resolved photoemission studies of flat liquid jets.\cite{Yamamoto_BChemSocJap2023}
The extension of CE-HHG to higher photon energies (and thus, higher ejected photoelectron kinetic energies) combined with the zero-field ToF k-mic operation mode\cite{Tkach_arXiv2024} also brings the possibility to extend time-resolved photoemission studies to the study of surface chemical reactions in elevated, near-ambient pressure environments surrounding the sample.

The range of exciting science now possible, and soon to be enabled, with high-performance time-resolved photoemission beamlines spans the fields of physics, chemistry, materials science, and electrical engineering.
Just as photoemission measurements have long been used as the "gold standard" for ground-state electronic structure, we expect that with the technical developments described herein, photoemission measurements can also soon become the preeminent method for characterizing excited states as well, with routine application to a wide variety of systems.

\section*{Acknowledgements}
This work has been made possible via steady support from the U.S. Dept. of Energy and the U.S. Air Force Office of Scientific Research, currently via award numbers DE-SC0022004 and FA9550-20-1-0259, respectively.
G. Sch\"onhense acknowledges support from BMBF (Projects 05K22UM1, 05K22UM2) and DFG (German Research Foundation) through TRR 173-268565370 Spin+X (project A02) and Project No.Scho341/16-1

\section*{References}

\bibliography{APLPhotBib.bib}

\providecommand{\noopsort}[1]{}\providecommand{\singleletter}[1]{#1}%
\begin{thebibliography}{163}%
\makeatletter
\providecommand \@ifxundefined [1]{%
 \@ifx{#1\undefined}
}%
\providecommand \@ifnum [1]{%
 \ifnum #1\expandafter \@firstoftwo
 \else \expandafter \@secondoftwo
 \fi
}%
\providecommand \@ifx [1]{%
 \ifx #1\expandafter \@firstoftwo
 \else \expandafter \@secondoftwo
 \fi
}%
\providecommand \natexlab [1]{#1}%
\providecommand \enquote  [1]{``#1''}%
\providecommand \bibnamefont  [1]{#1}%
\providecommand \bibfnamefont [1]{#1}%
\providecommand \citenamefont [1]{#1}%
\providecommand \href@noop [0]{\@secondoftwo}%
\providecommand \href [0]{\begingroup \@sanitize@url \@href}%
\providecommand \@href[1]{\@@startlink{#1}\@@href}%
\providecommand \@@href[1]{\endgroup#1\@@endlink}%
\providecommand \@sanitize@url [0]{\catcode `\\12\catcode `\$12\catcode
  `\&12\catcode `\#12\catcode `\^12\catcode `\_12\catcode `\%12\relax}%
\providecommand \@@startlink[1]{}%
\providecommand \@@endlink[0]{}%
\providecommand \url  [0]{\begingroup\@sanitize@url \@url }%
\providecommand \@url [1]{\endgroup\@href {#1}{\urlprefix }}%
\providecommand \urlprefix  [0]{URL }%
\providecommand \Eprint [0]{\href }%
\providecommand \doibase [0]{http://dx.doi.org/}%
\providecommand \selectlanguage [0]{\@gobble}%
\providecommand \bibinfo  [0]{\@secondoftwo}%
\providecommand \bibfield  [0]{\@secondoftwo}%
\providecommand \translation [1]{[#1]}%
\providecommand \BibitemOpen [0]{}%
\providecommand \bibitemStop [0]{}%
\providecommand \bibitemNoStop [0]{.\EOS\space}%
\providecommand \EOS [0]{\spacefactor3000\relax}%
\providecommand \BibitemShut  [1]{\csname bibitem#1\endcsname}%
\let\auto@bib@innerbib\@empty
\bibitem [{\citenamefont {Pupeza}\ \emph {et~al.}(2021)\citenamefont {Pupeza},
  \citenamefont {Zhang}, \citenamefont {H{\"o}gner},\ and\ \citenamefont
  {Ye}}]{Pupeza_NatPhot2021}%
  \BibitemOpen
  \bibfield  {author} {\bibinfo {author} {\bibfnamefont {I.}~\bibnamefont
  {Pupeza}}, \bibinfo {author} {\bibfnamefont {C.}~\bibnamefont {Zhang}},
  \bibinfo {author} {\bibfnamefont {M.}~\bibnamefont {H{\"o}gner}}, \ and\
  \bibinfo {author} {\bibfnamefont {J.}~\bibnamefont {Ye}},\ }\bibfield
  {title} {\enquote {\bibinfo {title} {Extreme-ultraviolet frequency combs for
  precision metrology and attosecond science},}\ }\href {\doibase
  10.1038/s41566-020-00741-3} {\bibfield  {journal} {\bibinfo  {journal}
  {Nature Photonics}\ }\textbf {\bibinfo {volume} {15}},\ \bibinfo {pages}
  {175--186} (\bibinfo {year} {2021})}\BibitemShut {NoStop}%
\bibitem [{\citenamefont {H\"ufner}(2003)}]{Hufner_book2003}%
  \BibitemOpen
  \bibfield  {author} {\bibinfo {author} {\bibfnamefont {S.}~\bibnamefont
  {H\"ufner}},\ }\href@noop {} {\emph {\bibinfo {title} {Photoelectron
  Spectroscopy: Principles and Applications}}}\ (\bibinfo  {publisher}
  {Springer},\ \bibinfo {year} {2003})\BibitemShut {NoStop}%
\bibitem [{\citenamefont {Kolasinski}(2012)}]{Kolasinski_Book2012}%
  \BibitemOpen
  \bibfield  {author} {\bibinfo {author} {\bibfnamefont {K.}~\bibnamefont
  {Kolasinski}},\ }\href@noop {} {\emph {\bibinfo {title} {Surface Science:
  Foundations of Catalysis and Nanoscience}}}\ (\bibinfo  {publisher} {Wiley},\
  \bibinfo {year} {2012})\BibitemShut {NoStop}%
\bibitem [{\citenamefont {Siegbahn}\ and\ \citenamefont
  {Nordling}(1967)}]{Siegbahn_Book1967}%
  \BibitemOpen
  \bibfield  {author} {\bibinfo {author} {\bibfnamefont {K.}~\bibnamefont
  {Siegbahn}}\ and\ \bibinfo {author} {\bibfnamefont {C.}~\bibnamefont
  {Nordling}},\ }\href@noop {} {\emph {\bibinfo {title} {ESCA; atomic,
  molecular, and solid-state structure studied by means of electron
  spectroscopy}}}\ (\bibinfo  {publisher} {Nova Acta Upsaliensis},\ \bibinfo
  {year} {1967})\BibitemShut {NoStop}%
\bibitem [{\citenamefont {Sobota}, \citenamefont {He},\ and\ \citenamefont
  {Shen}(2021)}]{Sobota_RMP2021}%
  \BibitemOpen
  \bibfield  {author} {\bibinfo {author} {\bibfnamefont {J.~A.}\ \bibnamefont
  {Sobota}}, \bibinfo {author} {\bibfnamefont {Y.}~\bibnamefont {He}}, \ and\
  \bibinfo {author} {\bibfnamefont {Z.-X.}\ \bibnamefont {Shen}},\ }\bibfield
  {title} {\enquote {\bibinfo {title} {Angle-resolved photoemission studies of
  quantum materials},}\ }\href {\doibase 10.1103/RevModPhys.93.025006}
  {\bibfield  {journal} {\bibinfo  {journal} {Rev. Mod. Phys.}\ }\textbf
  {\bibinfo {volume} {93}},\ \bibinfo {pages} {025006} (\bibinfo {year}
  {2021})}\BibitemShut {NoStop}%
\bibitem [{\citenamefont {Basov}\ \emph {et~al.}(2011)\citenamefont {Basov},
  \citenamefont {Averitt}, \citenamefont {van~der Marel}, \citenamefont
  {Dressel},\ and\ \citenamefont {Haule}}]{Basov_RMP2011}%
  \BibitemOpen
  \bibfield  {author} {\bibinfo {author} {\bibfnamefont {D.~N.}\ \bibnamefont
  {Basov}}, \bibinfo {author} {\bibfnamefont {R.~D.}\ \bibnamefont {Averitt}},
  \bibinfo {author} {\bibfnamefont {D.}~\bibnamefont {van~der Marel}}, \bibinfo
  {author} {\bibfnamefont {M.}~\bibnamefont {Dressel}}, \ and\ \bibinfo
  {author} {\bibfnamefont {K.}~\bibnamefont {Haule}},\ }\bibfield  {title}
  {\enquote {\bibinfo {title} {Electrodynamics of correlated electron
  materials},}\ }\href {\doibase 10.1103/RevModPhys.83.471} {\bibfield
  {journal} {\bibinfo  {journal} {Rev. Mod. Phys.}\ }\textbf {\bibinfo {volume}
  {83}},\ \bibinfo {pages} {471--541} (\bibinfo {year} {2011})}\BibitemShut
  {NoStop}%
\bibitem [{\citenamefont {Mukamel}(1995)}]{Mukamel_book1995}%
  \BibitemOpen
  \bibfield  {author} {\bibinfo {author} {\bibfnamefont {S.}~\bibnamefont
  {Mukamel}},\ }\href@noop {} {\emph {\bibinfo {title} {Principles of Nonlinear
  Optical Spectroscopy}}}\ (\bibinfo  {publisher} {Oxford University Press},\
  \bibinfo {year} {1995})\BibitemShut {NoStop}%
\bibitem [{\citenamefont {Fanciulli}\ \emph {et~al.}(2020)\citenamefont
  {Fanciulli}, \citenamefont {Schusser}, \citenamefont {Lee}, \citenamefont
  {Youbi}, \citenamefont {Heckmann}, \citenamefont {Richter}, \citenamefont
  {Cacho}, \citenamefont {Spezzani}, \citenamefont {Bresteau}, \citenamefont
  {Hergott}, \citenamefont {D'Oliveira}, \citenamefont {Tcherbakoff},
  \citenamefont {Ruchon}, \citenamefont {Min\'ar},\ and\ \citenamefont
  {Hricovini}}]{Fanciulli_PhysRevRes2020}%
  \BibitemOpen
  \bibfield  {author} {\bibinfo {author} {\bibfnamefont {M.}~\bibnamefont
  {Fanciulli}}, \bibinfo {author} {\bibfnamefont {J.}~\bibnamefont {Schusser}},
  \bibinfo {author} {\bibfnamefont {M.-I.}\ \bibnamefont {Lee}}, \bibinfo
  {author} {\bibfnamefont {Z.~E.}\ \bibnamefont {Youbi}}, \bibinfo {author}
  {\bibfnamefont {O.}~\bibnamefont {Heckmann}}, \bibinfo {author}
  {\bibfnamefont {M.~C.}\ \bibnamefont {Richter}}, \bibinfo {author}
  {\bibfnamefont {C.}~\bibnamefont {Cacho}}, \bibinfo {author} {\bibfnamefont
  {C.}~\bibnamefont {Spezzani}}, \bibinfo {author} {\bibfnamefont
  {D.}~\bibnamefont {Bresteau}}, \bibinfo {author} {\bibfnamefont {J.-F.
  m.~c.}\ \bibnamefont {Hergott}}, \bibinfo {author} {\bibfnamefont
  {P.}~\bibnamefont {D'Oliveira}}, \bibinfo {author} {\bibfnamefont
  {O.}~\bibnamefont {Tcherbakoff}}, \bibinfo {author} {\bibfnamefont
  {T.}~\bibnamefont {Ruchon}}, \bibinfo {author} {\bibfnamefont
  {J.}~\bibnamefont {Min\'ar}}, \ and\ \bibinfo {author} {\bibfnamefont
  {K.}~\bibnamefont {Hricovini}},\ }\bibfield  {title} {\enquote {\bibinfo
  {title} {Spin, time, and angle resolved photoemission spectroscopy on
  ${\mathrm{wte}}_{2}$},}\ }\href {\doibase 10.1103/PhysRevResearch.2.013261}
  {\bibfield  {journal} {\bibinfo  {journal} {Phys. Rev. Res.}\ }\textbf
  {\bibinfo {volume} {2}},\ \bibinfo {pages} {013261} (\bibinfo {year}
  {2020})}\BibitemShut {NoStop}%
\bibitem [{\citenamefont {Plotzing}\ \emph {et~al.}(2016)\citenamefont
  {Plotzing}, \citenamefont {Adam}, \citenamefont {Weier}, \citenamefont
  {Plucinski}, \citenamefont {Eich}, \citenamefont {Emmerich}, \citenamefont
  {Rollinger}, \citenamefont {Aeschlimann}, \citenamefont {Mathias},\ and\
  \citenamefont {Schneider}}]{Plotzing_RSI2016}%
  \BibitemOpen
  \bibfield  {author} {\bibinfo {author} {\bibfnamefont {M.}~\bibnamefont
  {Plotzing}}, \bibinfo {author} {\bibfnamefont {R.}~\bibnamefont {Adam}},
  \bibinfo {author} {\bibfnamefont {C.}~\bibnamefont {Weier}}, \bibinfo
  {author} {\bibfnamefont {L.}~\bibnamefont {Plucinski}}, \bibinfo {author}
  {\bibfnamefont {S.}~\bibnamefont {Eich}}, \bibinfo {author} {\bibfnamefont
  {S.}~\bibnamefont {Emmerich}}, \bibinfo {author} {\bibfnamefont
  {M.}~\bibnamefont {Rollinger}}, \bibinfo {author} {\bibfnamefont
  {M.}~\bibnamefont {Aeschlimann}}, \bibinfo {author} {\bibfnamefont
  {S.}~\bibnamefont {Mathias}}, \ and\ \bibinfo {author} {\bibfnamefont
  {C.~M.}\ \bibnamefont {Schneider}},\ }\bibfield  {title} {\enquote {\bibinfo
  {title} {Spin-resolved photoelectron spectroscopy using femtosecond extreme
  ultraviolet light pulses from high-order harmonic generation},}\ }\href
  {\doibase http://dx.doi.org/10.1063/1.4946782} {\bibfield  {journal}
  {\bibinfo  {journal} {Review of Scientific Instruments}\ }\textbf {\bibinfo
  {volume} {87}},\ \bibinfo {eid} {043903} (\bibinfo {year} {2016}),\
  http://dx.doi.org/10.1063/1.4946782}\BibitemShut {NoStop}%
\bibitem [{\citenamefont {Suzuki}(2019)}]{Suzuki_JChemPhys2019}%
  \BibitemOpen
  \bibfield  {author} {\bibinfo {author} {\bibfnamefont {T.}~\bibnamefont
  {Suzuki}},\ }\bibfield  {title} {\enquote {\bibinfo {title} {Ultrafast
  photoelectron spectroscopy of aqueous solutions},}\ }\href {\doibase
  10.1063/1.5098402} {\bibfield  {journal} {\bibinfo  {journal} {The Journal of
  Chemical Physics}\ }\textbf {\bibinfo {volume} {151}},\ \bibinfo {pages}
  {090901} (\bibinfo {year} {2019})},\ \Eprint
  {http://arxiv.org/abs/https://doi.org/10.1063/1.5098402}
  {https://doi.org/10.1063/1.5098402} \BibitemShut {NoStop}%
\bibitem [{\citenamefont {H{\"a}drich}\ \emph {et~al.}(2016)\citenamefont
  {H{\"a}drich}, \citenamefont {Rothhardt}, \citenamefont {Krebs},
  \citenamefont {Demmler}, \citenamefont {Klenke}, \citenamefont
  {T{\"u}nnermann},\ and\ \citenamefont {Limpert}}]{Hadrich_JPhysB2016}%
  \BibitemOpen
  \bibfield  {author} {\bibinfo {author} {\bibfnamefont {S.}~\bibnamefont
  {H{\"a}drich}}, \bibinfo {author} {\bibfnamefont {J.}~\bibnamefont
  {Rothhardt}}, \bibinfo {author} {\bibfnamefont {M.}~\bibnamefont {Krebs}},
  \bibinfo {author} {\bibfnamefont {S.}~\bibnamefont {Demmler}}, \bibinfo
  {author} {\bibfnamefont {A.}~\bibnamefont {Klenke}}, \bibinfo {author}
  {\bibfnamefont {A.}~\bibnamefont {T{\"u}nnermann}}, \ and\ \bibinfo {author}
  {\bibfnamefont {J.}~\bibnamefont {Limpert}},\ }\bibfield  {title} {\enquote
  {\bibinfo {title} {Single-pass high harmonic generation at high repetition
  rate and photon flux},}\ }\href
  {http://stacks.iop.org/0953-4075/49/i=17/a=172002} {\bibfield  {journal}
  {\bibinfo  {journal} {Journal of Physics B: Atomic, Molecular and Optical
  Physics}\ }\textbf {\bibinfo {volume} {49}},\ \bibinfo {pages} {172002}
  (\bibinfo {year} {2016})}\BibitemShut {NoStop}%
\bibitem [{\citenamefont {Madeo}\ \emph {et~al.}(2020)\citenamefont {Madeo},
  \citenamefont {Man}, \citenamefont {Sahoo}, \citenamefont {Campbell},
  \citenamefont {Pareek}, \citenamefont {Wong}, \citenamefont {Al-Mahboob},
  \citenamefont {Chan}, \citenamefont {Karmakar}, \citenamefont {Mariserla},
  \citenamefont {Li}, \citenamefont {Heinz}, \citenamefont {Cao},\ and\
  \citenamefont {Dani}}]{Madeo_Science2020}%
  \BibitemOpen
  \bibfield  {author} {\bibinfo {author} {\bibfnamefont {J.}~\bibnamefont
  {Madeo}}, \bibinfo {author} {\bibfnamefont {M.~K.~L.}\ \bibnamefont {Man}},
  \bibinfo {author} {\bibfnamefont {C.}~\bibnamefont {Sahoo}}, \bibinfo
  {author} {\bibfnamefont {M.}~\bibnamefont {Campbell}}, \bibinfo {author}
  {\bibfnamefont {V.}~\bibnamefont {Pareek}}, \bibinfo {author} {\bibfnamefont
  {E.~L.}\ \bibnamefont {Wong}}, \bibinfo {author} {\bibfnamefont
  {A.}~\bibnamefont {Al-Mahboob}}, \bibinfo {author} {\bibfnamefont {N.~S.}\
  \bibnamefont {Chan}}, \bibinfo {author} {\bibfnamefont {A.}~\bibnamefont
  {Karmakar}}, \bibinfo {author} {\bibfnamefont {B.~M.~K.}\ \bibnamefont
  {Mariserla}}, \bibinfo {author} {\bibfnamefont {X.}~\bibnamefont {Li}},
  \bibinfo {author} {\bibfnamefont {T.~F.}\ \bibnamefont {Heinz}}, \bibinfo
  {author} {\bibfnamefont {T.}~\bibnamefont {Cao}}, \ and\ \bibinfo {author}
  {\bibfnamefont {K.~M.}\ \bibnamefont {Dani}},\ }\bibfield  {title} {\enquote
  {\bibinfo {title} {Directly visualizing the momentum-forbidden dark excitons
  and their dynamics in atomically thin semiconductors},}\ }\href {\doibase
  10.1126/science.aba1029} {\bibfield  {journal} {\bibinfo  {journal}
  {Science}\ }\textbf {\bibinfo {volume} {370}},\ \bibinfo {pages} {1199--1204}
  (\bibinfo {year} {2020})},\ \Eprint
  {http://arxiv.org/abs/https://science.sciencemag.org/content/370/6521/1199.full.pdf}
  {https://science.sciencemag.org/content/370/6521/1199.full.pdf} \BibitemShut
  {NoStop}%
\bibitem [{\citenamefont {Puppin}\ \emph {et~al.}(2019)\citenamefont {Puppin},
  \citenamefont {Deng}, \citenamefont {Nicholson}, \citenamefont {Feldl},
  \citenamefont {Schr{\"o}ter}, \citenamefont {Vita}, \citenamefont
  {Kirchmann}, \citenamefont {Monney}, \citenamefont {Rettig}, \citenamefont
  {Wolf},\ and\ \citenamefont {Ernstorfer}}]{Puppin_RSI2019}%
  \BibitemOpen
  \bibfield  {author} {\bibinfo {author} {\bibfnamefont {M.}~\bibnamefont
  {Puppin}}, \bibinfo {author} {\bibfnamefont {Y.}~\bibnamefont {Deng}},
  \bibinfo {author} {\bibfnamefont {C.~W.}\ \bibnamefont {Nicholson}}, \bibinfo
  {author} {\bibfnamefont {J.}~\bibnamefont {Feldl}}, \bibinfo {author}
  {\bibfnamefont {N.~B.~M.}\ \bibnamefont {Schr{\"o}ter}}, \bibinfo {author}
  {\bibfnamefont {H.}~\bibnamefont {Vita}}, \bibinfo {author} {\bibfnamefont
  {P.~S.}\ \bibnamefont {Kirchmann}}, \bibinfo {author} {\bibfnamefont
  {C.}~\bibnamefont {Monney}}, \bibinfo {author} {\bibfnamefont
  {L.}~\bibnamefont {Rettig}}, \bibinfo {author} {\bibfnamefont
  {M.}~\bibnamefont {Wolf}}, \ and\ \bibinfo {author} {\bibfnamefont
  {R.}~\bibnamefont {Ernstorfer}},\ }\bibfield  {title} {\enquote {\bibinfo
  {title} {Time- and angle-resolved photoemission spectroscopy of solids in the
  extreme ultraviolet at 500 khz repetition rate},}\ }\bibfield  {booktitle}
  {\emph {\bibinfo {booktitle} {Review of Scientific Instruments}},\ }\href
  {\doibase 10.1063/1.5081938} {\bibfield  {journal} {\bibinfo  {journal}
  {Review of Scientific Instruments}\ }\textbf {\bibinfo {volume} {90}},\
  \bibinfo {pages} {023104} (\bibinfo {year} {2019})}\BibitemShut {NoStop}%
\bibitem [{\citenamefont {Keunecke}\ \emph {et~al.}(2020)\citenamefont
  {Keunecke}, \citenamefont {M{\"o}ller}, \citenamefont {Schmitt},
  \citenamefont {Nolte}, \citenamefont {Jansen}, \citenamefont {Reutzel},
  \citenamefont {Gutberlet}, \citenamefont {Halasi}, \citenamefont {Steil},
  \citenamefont {Steil},\ and\ \citenamefont {Mathias}}]{Keunecke_RSI2020}%
  \BibitemOpen
  \bibfield  {author} {\bibinfo {author} {\bibfnamefont {M.}~\bibnamefont
  {Keunecke}}, \bibinfo {author} {\bibfnamefont {C.}~\bibnamefont
  {M{\"o}ller}}, \bibinfo {author} {\bibfnamefont {D.}~\bibnamefont {Schmitt}},
  \bibinfo {author} {\bibfnamefont {H.}~\bibnamefont {Nolte}}, \bibinfo
  {author} {\bibfnamefont {G.~S.~M.}\ \bibnamefont {Jansen}}, \bibinfo {author}
  {\bibfnamefont {M.}~\bibnamefont {Reutzel}}, \bibinfo {author} {\bibfnamefont
  {M.}~\bibnamefont {Gutberlet}}, \bibinfo {author} {\bibfnamefont
  {G.}~\bibnamefont {Halasi}}, \bibinfo {author} {\bibfnamefont
  {D.}~\bibnamefont {Steil}}, \bibinfo {author} {\bibfnamefont
  {S.}~\bibnamefont {Steil}}, \ and\ \bibinfo {author} {\bibfnamefont
  {S.}~\bibnamefont {Mathias}},\ }\bibfield  {title} {\enquote {\bibinfo
  {title} {Time-resolved momentum microscopy with a 1 mhz high-harmonic extreme
  ultraviolet beamline},}\ }\href {\doibase 10.1063/5.0006531} {\bibfield
  {journal} {\bibinfo  {journal} {Review of Scientific Instruments}\ }\textbf
  {\bibinfo {volume} {91}},\ \bibinfo {pages} {063905} (\bibinfo {year}
  {2020})},\ \Eprint {http://arxiv.org/abs/https://doi.org/10.1063/5.0006531}
  {https://doi.org/10.1063/5.0006531} \BibitemShut {NoStop}%
\bibitem [{AFS()}]{AFS_web}%
  \BibitemOpen
  \href@noop {} {\enquote {\bibinfo {title} {https://www.afs-jena.de/},}\
  }\BibitemShut {NoStop}%
\bibitem [{cla()}]{class5_web}%
  \BibitemOpen
  \href@noop {} {\enquote {\bibinfo {title}
  {https://www.class5photonics.com/},}\ }\BibitemShut {NoStop}%
\bibitem [{KML()}]{KMLabs_web}%
  \BibitemOpen
  \href@noop {} {\enquote {\bibinfo {title} {https://www.kmlabs.com/},}\
  }\BibitemShut {NoStop}%
\bibitem [{\citenamefont {Chernov}\ \emph {et~al.}(2015)\citenamefont
  {Chernov}, \citenamefont {Medjanik}, \citenamefont {Tusche}, \citenamefont
  {Kutnyakhov}, \citenamefont {Nepijko}, \citenamefont {Oelsner}, \citenamefont
  {Braun}, \citenamefont {Min{\'a}r}, \citenamefont {Borek}, \citenamefont
  {Ebert}, \citenamefont {Elmers}, \citenamefont {Kirschner},\ and\
  \citenamefont {Sch{\"o}nhense}}]{Chernov_Ultramicroscopy2015}%
  \BibitemOpen
  \bibfield  {author} {\bibinfo {author} {\bibfnamefont {S.}~\bibnamefont
  {Chernov}}, \bibinfo {author} {\bibfnamefont {K.}~\bibnamefont {Medjanik}},
  \bibinfo {author} {\bibfnamefont {C.}~\bibnamefont {Tusche}}, \bibinfo
  {author} {\bibfnamefont {D.}~\bibnamefont {Kutnyakhov}}, \bibinfo {author}
  {\bibfnamefont {S.}~\bibnamefont {Nepijko}}, \bibinfo {author} {\bibfnamefont
  {A.}~\bibnamefont {Oelsner}}, \bibinfo {author} {\bibfnamefont
  {J.}~\bibnamefont {Braun}}, \bibinfo {author} {\bibfnamefont
  {J.}~\bibnamefont {Min{\'a}r}}, \bibinfo {author} {\bibfnamefont
  {S.}~\bibnamefont {Borek}}, \bibinfo {author} {\bibfnamefont
  {H.}~\bibnamefont {Ebert}}, \bibinfo {author} {\bibfnamefont
  {H.}~\bibnamefont {Elmers}}, \bibinfo {author} {\bibfnamefont
  {J.}~\bibnamefont {Kirschner}}, \ and\ \bibinfo {author} {\bibfnamefont
  {G.}~\bibnamefont {Sch{\"o}nhense}},\ }\bibfield  {title} {\enquote {\bibinfo
  {title} {Anomalous d-like surface resonances on mo(110) analyzed by
  time-of-flight momentum microscopy},}\ }\href {\doibase
  https://doi.org/10.1016/j.ultramic.2015.07.008} {\bibfield  {journal}
  {\bibinfo  {journal} {Ultramicroscopy}\ }\textbf {\bibinfo {volume} {159}},\
  \bibinfo {pages} {453--463} (\bibinfo {year} {2015})},\ \bibinfo {note}
  {special Issue: LEEM-PEEM 9}\BibitemShut {NoStop}%
\bibitem [{\citenamefont {Medjanik}\ \emph {et~al.}(2017)\citenamefont
  {Medjanik}, \citenamefont {Fedchenko}, \citenamefont {Chernov}, \citenamefont
  {Kutnyakhov}, \citenamefont {Ellguth}, \citenamefont {Oelsner}, \citenamefont
  {Schonhense}, \citenamefont {Peixoto}, \citenamefont {Lutz}, \citenamefont
  {Min}, \citenamefont {Reinert}, \citenamefont {Daster}, \citenamefont
  {Acremann}, \citenamefont {Viefhaus}, \citenamefont {Wurth}, \citenamefont
  {Elmers},\ and\ \citenamefont {Schonhense}}]{Medjanik_NatMat2017}%
  \BibitemOpen
  \bibfield  {author} {\bibinfo {author} {\bibfnamefont {K.}~\bibnamefont
  {Medjanik}}, \bibinfo {author} {\bibfnamefont {O.}~\bibnamefont {Fedchenko}},
  \bibinfo {author} {\bibfnamefont {S.}~\bibnamefont {Chernov}}, \bibinfo
  {author} {\bibfnamefont {D.}~\bibnamefont {Kutnyakhov}}, \bibinfo {author}
  {\bibfnamefont {M.}~\bibnamefont {Ellguth}}, \bibinfo {author} {\bibfnamefont
  {A.}~\bibnamefont {Oelsner}}, \bibinfo {author} {\bibfnamefont
  {B.}~\bibnamefont {Schonhense}}, \bibinfo {author} {\bibfnamefont {T.~R.~F.}\
  \bibnamefont {Peixoto}}, \bibinfo {author} {\bibfnamefont {P.}~\bibnamefont
  {Lutz}}, \bibinfo {author} {\bibfnamefont {C.~H.}\ \bibnamefont {Min}},
  \bibinfo {author} {\bibfnamefont {F.}~\bibnamefont {Reinert}}, \bibinfo
  {author} {\bibfnamefont {S.}~\bibnamefont {Daster}}, \bibinfo {author}
  {\bibfnamefont {Y.}~\bibnamefont {Acremann}}, \bibinfo {author}
  {\bibfnamefont {J.}~\bibnamefont {Viefhaus}}, \bibinfo {author}
  {\bibfnamefont {W.}~\bibnamefont {Wurth}}, \bibinfo {author} {\bibfnamefont
  {H.~J.}\ \bibnamefont {Elmers}}, \ and\ \bibinfo {author} {\bibfnamefont
  {G.}~\bibnamefont {Schonhense}},\ }\bibfield  {title} {\enquote {\bibinfo
  {title} {Direct 3d mapping of the fermi surface and fermi velocity},}\ }\href
  {http://dx.doi.org/10.1038/nmat4875} {\bibfield  {journal} {\bibinfo
  {journal} {Nat Mater}\ }\textbf {\bibinfo {volume} {16}},\ \bibinfo {pages}
  {615--621} (\bibinfo {year} {2017})}\BibitemShut {NoStop}%
\bibitem [{\citenamefont {Schmitt}\ \emph {et~al.}(2022)\citenamefont
  {Schmitt}, \citenamefont {Bange}, \citenamefont {Bennecke}, \citenamefont
  {AlMutairi}, \citenamefont {Meneghini}, \citenamefont {Watanabe},
  \citenamefont {Taniguchi}, \citenamefont {Steil}, \citenamefont {Luke},
  \citenamefont {Weitz}, \citenamefont {Steil}, \citenamefont {Jansen},
  \citenamefont {Brem}, \citenamefont {Malic}, \citenamefont {Hofmann},
  \citenamefont {Reutzel},\ and\ \citenamefont {Mathias}}]{Schmitt_Nature2022}%
  \BibitemOpen
  \bibfield  {author} {\bibinfo {author} {\bibfnamefont {D.}~\bibnamefont
  {Schmitt}}, \bibinfo {author} {\bibfnamefont {J.~P.}\ \bibnamefont {Bange}},
  \bibinfo {author} {\bibfnamefont {W.}~\bibnamefont {Bennecke}}, \bibinfo
  {author} {\bibfnamefont {A.}~\bibnamefont {AlMutairi}}, \bibinfo {author}
  {\bibfnamefont {G.}~\bibnamefont {Meneghini}}, \bibinfo {author}
  {\bibfnamefont {K.}~\bibnamefont {Watanabe}}, \bibinfo {author}
  {\bibfnamefont {T.}~\bibnamefont {Taniguchi}}, \bibinfo {author}
  {\bibfnamefont {D.}~\bibnamefont {Steil}}, \bibinfo {author} {\bibfnamefont
  {D.~R.}\ \bibnamefont {Luke}}, \bibinfo {author} {\bibfnamefont {R.~T.}\
  \bibnamefont {Weitz}}, \bibinfo {author} {\bibfnamefont {S.}~\bibnamefont
  {Steil}}, \bibinfo {author} {\bibfnamefont {G.~S.~M.}\ \bibnamefont
  {Jansen}}, \bibinfo {author} {\bibfnamefont {S.}~\bibnamefont {Brem}},
  \bibinfo {author} {\bibfnamefont {E.}~\bibnamefont {Malic}}, \bibinfo
  {author} {\bibfnamefont {S.}~\bibnamefont {Hofmann}}, \bibinfo {author}
  {\bibfnamefont {M.}~\bibnamefont {Reutzel}}, \ and\ \bibinfo {author}
  {\bibfnamefont {S.}~\bibnamefont {Mathias}},\ }\bibfield  {title} {\enquote
  {\bibinfo {title} {Formation of moir\'e interlayer excitons in space and
  time},}\ }\href {\doibase 10.1038/s41586-022-04977-7} {\bibfield  {journal}
  {\bibinfo  {journal} {Nature}\ }\textbf {\bibinfo {volume} {608}},\ \bibinfo
  {pages} {499--503} (\bibinfo {year} {2022})}\BibitemShut {NoStop}%
\bibitem [{\citenamefont {Karni}\ \emph {et~al.}(2022)\citenamefont {Karni},
  \citenamefont {Barr{\'e}}, \citenamefont {Pareek}, \citenamefont {Georgaras},
  \citenamefont {Man}, \citenamefont {Sahoo}, \citenamefont {Bacon},
  \citenamefont {Zhu}, \citenamefont {Ribeiro}, \citenamefont {O'Beirne},
  \citenamefont {Hu}, \citenamefont {Al-Mahboob}, \citenamefont {Abdelrasoul},
  \citenamefont {Chan}, \citenamefont {Karmakar}, \citenamefont {Winchester},
  \citenamefont {Kim}, \citenamefont {Watanabe}, \citenamefont {Taniguchi},
  \citenamefont {Barmak}, \citenamefont {Mad{\'e}o}, \citenamefont
  {da~Jornada}, \citenamefont {Heinz},\ and\ \citenamefont
  {Dani}}]{Karni_Nature2022}%
  \BibitemOpen
  \bibfield  {author} {\bibinfo {author} {\bibfnamefont {O.}~\bibnamefont
  {Karni}}, \bibinfo {author} {\bibfnamefont {E.}~\bibnamefont {Barr{\'e}}},
  \bibinfo {author} {\bibfnamefont {V.}~\bibnamefont {Pareek}}, \bibinfo
  {author} {\bibfnamefont {J.~D.}\ \bibnamefont {Georgaras}}, \bibinfo {author}
  {\bibfnamefont {M.~K.~L.}\ \bibnamefont {Man}}, \bibinfo {author}
  {\bibfnamefont {C.}~\bibnamefont {Sahoo}}, \bibinfo {author} {\bibfnamefont
  {D.~R.}\ \bibnamefont {Bacon}}, \bibinfo {author} {\bibfnamefont
  {X.}~\bibnamefont {Zhu}}, \bibinfo {author} {\bibfnamefont {H.~B.}\
  \bibnamefont {Ribeiro}}, \bibinfo {author} {\bibfnamefont {A.~L.}\
  \bibnamefont {O'Beirne}}, \bibinfo {author} {\bibfnamefont {J.}~\bibnamefont
  {Hu}}, \bibinfo {author} {\bibfnamefont {A.}~\bibnamefont {Al-Mahboob}},
  \bibinfo {author} {\bibfnamefont {M.~M.~M.}\ \bibnamefont {Abdelrasoul}},
  \bibinfo {author} {\bibfnamefont {N.~S.}\ \bibnamefont {Chan}}, \bibinfo
  {author} {\bibfnamefont {A.}~\bibnamefont {Karmakar}}, \bibinfo {author}
  {\bibfnamefont {A.~J.}\ \bibnamefont {Winchester}}, \bibinfo {author}
  {\bibfnamefont {B.}~\bibnamefont {Kim}}, \bibinfo {author} {\bibfnamefont
  {K.}~\bibnamefont {Watanabe}}, \bibinfo {author} {\bibfnamefont
  {T.}~\bibnamefont {Taniguchi}}, \bibinfo {author} {\bibfnamefont
  {K.}~\bibnamefont {Barmak}}, \bibinfo {author} {\bibfnamefont
  {J.}~\bibnamefont {Mad{\'e}o}}, \bibinfo {author} {\bibfnamefont {F.~H.}\
  \bibnamefont {da~Jornada}}, \bibinfo {author} {\bibfnamefont {T.~F.}\
  \bibnamefont {Heinz}}, \ and\ \bibinfo {author} {\bibfnamefont {K.~M.}\
  \bibnamefont {Dani}},\ }\bibfield  {title} {\enquote {\bibinfo {title}
  {Structure of the moir{\'e}exciton captured by imaging its electron and
  hole},}\ }\href {\doibase 10.1038/s41586-021-04360-y} {\bibfield  {journal}
  {\bibinfo  {journal} {Nature}\ }\textbf {\bibinfo {volume} {603}},\ \bibinfo
  {pages} {247--252} (\bibinfo {year} {2022})}\BibitemShut {NoStop}%
\bibitem [{\citenamefont {Wallauer}\ \emph {et~al.}(2021)\citenamefont
  {Wallauer}, \citenamefont {Raths}, \citenamefont {Stallberg}, \citenamefont
  {M{\"u}nster}, \citenamefont {Brandstetter}, \citenamefont {Yang},
  \citenamefont {G{\"u}dde}, \citenamefont {Puschnig}, \citenamefont
  {Soubatch}, \citenamefont {Kumpf}, \citenamefont {Bocquet}, \citenamefont
  {Tautz},\ and\ \citenamefont {H{\"o}fer}}]{Wallauer_Science2021}%
  \BibitemOpen
  \bibfield  {author} {\bibinfo {author} {\bibfnamefont {R.}~\bibnamefont
  {Wallauer}}, \bibinfo {author} {\bibfnamefont {M.}~\bibnamefont {Raths}},
  \bibinfo {author} {\bibfnamefont {K.}~\bibnamefont {Stallberg}}, \bibinfo
  {author} {\bibfnamefont {L.}~\bibnamefont {M{\"u}nster}}, \bibinfo {author}
  {\bibfnamefont {D.}~\bibnamefont {Brandstetter}}, \bibinfo {author}
  {\bibfnamefont {X.}~\bibnamefont {Yang}}, \bibinfo {author} {\bibfnamefont
  {J.}~\bibnamefont {G{\"u}dde}}, \bibinfo {author} {\bibfnamefont
  {P.}~\bibnamefont {Puschnig}}, \bibinfo {author} {\bibfnamefont
  {S.}~\bibnamefont {Soubatch}}, \bibinfo {author} {\bibfnamefont
  {C.}~\bibnamefont {Kumpf}}, \bibinfo {author} {\bibfnamefont {F.~C.}\
  \bibnamefont {Bocquet}}, \bibinfo {author} {\bibfnamefont {F.~S.}\
  \bibnamefont {Tautz}}, \ and\ \bibinfo {author} {\bibfnamefont
  {U.}~\bibnamefont {H{\"o}fer}},\ }\bibfield  {title} {\enquote {\bibinfo
  {title} {Tracing orbital images on ultrafast time scales},}\ }\href {\doibase
  10.1126/science.abf3286} {\bibfield  {journal} {\bibinfo  {journal}
  {Science}\ } (\bibinfo {year} {2021}),\ 10.1126/science.abf3286},\ \Eprint
  {http://arxiv.org/abs/https://science.sciencemag.org/content/early/2021/02/17/science.abf3286.full.pdf}
  {https://science.sciencemag.org/content/early/2021/02/17/science.abf3286.full.pdf}
  \BibitemShut {NoStop}%
\bibitem [{\citenamefont {Neef}\ \emph {et~al.}(2023)\citenamefont {Neef},
  \citenamefont {Beaulieu}, \citenamefont {Hammer}, \citenamefont {Dong},
  \citenamefont {Maklar}, \citenamefont {Pincelli}, \citenamefont {Xian},
  \citenamefont {Wolf}, \citenamefont {Rettig}, \citenamefont {Pflaum},\ and\
  \citenamefont {Ernstorfer}}]{Neef_Nature2023}%
  \BibitemOpen
  \bibfield  {author} {\bibinfo {author} {\bibfnamefont {A.}~\bibnamefont
  {Neef}}, \bibinfo {author} {\bibfnamefont {S.}~\bibnamefont {Beaulieu}},
  \bibinfo {author} {\bibfnamefont {S.}~\bibnamefont {Hammer}}, \bibinfo
  {author} {\bibfnamefont {S.}~\bibnamefont {Dong}}, \bibinfo {author}
  {\bibfnamefont {J.}~\bibnamefont {Maklar}}, \bibinfo {author} {\bibfnamefont
  {T.}~\bibnamefont {Pincelli}}, \bibinfo {author} {\bibfnamefont {R.~P.}\
  \bibnamefont {Xian}}, \bibinfo {author} {\bibfnamefont {M.}~\bibnamefont
  {Wolf}}, \bibinfo {author} {\bibfnamefont {L.}~\bibnamefont {Rettig}},
  \bibinfo {author} {\bibfnamefont {J.}~\bibnamefont {Pflaum}}, \ and\ \bibinfo
  {author} {\bibfnamefont {R.}~\bibnamefont {Ernstorfer}},\ }\bibfield  {title}
  {\enquote {\bibinfo {title} {Orbital-resolved observation of singlet
  fission},}\ }\href {\doibase 10.1038/s41586-023-05814-1} {\bibfield
  {journal} {\bibinfo  {journal} {Nature}\ }\textbf {\bibinfo {volume} {616}},\
  \bibinfo {pages} {275--279} (\bibinfo {year} {2023})}\BibitemShut {NoStop}%
\bibitem [{\citenamefont {Man}\ \emph {et~al.}(2021)\citenamefont {Man},
  \citenamefont {Mad{\'e}o}, \citenamefont {Sahoo}, \citenamefont {Xie},
  \citenamefont {Campbell}, \citenamefont {Pareek}, \citenamefont {Karmakar},
  \citenamefont {Wong}, \citenamefont {Al-Mahboob}, \citenamefont {Chan},
  \citenamefont {Bacon}, \citenamefont {Zhu}, \citenamefont {Abdelrasoul},
  \citenamefont {Li}, \citenamefont {Heinz}, \citenamefont {da~Jornada},
  \citenamefont {Cao},\ and\ \citenamefont {Dani}}]{Man_SciAdv2021}%
  \BibitemOpen
  \bibfield  {author} {\bibinfo {author} {\bibfnamefont {M.~K.~L.}\
  \bibnamefont {Man}}, \bibinfo {author} {\bibfnamefont {J.}~\bibnamefont
  {Mad{\'e}o}}, \bibinfo {author} {\bibfnamefont {C.}~\bibnamefont {Sahoo}},
  \bibinfo {author} {\bibfnamefont {K.}~\bibnamefont {Xie}}, \bibinfo {author}
  {\bibfnamefont {M.}~\bibnamefont {Campbell}}, \bibinfo {author}
  {\bibfnamefont {V.}~\bibnamefont {Pareek}}, \bibinfo {author} {\bibfnamefont
  {A.}~\bibnamefont {Karmakar}}, \bibinfo {author} {\bibfnamefont {E.~L.}\
  \bibnamefont {Wong}}, \bibinfo {author} {\bibfnamefont {A.}~\bibnamefont
  {Al-Mahboob}}, \bibinfo {author} {\bibfnamefont {N.~S.}\ \bibnamefont
  {Chan}}, \bibinfo {author} {\bibfnamefont {D.~R.}\ \bibnamefont {Bacon}},
  \bibinfo {author} {\bibfnamefont {X.}~\bibnamefont {Zhu}}, \bibinfo {author}
  {\bibfnamefont {M.~M.~M.}\ \bibnamefont {Abdelrasoul}}, \bibinfo {author}
  {\bibfnamefont {X.}~\bibnamefont {Li}}, \bibinfo {author} {\bibfnamefont
  {T.~F.}\ \bibnamefont {Heinz}}, \bibinfo {author} {\bibfnamefont {F.~H.}\
  \bibnamefont {da~Jornada}}, \bibinfo {author} {\bibfnamefont
  {T.}~\bibnamefont {Cao}}, \ and\ \bibinfo {author} {\bibfnamefont {K.~M.}\
  \bibnamefont {Dani}},\ }\bibfield  {title} {\enquote {\bibinfo {title}
  {Experimental measurement of the intrinsic excitonic wave function},}\ }\href
  {\doibase 10.1126/sciadv.abg0192} {\bibfield  {journal} {\bibinfo  {journal}
  {Science Advances}\ }\textbf {\bibinfo {volume} {7}} (\bibinfo {year}
  {2021}),\ 10.1126/sciadv.abg0192},\ \Eprint
  {http://arxiv.org/abs/https://advances.sciencemag.org/content/7/17/eabg0192.full.pdf}
  {https://advances.sciencemag.org/content/7/17/eabg0192.full.pdf} \BibitemShut
  {NoStop}%
\bibitem [{\citenamefont {Kolbe}\ \emph {et~al.}(2011)\citenamefont {Kolbe},
  \citenamefont {Lushchyk}, \citenamefont {Petereit}, \citenamefont {Elmers},
  \citenamefont {Sch\"onhense}, \citenamefont {Oelsner}, \citenamefont
  {Tusche},\ and\ \citenamefont {Kirschner}}]{Kolbe_PRL2011}%
  \BibitemOpen
  \bibfield  {author} {\bibinfo {author} {\bibfnamefont {M.}~\bibnamefont
  {Kolbe}}, \bibinfo {author} {\bibfnamefont {P.}~\bibnamefont {Lushchyk}},
  \bibinfo {author} {\bibfnamefont {B.}~\bibnamefont {Petereit}}, \bibinfo
  {author} {\bibfnamefont {H.~J.}\ \bibnamefont {Elmers}}, \bibinfo {author}
  {\bibfnamefont {G.}~\bibnamefont {Sch\"onhense}}, \bibinfo {author}
  {\bibfnamefont {A.}~\bibnamefont {Oelsner}}, \bibinfo {author} {\bibfnamefont
  {C.}~\bibnamefont {Tusche}}, \ and\ \bibinfo {author} {\bibfnamefont
  {J.}~\bibnamefont {Kirschner}},\ }\bibfield  {title} {\enquote {\bibinfo
  {title} {Highly efficient multichannel spin-polarization detection},}\ }\href
  {\doibase 10.1103/PhysRevLett.107.207601} {\bibfield  {journal} {\bibinfo
  {journal} {Phys. Rev. Lett.}\ }\textbf {\bibinfo {volume} {107}},\ \bibinfo
  {pages} {207601} (\bibinfo {year} {2011})}\BibitemShut {NoStop}%
\bibitem [{\citenamefont {Sch\"onhense}\ \emph {et~al.}(2017)\citenamefont
  {Sch\"onhense}, \citenamefont {Medjanik}, \citenamefont {Chernov},
  \citenamefont {Kutnyakhov}, \citenamefont {Fedchenko}, \citenamefont
  {Ellguth}, \citenamefont {Vasilyev}, \citenamefont
  {Zaporozhchenko-Zymakov{\'a}}, \citenamefont {Panzer}, \citenamefont
  {Oelsner}, \citenamefont {Tusche}, \citenamefont {Sch{\"o}nhense},
  \citenamefont {Braun}, \citenamefont {Min{\'a}r}, \citenamefont {Ebert},
  \citenamefont {Viefhaus}, \citenamefont {Wurth},\ and\ \citenamefont
  {Elmers}}]{Schoenhense_Ultramicroscopy2017}%
  \BibitemOpen
  \bibfield  {author} {\bibinfo {author} {\bibfnamefont {G.}~\bibnamefont
  {Sch\"onhense}}, \bibinfo {author} {\bibfnamefont {K.}~\bibnamefont
  {Medjanik}}, \bibinfo {author} {\bibfnamefont {S.}~\bibnamefont {Chernov}},
  \bibinfo {author} {\bibfnamefont {D.}~\bibnamefont {Kutnyakhov}}, \bibinfo
  {author} {\bibfnamefont {O.}~\bibnamefont {Fedchenko}}, \bibinfo {author}
  {\bibfnamefont {M.}~\bibnamefont {Ellguth}}, \bibinfo {author} {\bibfnamefont
  {D.}~\bibnamefont {Vasilyev}}, \bibinfo {author} {\bibfnamefont
  {A.}~\bibnamefont {Zaporozhchenko-Zymakov{\'a}}}, \bibinfo {author}
  {\bibfnamefont {D.}~\bibnamefont {Panzer}}, \bibinfo {author} {\bibfnamefont
  {A.}~\bibnamefont {Oelsner}}, \bibinfo {author} {\bibfnamefont
  {C.}~\bibnamefont {Tusche}}, \bibinfo {author} {\bibfnamefont
  {B.}~\bibnamefont {Sch{\"o}nhense}}, \bibinfo {author} {\bibfnamefont
  {J.}~\bibnamefont {Braun}}, \bibinfo {author} {\bibfnamefont
  {J.}~\bibnamefont {Min{\'a}r}}, \bibinfo {author} {\bibfnamefont
  {H.}~\bibnamefont {Ebert}}, \bibinfo {author} {\bibfnamefont
  {J.}~\bibnamefont {Viefhaus}}, \bibinfo {author} {\bibfnamefont
  {W.}~\bibnamefont {Wurth}}, \ and\ \bibinfo {author} {\bibfnamefont
  {H.}~\bibnamefont {Elmers}},\ }\bibfield  {title} {\enquote {\bibinfo {title}
  {Spin-filtered time-of-flight k-space microscopy of ir -- towards the
  ``complete'' photoemission experiment},}\ }\href {\doibase
  https://doi.org/10.1016/j.ultramic.2017.06.025} {\bibfield  {journal}
  {\bibinfo  {journal} {Ultramicroscopy}\ }\textbf {\bibinfo {volume} {183}},\
  \bibinfo {pages} {19--29} (\bibinfo {year} {2017})},\ \bibinfo {note}
  {lEEM/PEEM-10}\BibitemShut {NoStop}%
\bibitem [{NSF()}]{NSFNeXUS}%
  \BibitemOpen
  \href@noop {} {\enquote {\bibinfo {title} {https://nsf-nexus.osu.edu/},}\
  }\BibitemShut {NoStop}%
\bibitem [{\citenamefont {Mills}\ \emph {et~al.}(2012)\citenamefont {Mills},
  \citenamefont {Hammond}, \citenamefont {Lam},\ and\ \citenamefont
  {Jones}}]{Mills_JPhysB2012}%
  \BibitemOpen
  \bibfield  {author} {\bibinfo {author} {\bibfnamefont {A.~K.}\ \bibnamefont
  {Mills}}, \bibinfo {author} {\bibfnamefont {T.~J.}\ \bibnamefont {Hammond}},
  \bibinfo {author} {\bibfnamefont {M.~H.~C.}\ \bibnamefont {Lam}}, \ and\
  \bibinfo {author} {\bibfnamefont {D.~J.}\ \bibnamefont {Jones}},\ }\bibfield
  {title} {\enquote {\bibinfo {title} {Xuv frequency combs via femtosecond
  enhancement cavities},}\ }\href
  {http://stacks.iop.org/0953-4075/45/i=14/a=142001} {\bibfield  {journal}
  {\bibinfo  {journal} {Journal of Physics B: Atomic, Molecular and Optical
  Physics}\ }\textbf {\bibinfo {volume} {45}},\ \bibinfo {pages} {142001}
  (\bibinfo {year} {2012})}\BibitemShut {NoStop}%
\bibitem [{\citenamefont {Gherman}\ and\ \citenamefont
  {Romanini}(2002)}]{Gherman_OptExp2002}%
  \BibitemOpen
  \bibfield  {author} {\bibinfo {author} {\bibfnamefont {T.}~\bibnamefont
  {Gherman}}\ and\ \bibinfo {author} {\bibfnamefont {D.}~\bibnamefont
  {Romanini}},\ }\bibfield  {title} {\enquote {\bibinfo {title} {Modelocked
  cavity--enhanced absorption spectroscopy},}\ }\href {\doibase
  10.1364/OE.10.001033} {\bibfield  {journal} {\bibinfo  {journal} {Opt.
  Express}\ }\textbf {\bibinfo {volume} {10}},\ \bibinfo {pages} {1033--1042}
  (\bibinfo {year} {2002})}\BibitemShut {NoStop}%
\bibitem [{\citenamefont {Jones}\ and\ \citenamefont
  {Ye}(2002)}]{Jones_OptLett2002}%
  \BibitemOpen
  \bibfield  {author} {\bibinfo {author} {\bibfnamefont {R.~J.}\ \bibnamefont
  {Jones}}\ and\ \bibinfo {author} {\bibfnamefont {J.}~\bibnamefont {Ye}},\
  }\bibfield  {title} {\enquote {\bibinfo {title} {Femtosecond pulse
  amplification by coherent addition in a passive optical cavity},}\ }\href
  {http://ol.osa.org/abstract.cfm?URI=ol-27-20-1848} {\bibfield  {journal}
  {\bibinfo  {journal} {Opt. Lett.}\ }\textbf {\bibinfo {volume} {27}},\
  \bibinfo {pages} {1848--1850} (\bibinfo {year} {2002})}\BibitemShut {NoStop}%
\bibitem [{\citenamefont {Jones}\ and\ \citenamefont
  {Ye}(2004)}]{Jones_OptLett2004}%
  \BibitemOpen
  \bibfield  {author} {\bibinfo {author} {\bibfnamefont {R.~J.}\ \bibnamefont
  {Jones}}\ and\ \bibinfo {author} {\bibfnamefont {J.}~\bibnamefont {Ye}},\
  }\bibfield  {title} {\enquote {\bibinfo {title} {High-repetition-rate
  coherent femtosecond pulse amplification withan external passive optical
  cavity},}\ }\href {\doibase 10.1364/OL.29.002812} {\bibfield  {journal}
  {\bibinfo  {journal} {Opt. Lett.}\ }\textbf {\bibinfo {volume} {29}},\
  \bibinfo {pages} {2812--2814} (\bibinfo {year} {2004})}\BibitemShut {NoStop}%
\bibitem [{\citenamefont {Carstens}\ \emph {et~al.}(2014)\citenamefont
  {Carstens}, \citenamefont {Lilienfein}, \citenamefont {Holzberger},
  \citenamefont {Jocher}, \citenamefont {Eidam}, \citenamefont {Limpert},
  \citenamefont {T\"{u}nnermann}, \citenamefont {Weitenberg}, \citenamefont
  {Yost}, \citenamefont {Alghamdi}, \citenamefont {Alahmed}, \citenamefont
  {Azzeer}, \citenamefont {Apolonski}, \citenamefont {Fill}, \citenamefont
  {Krausz},\ and\ \citenamefont {Pupeza}}]{Carstens_OptLett2014}%
  \BibitemOpen
  \bibfield  {author} {\bibinfo {author} {\bibfnamefont {H.}~\bibnamefont
  {Carstens}}, \bibinfo {author} {\bibfnamefont {N.}~\bibnamefont
  {Lilienfein}}, \bibinfo {author} {\bibfnamefont {S.}~\bibnamefont
  {Holzberger}}, \bibinfo {author} {\bibfnamefont {C.}~\bibnamefont {Jocher}},
  \bibinfo {author} {\bibfnamefont {T.}~\bibnamefont {Eidam}}, \bibinfo
  {author} {\bibfnamefont {J.}~\bibnamefont {Limpert}}, \bibinfo {author}
  {\bibfnamefont {A.}~\bibnamefont {T\"{u}nnermann}}, \bibinfo {author}
  {\bibfnamefont {J.}~\bibnamefont {Weitenberg}}, \bibinfo {author}
  {\bibfnamefont {D.~C.}\ \bibnamefont {Yost}}, \bibinfo {author}
  {\bibfnamefont {A.}~\bibnamefont {Alghamdi}}, \bibinfo {author}
  {\bibfnamefont {Z.}~\bibnamefont {Alahmed}}, \bibinfo {author} {\bibfnamefont
  {A.}~\bibnamefont {Azzeer}}, \bibinfo {author} {\bibfnamefont
  {A.}~\bibnamefont {Apolonski}}, \bibinfo {author} {\bibfnamefont
  {E.}~\bibnamefont {Fill}}, \bibinfo {author} {\bibfnamefont {F.}~\bibnamefont
  {Krausz}}, \ and\ \bibinfo {author} {\bibfnamefont {I.}~\bibnamefont
  {Pupeza}},\ }\bibfield  {title} {\enquote {\bibinfo {title} {Megawatt-scale
  average-power ultrashort pulses in an enhancement cavity},}\ }\href {\doibase
  10.1364/OL.39.002595} {\bibfield  {journal} {\bibinfo  {journal} {Opt.
  Lett.}\ }\textbf {\bibinfo {volume} {39}},\ \bibinfo {pages} {2595--2598}
  (\bibinfo {year} {2014})}\BibitemShut {NoStop}%
\bibitem [{\citenamefont {Corder}\ \emph
  {et~al.}(2018{\natexlab{a}})\citenamefont {Corder}, \citenamefont {Zhao},
  \citenamefont {Bakalis}, \citenamefont {Li}, \citenamefont {Kershis},
  \citenamefont {Muraca}, \citenamefont {White},\ and\ \citenamefont
  {Allison}}]{Corder_StructDyn2018}%
  \BibitemOpen
  \bibfield  {author} {\bibinfo {author} {\bibfnamefont {C.}~\bibnamefont
  {Corder}}, \bibinfo {author} {\bibfnamefont {P.}~\bibnamefont {Zhao}},
  \bibinfo {author} {\bibfnamefont {J.}~\bibnamefont {Bakalis}}, \bibinfo
  {author} {\bibfnamefont {X.}~\bibnamefont {Li}}, \bibinfo {author}
  {\bibfnamefont {M.~D.}\ \bibnamefont {Kershis}}, \bibinfo {author}
  {\bibfnamefont {A.~R.}\ \bibnamefont {Muraca}}, \bibinfo {author}
  {\bibfnamefont {M.~G.}\ \bibnamefont {White}}, \ and\ \bibinfo {author}
  {\bibfnamefont {T.~K.}\ \bibnamefont {Allison}},\ }\bibfield  {title}
  {\enquote {\bibinfo {title} {Ultrafast extreme ultraviolet photoemission
  without space charge},}\ }\href {\doibase 10.1063/1.5045578} {\bibfield
  {journal} {\bibinfo  {journal} {Structural Dynamics}\ }\textbf {\bibinfo
  {volume} {5}},\ \bibinfo {pages} {054301} (\bibinfo {year}
  {2018}{\natexlab{a}})},\ \Eprint
  {http://arxiv.org/abs/https://doi.org/10.1063/1.5045578}
  {https://doi.org/10.1063/1.5045578} \BibitemShut {NoStop}%
\bibitem [{\citenamefont {Mills}\ \emph {et~al.}(2019)\citenamefont {Mills},
  \citenamefont {Zhdanovich}, \citenamefont {Na}, \citenamefont {Boschini},
  \citenamefont {Razzoli}, \citenamefont {Michiardi}, \citenamefont
  {Sheyerman}, \citenamefont {Schneider}, \citenamefont {Hammond},
  \citenamefont {S{\"u}ss}, \citenamefont {Felser}, \citenamefont
  {Damascelli},\ and\ \citenamefont {Jones}}]{Mills_RSI2019}%
  \BibitemOpen
  \bibfield  {author} {\bibinfo {author} {\bibfnamefont {A.~K.}\ \bibnamefont
  {Mills}}, \bibinfo {author} {\bibfnamefont {S.}~\bibnamefont {Zhdanovich}},
  \bibinfo {author} {\bibfnamefont {M.~X.}\ \bibnamefont {Na}}, \bibinfo
  {author} {\bibfnamefont {F.}~\bibnamefont {Boschini}}, \bibinfo {author}
  {\bibfnamefont {E.}~\bibnamefont {Razzoli}}, \bibinfo {author} {\bibfnamefont
  {M.}~\bibnamefont {Michiardi}}, \bibinfo {author} {\bibfnamefont
  {A.}~\bibnamefont {Sheyerman}}, \bibinfo {author} {\bibfnamefont
  {M.}~\bibnamefont {Schneider}}, \bibinfo {author} {\bibfnamefont {T.~J.}\
  \bibnamefont {Hammond}}, \bibinfo {author} {\bibfnamefont {V.}~\bibnamefont
  {S{\"u}ss}}, \bibinfo {author} {\bibfnamefont {C.}~\bibnamefont {Felser}},
  \bibinfo {author} {\bibfnamefont {A.}~\bibnamefont {Damascelli}}, \ and\
  \bibinfo {author} {\bibfnamefont {D.~J.}\ \bibnamefont {Jones}},\ }\bibfield
  {title} {\enquote {\bibinfo {title} {Cavity-enhanced high harmonic generation
  for extreme ultraviolet time- and angle-resolved photoemission
  spectroscopy},}\ }\href {\doibase 10.1063/1.5090507} {\bibfield  {journal}
  {\bibinfo  {journal} {Review of Scientific Instruments}\ }\textbf {\bibinfo
  {volume} {90}},\ \bibinfo {pages} {083001} (\bibinfo {year} {2019})},\
  \Eprint {http://arxiv.org/abs/https://doi.org/10.1063/1.5090507}
  {https://doi.org/10.1063/1.5090507} \BibitemShut {NoStop}%
\bibitem [{\citenamefont {Saule}\ \emph {et~al.}(2019)\citenamefont {Saule},
  \citenamefont {Heinrich}, \citenamefont {Sch{\"o}tz}, \citenamefont
  {Lilienfein}, \citenamefont {H{\"o}gner}, \citenamefont {deVries},
  \citenamefont {Pl{\"o}tner}, \citenamefont {Weitenberg}, \citenamefont
  {Esser}, \citenamefont {Schulte}, \citenamefont {Russbueldt}, \citenamefont
  {Limpert}, \citenamefont {Kling}, \citenamefont {Kleineberg},\ and\
  \citenamefont {Pupeza}}]{Saule_NatComm2019}%
  \BibitemOpen
  \bibfield  {author} {\bibinfo {author} {\bibfnamefont {T.}~\bibnamefont
  {Saule}}, \bibinfo {author} {\bibfnamefont {S.}~\bibnamefont {Heinrich}},
  \bibinfo {author} {\bibfnamefont {J.}~\bibnamefont {Sch{\"o}tz}}, \bibinfo
  {author} {\bibfnamefont {N.}~\bibnamefont {Lilienfein}}, \bibinfo {author}
  {\bibfnamefont {M.}~\bibnamefont {H{\"o}gner}}, \bibinfo {author}
  {\bibfnamefont {O.}~\bibnamefont {deVries}}, \bibinfo {author} {\bibfnamefont
  {M.}~\bibnamefont {Pl{\"o}tner}}, \bibinfo {author} {\bibfnamefont
  {J.}~\bibnamefont {Weitenberg}}, \bibinfo {author} {\bibfnamefont
  {D.}~\bibnamefont {Esser}}, \bibinfo {author} {\bibfnamefont
  {J.}~\bibnamefont {Schulte}}, \bibinfo {author} {\bibfnamefont
  {P.}~\bibnamefont {Russbueldt}}, \bibinfo {author} {\bibfnamefont
  {J.}~\bibnamefont {Limpert}}, \bibinfo {author} {\bibfnamefont {M.~F.}\
  \bibnamefont {Kling}}, \bibinfo {author} {\bibfnamefont {U.}~\bibnamefont
  {Kleineberg}}, \ and\ \bibinfo {author} {\bibfnamefont {I.}~\bibnamefont
  {Pupeza}},\ }\bibfield  {title} {\enquote {\bibinfo {title} {High-flux
  ultrafast extreme-ultraviolet photoemission spectroscopy at 18.4 mhz pulse
  repetition rate},}\ }\href {\doibase 10.1038/s41467-019-08367-y} {\bibfield
  {journal} {\bibinfo  {journal} {Nature Communications}\ }\textbf {\bibinfo
  {volume} {10}},\ \bibinfo {pages} {458} (\bibinfo {year} {2019})}\BibitemShut
  {NoStop}%
\bibitem [{\citenamefont {Na}, \citenamefont {Mills},\ and\ \citenamefont
  {Jones}(2023)}]{Na_PhysicsReports2023}%
  \BibitemOpen
  \bibfield  {author} {\bibinfo {author} {\bibfnamefont {M.}~\bibnamefont
  {Na}}, \bibinfo {author} {\bibfnamefont {A.~K.}\ \bibnamefont {Mills}}, \
  and\ \bibinfo {author} {\bibfnamefont {D.~J.}\ \bibnamefont {Jones}},\
  }\bibfield  {title} {\enquote {\bibinfo {title} {Advancing time- and
  angle-resolved photoemission spectroscopy: The role of ultrafast laser
  development},}\ }\href {\doibase
  https://doi.org/10.1016/j.physrep.2023.09.005} {\bibfield  {journal}
  {\bibinfo  {journal} {Physics Reports}\ }\textbf {\bibinfo {volume} {1036}},\
  \bibinfo {pages} {1--47} (\bibinfo {year} {2023})}\BibitemShut {NoStop}%
\bibitem [{\citenamefont {Heinrich}\ \emph {et~al.}(2021)\citenamefont
  {Heinrich}, \citenamefont {Saule}, \citenamefont {H{\"o}gner}, \citenamefont
  {Cui}, \citenamefont {Yakovlev}, \citenamefont {Pupeza},\ and\ \citenamefont
  {Kleineberg}}]{Heinrich_NatComm2021}%
  \BibitemOpen
  \bibfield  {author} {\bibinfo {author} {\bibfnamefont {S.}~\bibnamefont
  {Heinrich}}, \bibinfo {author} {\bibfnamefont {T.}~\bibnamefont {Saule}},
  \bibinfo {author} {\bibfnamefont {M.}~\bibnamefont {H{\"o}gner}}, \bibinfo
  {author} {\bibfnamefont {Y.}~\bibnamefont {Cui}}, \bibinfo {author}
  {\bibfnamefont {V.~S.}\ \bibnamefont {Yakovlev}}, \bibinfo {author}
  {\bibfnamefont {I.}~\bibnamefont {Pupeza}}, \ and\ \bibinfo {author}
  {\bibfnamefont {U.}~\bibnamefont {Kleineberg}},\ }\bibfield  {title}
  {\enquote {\bibinfo {title} {Attosecond intra-valence band dynamics and
  resonant-photoemission delays in w(110)},}\ }\href {\doibase
  10.1038/s41467-021-23650-7} {\bibfield  {journal} {\bibinfo  {journal}
  {Nature Communications}\ }\textbf {\bibinfo {volume} {12}},\ \bibinfo {pages}
  {3404} (\bibinfo {year} {2021})}\BibitemShut {NoStop}%
\bibitem [{\citenamefont {Kunin}\ \emph {et~al.}(2023)\citenamefont {Kunin},
  \citenamefont {Chernov}, \citenamefont {Bakalis}, \citenamefont {Li},
  \citenamefont {Cheng}, \citenamefont {Withers}, \citenamefont {White},
  \citenamefont {Sch\"onhense}, \citenamefont {Du}, \citenamefont {Kawakami},\
  and\ \citenamefont {Allison}}]{Kunin_PRL2023}%
  \BibitemOpen
  \bibfield  {author} {\bibinfo {author} {\bibfnamefont {A.}~\bibnamefont
  {Kunin}}, \bibinfo {author} {\bibfnamefont {S.}~\bibnamefont {Chernov}},
  \bibinfo {author} {\bibfnamefont {J.}~\bibnamefont {Bakalis}}, \bibinfo
  {author} {\bibfnamefont {Z.}~\bibnamefont {Li}}, \bibinfo {author}
  {\bibfnamefont {S.}~\bibnamefont {Cheng}}, \bibinfo {author} {\bibfnamefont
  {Z.~H.}\ \bibnamefont {Withers}}, \bibinfo {author} {\bibfnamefont {M.~G.}\
  \bibnamefont {White}}, \bibinfo {author} {\bibfnamefont {G.}~\bibnamefont
  {Sch\"onhense}}, \bibinfo {author} {\bibfnamefont {X.}~\bibnamefont {Du}},
  \bibinfo {author} {\bibfnamefont {R.~K.}\ \bibnamefont {Kawakami}}, \ and\
  \bibinfo {author} {\bibfnamefont {T.~K.}\ \bibnamefont {Allison}},\
  }\bibfield  {title} {\enquote {\bibinfo {title} {Momentum-resolved exciton
  coupling and valley polarization dynamics in monolayer
  ${\mathrm{ws}}_{2}$},}\ }\href {\doibase 10.1103/PhysRevLett.130.046202}
  {\bibfield  {journal} {\bibinfo  {journal} {Phys. Rev. Lett.}\ }\textbf
  {\bibinfo {volume} {130}},\ \bibinfo {pages} {046202} (\bibinfo {year}
  {2023})}\BibitemShut {NoStop}%
\bibitem [{\citenamefont {Na}\ \emph {et~al.}(2019)\citenamefont {Na},
  \citenamefont {Mills}, \citenamefont {Boschini}, \citenamefont {Michiardi},
  \citenamefont {Nosarzewski}, \citenamefont {Day}, \citenamefont {Razzoli},
  \citenamefont {Sheyerman}, \citenamefont {Schneider}, \citenamefont {Levy},
  \citenamefont {Zhdanovich}, \citenamefont {Devereaux}, \citenamefont
  {Kemper}, \citenamefont {Jones},\ and\ \citenamefont
  {Damascelli}}]{Na_Science2019}%
  \BibitemOpen
  \bibfield  {author} {\bibinfo {author} {\bibfnamefont {M.~X.}\ \bibnamefont
  {Na}}, \bibinfo {author} {\bibfnamefont {A.~K.}\ \bibnamefont {Mills}},
  \bibinfo {author} {\bibfnamefont {F.}~\bibnamefont {Boschini}}, \bibinfo
  {author} {\bibfnamefont {M.}~\bibnamefont {Michiardi}}, \bibinfo {author}
  {\bibfnamefont {B.}~\bibnamefont {Nosarzewski}}, \bibinfo {author}
  {\bibfnamefont {R.~P.}\ \bibnamefont {Day}}, \bibinfo {author} {\bibfnamefont
  {E.}~\bibnamefont {Razzoli}}, \bibinfo {author} {\bibfnamefont
  {A.}~\bibnamefont {Sheyerman}}, \bibinfo {author} {\bibfnamefont
  {M.}~\bibnamefont {Schneider}}, \bibinfo {author} {\bibfnamefont
  {G.}~\bibnamefont {Levy}}, \bibinfo {author} {\bibfnamefont {S.}~\bibnamefont
  {Zhdanovich}}, \bibinfo {author} {\bibfnamefont {T.~P.}\ \bibnamefont
  {Devereaux}}, \bibinfo {author} {\bibfnamefont {A.~F.}\ \bibnamefont
  {Kemper}}, \bibinfo {author} {\bibfnamefont {D.~J.}\ \bibnamefont {Jones}}, \
  and\ \bibinfo {author} {\bibfnamefont {A.}~\bibnamefont {Damascelli}},\
  }\bibfield  {title} {\enquote {\bibinfo {title} {Direct determination of
  mode-projected electron-phonon coupling in the time domain},}\ }\href
  {\doibase 10.1126/science.aaw1662} {\bibfield  {journal} {\bibinfo  {journal}
  {Science}\ }\textbf {\bibinfo {volume} {366}},\ \bibinfo {pages} {1231--1236}
  (\bibinfo {year} {2019})},\ \Eprint
  {http://arxiv.org/abs/https://science.sciencemag.org/content/366/6470/1231.full.pdf}
  {https://science.sciencemag.org/content/366/6470/1231.full.pdf} \BibitemShut
  {NoStop}%
\bibitem [{\citenamefont {Wang}\ \emph {et~al.}(2018)\citenamefont {Wang},
  \citenamefont {Chernikov}, \citenamefont {Glazov}, \citenamefont {Heinz},
  \citenamefont {Marie}, \citenamefont {Amand},\ and\ \citenamefont
  {Urbaszek}}]{Wang_RMP2018}%
  \BibitemOpen
  \bibfield  {author} {\bibinfo {author} {\bibfnamefont {G.}~\bibnamefont
  {Wang}}, \bibinfo {author} {\bibfnamefont {A.}~\bibnamefont {Chernikov}},
  \bibinfo {author} {\bibfnamefont {M.~M.}\ \bibnamefont {Glazov}}, \bibinfo
  {author} {\bibfnamefont {T.~F.}\ \bibnamefont {Heinz}}, \bibinfo {author}
  {\bibfnamefont {X.}~\bibnamefont {Marie}}, \bibinfo {author} {\bibfnamefont
  {T.}~\bibnamefont {Amand}}, \ and\ \bibinfo {author} {\bibfnamefont
  {B.}~\bibnamefont {Urbaszek}},\ }\bibfield  {title} {\enquote {\bibinfo
  {title} {Colloquium: Excitons in atomically thin transition metal
  dichalcogenides},}\ }\href {\doibase 10.1103/RevModPhys.90.021001} {\bibfield
   {journal} {\bibinfo  {journal} {Rev. Mod. Phys.}\ }\textbf {\bibinfo
  {volume} {90}},\ \bibinfo {pages} {021001} (\bibinfo {year}
  {2018})}\BibitemShut {NoStop}%
\bibitem [{\citenamefont {Miaja-Avila}\ \emph {et~al.}(2006)\citenamefont
  {Miaja-Avila}, \citenamefont {Lei}, \citenamefont {Aeschlimann},
  \citenamefont {Gland}, \citenamefont {Murnane}, \citenamefont {Kapteyn},\
  and\ \citenamefont {Saathoff}}]{Miaja-Avila_PRL2006}%
  \BibitemOpen
  \bibfield  {author} {\bibinfo {author} {\bibfnamefont {L.}~\bibnamefont
  {Miaja-Avila}}, \bibinfo {author} {\bibfnamefont {C.}~\bibnamefont {Lei}},
  \bibinfo {author} {\bibfnamefont {M.}~\bibnamefont {Aeschlimann}}, \bibinfo
  {author} {\bibfnamefont {J.~L.}\ \bibnamefont {Gland}}, \bibinfo {author}
  {\bibfnamefont {M.~M.}\ \bibnamefont {Murnane}}, \bibinfo {author}
  {\bibfnamefont {H.~C.}\ \bibnamefont {Kapteyn}}, \ and\ \bibinfo {author}
  {\bibfnamefont {G.}~\bibnamefont {Saathoff}},\ }\bibfield  {title} {\enquote
  {\bibinfo {title} {Laser-assisted photoelectric effect from surfaces},}\
  }\href {\doibase 10.1103/PhysRevLett.97.113604} {\bibfield  {journal}
  {\bibinfo  {journal} {Phys. Rev. Lett.}\ }\textbf {\bibinfo {volume} {97}},\
  \bibinfo {pages} {113604} (\bibinfo {year} {2006})}\BibitemShut {NoStop}%
\bibitem [{\citenamefont {Saathoff}\ \emph {et~al.}(2008)\citenamefont
  {Saathoff}, \citenamefont {Miaja-Avila}, \citenamefont {Aeschlimann},
  \citenamefont {Murnane},\ and\ \citenamefont {Kapteyn}}]{Saathoff_PRA2008}%
  \BibitemOpen
  \bibfield  {author} {\bibinfo {author} {\bibfnamefont {G.}~\bibnamefont
  {Saathoff}}, \bibinfo {author} {\bibfnamefont {L.}~\bibnamefont
  {Miaja-Avila}}, \bibinfo {author} {\bibfnamefont {M.}~\bibnamefont
  {Aeschlimann}}, \bibinfo {author} {\bibfnamefont {M.~M.}\ \bibnamefont
  {Murnane}}, \ and\ \bibinfo {author} {\bibfnamefont {H.~C.}\ \bibnamefont
  {Kapteyn}},\ }\bibfield  {title} {\enquote {\bibinfo {title} {Laser-assisted
  photoemission from surfaces},}\ }\href {\doibase 10.1103/PhysRevA.77.022903}
  {\bibfield  {journal} {\bibinfo  {journal} {Phys. Rev. A}\ }\textbf {\bibinfo
  {volume} {77}},\ \bibinfo {pages} {022903} (\bibinfo {year}
  {2008})}\BibitemShut {NoStop}%
\bibitem [{\citenamefont {Paul}\ \emph {et~al.}(2001)\citenamefont {Paul},
  \citenamefont {Toma}, \citenamefont {Breger}, \citenamefont {Mullot},
  \citenamefont {Auge}, \citenamefont {Balcou}, \citenamefont {Muller},\ and\
  \citenamefont {Agostini}}]{Paul_Science2001}%
  \BibitemOpen
  \bibfield  {author} {\bibinfo {author} {\bibfnamefont {P.~M.}\ \bibnamefont
  {Paul}}, \bibinfo {author} {\bibfnamefont {E.~S.}\ \bibnamefont {Toma}},
  \bibinfo {author} {\bibfnamefont {P.}~\bibnamefont {Breger}}, \bibinfo
  {author} {\bibfnamefont {G.}~\bibnamefont {Mullot}}, \bibinfo {author}
  {\bibfnamefont {F.}~\bibnamefont {Auge}}, \bibinfo {author} {\bibfnamefont
  {P.}~\bibnamefont {Balcou}}, \bibinfo {author} {\bibfnamefont {H.~G.}\
  \bibnamefont {Muller}}, \ and\ \bibinfo {author} {\bibfnamefont
  {P.}~\bibnamefont {Agostini}},\ }\bibfield  {title} {\enquote {\bibinfo
  {title} {Observation of a train of attosecond pulses from high harmonic
  generation},}\ }\href@noop {} {\bibfield  {journal} {\bibinfo  {journal}
  {Science}\ }\textbf {\bibinfo {volume} {292}},\ \bibinfo {pages} {1689--1692}
  (\bibinfo {year} {2001})}\BibitemShut {NoStop}%
\bibitem [{\citenamefont {Locher}\ \emph {et~al.}(2015)\citenamefont {Locher},
  \citenamefont {Castiglioni}, \citenamefont {Lucchini}, \citenamefont {Greif},
  \citenamefont {Gallmann}, \citenamefont {Osterwalder}, \citenamefont
  {Hengsberger},\ and\ \citenamefont {Keller}}]{Locher_Optica2015}%
  \BibitemOpen
  \bibfield  {author} {\bibinfo {author} {\bibfnamefont {R.}~\bibnamefont
  {Locher}}, \bibinfo {author} {\bibfnamefont {L.}~\bibnamefont {Castiglioni}},
  \bibinfo {author} {\bibfnamefont {M.}~\bibnamefont {Lucchini}}, \bibinfo
  {author} {\bibfnamefont {M.}~\bibnamefont {Greif}}, \bibinfo {author}
  {\bibfnamefont {L.}~\bibnamefont {Gallmann}}, \bibinfo {author}
  {\bibfnamefont {J.}~\bibnamefont {Osterwalder}}, \bibinfo {author}
  {\bibfnamefont {M.}~\bibnamefont {Hengsberger}}, \ and\ \bibinfo {author}
  {\bibfnamefont {U.}~\bibnamefont {Keller}},\ }\bibfield  {title} {\enquote
  {\bibinfo {title} {Energy-dependent photoemission delays from noble metal
  surfaces by attosecond interferometry},}\ }\href {\doibase
  10.1364/OPTICA.2.000405} {\bibfield  {journal} {\bibinfo  {journal} {Optica}\
  }\textbf {\bibinfo {volume} {2}},\ \bibinfo {pages} {405--410} (\bibinfo
  {year} {2015})}\BibitemShut {NoStop}%
\bibitem [{\citenamefont {Lucchini}\ \emph {et~al.}(2015)\citenamefont
  {Lucchini}, \citenamefont {Castiglioni}, \citenamefont {Kasmi}, \citenamefont
  {Kliuiev}, \citenamefont {Ludwig}, \citenamefont {Greif}, \citenamefont
  {Osterwalder}, \citenamefont {Hengsberger}, \citenamefont {Gallmann},\ and\
  \citenamefont {Keller}}]{Luchhini_PRL2015}%
  \BibitemOpen
  \bibfield  {author} {\bibinfo {author} {\bibfnamefont {M.}~\bibnamefont
  {Lucchini}}, \bibinfo {author} {\bibfnamefont {L.}~\bibnamefont
  {Castiglioni}}, \bibinfo {author} {\bibfnamefont {L.}~\bibnamefont {Kasmi}},
  \bibinfo {author} {\bibfnamefont {P.}~\bibnamefont {Kliuiev}}, \bibinfo
  {author} {\bibfnamefont {A.}~\bibnamefont {Ludwig}}, \bibinfo {author}
  {\bibfnamefont {M.}~\bibnamefont {Greif}}, \bibinfo {author} {\bibfnamefont
  {J.}~\bibnamefont {Osterwalder}}, \bibinfo {author} {\bibfnamefont
  {M.}~\bibnamefont {Hengsberger}}, \bibinfo {author} {\bibfnamefont
  {L.}~\bibnamefont {Gallmann}}, \ and\ \bibinfo {author} {\bibfnamefont
  {U.}~\bibnamefont {Keller}},\ }\bibfield  {title} {\enquote {\bibinfo {title}
  {Light-matter interaction at surfaces in the spatiotemporal limit of
  macroscopic models},}\ }\href {\doibase 10.1103/PhysRevLett.115.137401}
  {\bibfield  {journal} {\bibinfo  {journal} {Phys. Rev. Lett.}\ }\textbf
  {\bibinfo {volume} {115}},\ \bibinfo {pages} {137401} (\bibinfo {year}
  {2015})}\BibitemShut {NoStop}%
\bibitem [{\citenamefont {Tao}\ \emph {et~al.}(2016)\citenamefont {Tao},
  \citenamefont {Chen}, \citenamefont {Szilv{\'a}si}, \citenamefont {Keller},
  \citenamefont {Mavrikakis}, \citenamefont {Kapteyn},\ and\ \citenamefont
  {Murnane}}]{Tao_Science2016}%
  \BibitemOpen
  \bibfield  {author} {\bibinfo {author} {\bibfnamefont {Z.}~\bibnamefont
  {Tao}}, \bibinfo {author} {\bibfnamefont {C.}~\bibnamefont {Chen}}, \bibinfo
  {author} {\bibfnamefont {T.}~\bibnamefont {Szilv{\'a}si}}, \bibinfo {author}
  {\bibfnamefont {M.}~\bibnamefont {Keller}}, \bibinfo {author} {\bibfnamefont
  {M.}~\bibnamefont {Mavrikakis}}, \bibinfo {author} {\bibfnamefont
  {H.}~\bibnamefont {Kapteyn}}, \ and\ \bibinfo {author} {\bibfnamefont
  {M.}~\bibnamefont {Murnane}},\ }\bibfield  {title} {\enquote {\bibinfo
  {title} {Direct time-domain observation of attosecond final-state lifetimes
  in photoemission from solids},}\ }\href {\doibase 10.1126/science.aaf6793}
  {\bibfield  {journal} {\bibinfo  {journal} {Science}\ }\textbf {\bibinfo
  {volume} {353}},\ \bibinfo {pages} {62--67} (\bibinfo {year} {2016})},\
  \Eprint
  {http://arxiv.org/abs/http://science.sciencemag.org/content/353/6294/62.full.pdf}
  {http://science.sciencemag.org/content/353/6294/62.full.pdf} \BibitemShut
  {NoStop}%
\bibitem [{\citenamefont {Allison}\ \emph {et~al.}(2011)\citenamefont
  {Allison}, \citenamefont {Cing\"oz}, \citenamefont {Yost},\ and\
  \citenamefont {Ye}}]{Allison_PRL2011}%
  \BibitemOpen
  \bibfield  {author} {\bibinfo {author} {\bibfnamefont {T.~K.}\ \bibnamefont
  {Allison}}, \bibinfo {author} {\bibfnamefont {A.}~\bibnamefont {Cing\"oz}},
  \bibinfo {author} {\bibfnamefont {D.~C.}\ \bibnamefont {Yost}}, \ and\
  \bibinfo {author} {\bibfnamefont {J.}~\bibnamefont {Ye}},\ }\bibfield
  {title} {\enquote {\bibinfo {title} {Extreme nonlinear optics in a
  femtosecond enhancement cavity},}\ }\href {\doibase
  10.1103/PhysRevLett.107.183903} {\bibfield  {journal} {\bibinfo  {journal}
  {Phys. Rev. Lett.}\ }\textbf {\bibinfo {volume} {107}},\ \bibinfo {pages}
  {183903} (\bibinfo {year} {2011})}\BibitemShut {NoStop}%
\bibitem [{\citenamefont {L{\"u}ftner}\ \emph {et~al.}(2014)\citenamefont
  {L{\"u}ftner}, \citenamefont {Milko}, \citenamefont {Huppmann}, \citenamefont
  {Scholz}, \citenamefont {Ngyuen}, \citenamefont {Wie{\ss}ner}, \citenamefont
  {Sch{\"o}ll}, \citenamefont {Reinert},\ and\ \citenamefont
  {Puschnig}}]{Luftner_JElecSpec2014}%
  \BibitemOpen
  \bibfield  {author} {\bibinfo {author} {\bibfnamefont {D.}~\bibnamefont
  {L{\"u}ftner}}, \bibinfo {author} {\bibfnamefont {M.}~\bibnamefont {Milko}},
  \bibinfo {author} {\bibfnamefont {S.}~\bibnamefont {Huppmann}}, \bibinfo
  {author} {\bibfnamefont {M.}~\bibnamefont {Scholz}}, \bibinfo {author}
  {\bibfnamefont {N.}~\bibnamefont {Ngyuen}}, \bibinfo {author} {\bibfnamefont
  {M.}~\bibnamefont {Wie{\ss}ner}}, \bibinfo {author} {\bibfnamefont
  {A.}~\bibnamefont {Sch{\"o}ll}}, \bibinfo {author} {\bibfnamefont
  {F.}~\bibnamefont {Reinert}}, \ and\ \bibinfo {author} {\bibfnamefont
  {P.}~\bibnamefont {Puschnig}},\ }\bibfield  {title} {\enquote {\bibinfo
  {title} {Cupc/au(110): Determination of the azimuthal alignment by a
  combination of angle-resolved photoemission and density functional theory},}\
  }\href {\doibase https://doi.org/10.1016/j.elspec.2014.06.002} {\bibfield
  {journal} {\bibinfo  {journal} {Journal of Electron Spectroscopy and Related
  Phenomena}\ }\textbf {\bibinfo {volume} {195}},\ \bibinfo {pages} {293--300}
  (\bibinfo {year} {2014})}\BibitemShut {NoStop}%
\bibitem [{\citenamefont {Puschnig}\ \emph {et~al.}(2009)\citenamefont
  {Puschnig}, \citenamefont {Berkebile}, \citenamefont {Fleming}, \citenamefont
  {Koller}, \citenamefont {Emtsev}, \citenamefont {Seyller}, \citenamefont
  {Riley}, \citenamefont {Ambrosch-Draxl}, \citenamefont {Netzer},\ and\
  \citenamefont {Ramsey}}]{Puschnig_Science2009}%
  \BibitemOpen
  \bibfield  {author} {\bibinfo {author} {\bibfnamefont {P.}~\bibnamefont
  {Puschnig}}, \bibinfo {author} {\bibfnamefont {S.}~\bibnamefont {Berkebile}},
  \bibinfo {author} {\bibfnamefont {A.~J.}\ \bibnamefont {Fleming}}, \bibinfo
  {author} {\bibfnamefont {G.}~\bibnamefont {Koller}}, \bibinfo {author}
  {\bibfnamefont {K.}~\bibnamefont {Emtsev}}, \bibinfo {author} {\bibfnamefont
  {T.}~\bibnamefont {Seyller}}, \bibinfo {author} {\bibfnamefont {J.~D.}\
  \bibnamefont {Riley}}, \bibinfo {author} {\bibfnamefont {C.}~\bibnamefont
  {Ambrosch-Draxl}}, \bibinfo {author} {\bibfnamefont {F.~P.}\ \bibnamefont
  {Netzer}}, \ and\ \bibinfo {author} {\bibfnamefont {M.~G.}\ \bibnamefont
  {Ramsey}},\ }\bibfield  {title} {\enquote {\bibinfo {title} {Reconstruction
  of molecular orbital densities from photoemission data},}\ }\href {\doibase
  10.1126/science.1176105} {\bibfield  {journal} {\bibinfo  {journal}
  {Science}\ }\textbf {\bibinfo {volume} {326}},\ \bibinfo {pages} {702--706}
  (\bibinfo {year} {2009})},\ \Eprint
  {http://arxiv.org/abs/http://science.sciencemag.org/content/326/5953/702.full.pdf}
  {http://science.sciencemag.org/content/326/5953/702.full.pdf} \BibitemShut
  {NoStop}%
\bibitem [{\citenamefont {Damascelli}, \citenamefont {Hussain},\ and\
  \citenamefont {Shen}(2003)}]{Damascelli_RMP2003}%
  \BibitemOpen
  \bibfield  {author} {\bibinfo {author} {\bibfnamefont {A.}~\bibnamefont
  {Damascelli}}, \bibinfo {author} {\bibfnamefont {Z.}~\bibnamefont {Hussain}},
  \ and\ \bibinfo {author} {\bibfnamefont {Z.-X.}\ \bibnamefont {Shen}},\
  }\bibfield  {title} {\enquote {\bibinfo {title} {Angle-resolved photoemission
  studies of the cuprate superconductors},}\ }\href {\doibase
  10.1103/RevModPhys.75.473} {\bibfield  {journal} {\bibinfo  {journal} {Rev.
  Mod. Phys.}\ }\textbf {\bibinfo {volume} {75}},\ \bibinfo {pages} {473--541}
  (\bibinfo {year} {2003})}\BibitemShut {NoStop}%
\bibitem [{\citenamefont {Pollard}, \citenamefont {Lee},\ and\ \citenamefont
  {Mathies}(1990)}]{Pollard_JCP1990}%
  \BibitemOpen
  \bibfield  {author} {\bibinfo {author} {\bibfnamefont {W.~T.}\ \bibnamefont
  {Pollard}}, \bibinfo {author} {\bibfnamefont {S.-Y.}\ \bibnamefont {Lee}}, \
  and\ \bibinfo {author} {\bibfnamefont {R.~A.}\ \bibnamefont {Mathies}},\
  }\bibfield  {title} {\enquote {\bibinfo {title} {Wave packet theory of
  dynamic absorption spectra in femtosecond pump--probe experiments},}\ }\href
  {\doibase 10.1063/1.457815} {\bibfield  {journal} {\bibinfo  {journal} {The
  Journal of Chemical Physics}\ }\textbf {\bibinfo {volume} {92}},\ \bibinfo
  {pages} {4012--4029} (\bibinfo {year} {1990})}\BibitemShut {NoStop}%
\bibitem [{\citenamefont {Wang}\ \emph {et~al.}(2015)\citenamefont {Wang},
  \citenamefont {Xu}, \citenamefont {Ulonska}, \citenamefont {Robinson},
  \citenamefont {Ranitovic},\ and\ \citenamefont {Kaindl}}]{Wang_NatCom2015}%
  \BibitemOpen
  \bibfield  {author} {\bibinfo {author} {\bibfnamefont {H.}~\bibnamefont
  {Wang}}, \bibinfo {author} {\bibfnamefont {Y.}~\bibnamefont {Xu}}, \bibinfo
  {author} {\bibfnamefont {S.}~\bibnamefont {Ulonska}}, \bibinfo {author}
  {\bibfnamefont {J.~S.}\ \bibnamefont {Robinson}}, \bibinfo {author}
  {\bibfnamefont {P.}~\bibnamefont {Ranitovic}}, \ and\ \bibinfo {author}
  {\bibfnamefont {R.~A.}\ \bibnamefont {Kaindl}},\ }\bibfield  {title}
  {\enquote {\bibinfo {title} {Bright high-repetition-rate source of narrowband
  extreme-ultraviolet harmonics beyond 22 ev},}\ }\href
  {http://dx.doi.org/10.1038/ncomms8459} {\ \textbf {\bibinfo {volume} {6}},\
  \bibinfo {pages} {7459 EP --} (\bibinfo {year} {2015})}\BibitemShut {NoStop}%
\bibitem [{\citenamefont {Hellmann}\ \emph {et~al.}(2009)\citenamefont
  {Hellmann}, \citenamefont {Rossnagel}, \citenamefont {Marczynski-B\"uhlow},\
  and\ \citenamefont {Kipp}}]{Hellmann_PRB2009}%
  \BibitemOpen
  \bibfield  {author} {\bibinfo {author} {\bibfnamefont {S.}~\bibnamefont
  {Hellmann}}, \bibinfo {author} {\bibfnamefont {K.}~\bibnamefont {Rossnagel}},
  \bibinfo {author} {\bibfnamefont {M.}~\bibnamefont {Marczynski-B\"uhlow}}, \
  and\ \bibinfo {author} {\bibfnamefont {L.}~\bibnamefont {Kipp}},\ }\bibfield
  {title} {\enquote {\bibinfo {title} {Vacuum space-charge effects in
  solid-state photoemission},}\ }\href {\doibase 10.1103/PhysRevB.79.035402}
  {\bibfield  {journal} {\bibinfo  {journal} {Phys. Rev. B}\ }\textbf {\bibinfo
  {volume} {79}},\ \bibinfo {pages} {035402} (\bibinfo {year}
  {2009})}\BibitemShut {NoStop}%
\bibitem [{\citenamefont {Frietsch}\ \emph {et~al.}(2013)\citenamefont
  {Frietsch}, \citenamefont {Carley}, \citenamefont {D{\"o}brich},
  \citenamefont {Gahl}, \citenamefont {Teichmann}, \citenamefont {Schwarzkopf},
  \citenamefont {Wernet},\ and\ \citenamefont {Weinelt}}]{Frietsch_RSI2013}%
  \BibitemOpen
  \bibfield  {author} {\bibinfo {author} {\bibfnamefont {B.}~\bibnamefont
  {Frietsch}}, \bibinfo {author} {\bibfnamefont {R.}~\bibnamefont {Carley}},
  \bibinfo {author} {\bibfnamefont {K.}~\bibnamefont {D{\"o}brich}}, \bibinfo
  {author} {\bibfnamefont {C.}~\bibnamefont {Gahl}}, \bibinfo {author}
  {\bibfnamefont {M.}~\bibnamefont {Teichmann}}, \bibinfo {author}
  {\bibfnamefont {O.}~\bibnamefont {Schwarzkopf}}, \bibinfo {author}
  {\bibfnamefont {P.}~\bibnamefont {Wernet}}, \ and\ \bibinfo {author}
  {\bibfnamefont {M.}~\bibnamefont {Weinelt}},\ }\bibfield  {title} {\enquote
  {\bibinfo {title} {A high-order harmonic generation apparatus for time- and
  angle-resolved photoelectron spectroscopy},}\ }\href {\doibase
  10.1063/1.4812992} {\bibfield  {journal} {\bibinfo  {journal} {Review of
  Scientific Instruments}\ }\textbf {\bibinfo {volume} {84}},\ \bibinfo {pages}
  {075106} (\bibinfo {year} {2013})},\ \Eprint
  {http://arxiv.org/abs/https://doi.org/10.1063/1.4812992}
  {https://doi.org/10.1063/1.4812992} \BibitemShut {NoStop}%
\bibitem [{\citenamefont {Mathias}\ \emph {et~al.}(2010)\citenamefont
  {Mathias}, \citenamefont {Bauer}, \citenamefont {Aeschlimann}, \citenamefont
  {Kapteyn},\ and\ \citenamefont {Murnane}}]{Mathias_Collection2010}%
  \BibitemOpen
  \bibfield  {author} {\bibinfo {author} {\bibfnamefont {S.}~\bibnamefont
  {Mathias}}, \bibinfo {author} {\bibfnamefont {M.}~\bibnamefont {Bauer}},
  \bibinfo {author} {\bibfnamefont {L.}~\bibnamefont {Aeschlimann},
  \bibfnamefont {M.and Miaja-Avial}}, \bibinfo {author} {\bibfnamefont {H.~C.}\
  \bibnamefont {Kapteyn}}, \ and\ \bibinfo {author} {\bibfnamefont {M.~M.}\
  \bibnamefont {Murnane}},\ }\bibfield  {title} {\enquote {\bibinfo {title}
  {Time resolved photoelectron spectroscopy at surfaces using femtosecond xuv
  pulses},}\ }in\ \href@noop {} {\emph {\bibinfo {booktitle} {Dynamics at Solid
  State Surfaces and Interfaces}}},\ \bibinfo {editor} {edited by\ \bibinfo
  {editor} {\bibfnamefont {U.}~\bibnamefont {Bovensiepen}}, \bibinfo {editor}
  {\bibfnamefont {H.}~\bibnamefont {Petek}}, \ and\ \bibinfo {editor}
  {\bibfnamefont {M.}~\bibnamefont {Wolf}}}\ (\bibinfo  {publisher} {WILEY-VCH
  Verlag},\ \bibinfo {year} {2010})\BibitemShut {NoStop}%
\bibitem [{\citenamefont {Graf}\ \emph {et~al.}(2010)\citenamefont {Graf},
  \citenamefont {Hellmann}, \citenamefont {Jozwiak}, \citenamefont {Smallwood},
  \citenamefont {Hussain}, \citenamefont {Kaindl}, \citenamefont {Kipp},
  \citenamefont {Rossnagel},\ and\ \citenamefont
  {Lanzara}}]{Graf_JApplPhys2010}%
  \BibitemOpen
  \bibfield  {author} {\bibinfo {author} {\bibfnamefont {J.}~\bibnamefont
  {Graf}}, \bibinfo {author} {\bibfnamefont {S.}~\bibnamefont {Hellmann}},
  \bibinfo {author} {\bibfnamefont {C.}~\bibnamefont {Jozwiak}}, \bibinfo
  {author} {\bibfnamefont {C.~L.}\ \bibnamefont {Smallwood}}, \bibinfo {author}
  {\bibfnamefont {Z.}~\bibnamefont {Hussain}}, \bibinfo {author} {\bibfnamefont
  {R.~A.}\ \bibnamefont {Kaindl}}, \bibinfo {author} {\bibfnamefont
  {L.}~\bibnamefont {Kipp}}, \bibinfo {author} {\bibfnamefont {K.}~\bibnamefont
  {Rossnagel}}, \ and\ \bibinfo {author} {\bibfnamefont {A.}~\bibnamefont
  {Lanzara}},\ }\bibfield  {title} {\enquote {\bibinfo {title} {Vacuum space
  charge effect in laser-based solid-state photoemission spectroscopy},}\
  }\href {\doibase http://dx.doi.org/10.1063/1.3273487} {\bibfield  {journal}
  {\bibinfo  {journal} {Journal of Applied Physics}\ }\textbf {\bibinfo
  {volume} {107}},\ \bibinfo {eid} {014912} (\bibinfo {year}
  {2010})}\BibitemShut {NoStop}%
\bibitem [{\citenamefont {Passlack}\ \emph {et~al.}(2006)\citenamefont
  {Passlack}, \citenamefont {Mathias}, \citenamefont {Andreyev}, \citenamefont
  {Mittnacht}, \citenamefont {Aeschlimann},\ and\ \citenamefont
  {Bauer}}]{Passlack_JApplPhys2006}%
  \BibitemOpen
  \bibfield  {author} {\bibinfo {author} {\bibfnamefont {S.}~\bibnamefont
  {Passlack}}, \bibinfo {author} {\bibfnamefont {S.}~\bibnamefont {Mathias}},
  \bibinfo {author} {\bibfnamefont {O.}~\bibnamefont {Andreyev}}, \bibinfo
  {author} {\bibfnamefont {D.}~\bibnamefont {Mittnacht}}, \bibinfo {author}
  {\bibfnamefont {M.}~\bibnamefont {Aeschlimann}}, \ and\ \bibinfo {author}
  {\bibfnamefont {M.}~\bibnamefont {Bauer}},\ }\bibfield  {title} {\enquote
  {\bibinfo {title} {Space charge effects in photoemission with a low
  repetition, high intensity femtosecond laser source},}\ }\href {\doibase
  http://dx.doi.org/10.1063/1.2217985} {\bibfield  {journal} {\bibinfo
  {journal} {Journal of Applied Physics}\ }\textbf {\bibinfo {volume} {100}},\
  \bibinfo {eid} {024912} (\bibinfo {year} {2006})}\BibitemShut {NoStop}%
\bibitem [{\citenamefont {Zhou}\ \emph {et~al.}(2005)\citenamefont {Zhou},
  \citenamefont {Wannberg}, \citenamefont {Yang}, \citenamefont {Brouet},
  \citenamefont {Sun}, \citenamefont {Douglas}, \citenamefont {Dessau},
  \citenamefont {Hussain},\ and\ \citenamefont {Shen}}]{Zhou_JElecSpec2005}%
  \BibitemOpen
  \bibfield  {author} {\bibinfo {author} {\bibfnamefont {X.}~\bibnamefont
  {Zhou}}, \bibinfo {author} {\bibfnamefont {B.}~\bibnamefont {Wannberg}},
  \bibinfo {author} {\bibfnamefont {W.}~\bibnamefont {Yang}}, \bibinfo {author}
  {\bibfnamefont {V.}~\bibnamefont {Brouet}}, \bibinfo {author} {\bibfnamefont
  {Z.}~\bibnamefont {Sun}}, \bibinfo {author} {\bibfnamefont {J.}~\bibnamefont
  {Douglas}}, \bibinfo {author} {\bibfnamefont {D.}~\bibnamefont {Dessau}},
  \bibinfo {author} {\bibfnamefont {Z.}~\bibnamefont {Hussain}}, \ and\
  \bibinfo {author} {\bibfnamefont {Z.-X.}\ \bibnamefont {Shen}},\ }\bibfield
  {title} {\enquote {\bibinfo {title} {Space charge effect and mirror charge
  effect in photoemission spectroscopy},}\ }\href {\doibase
  http://dx.doi.org/10.1016/j.elspec.2004.08.004} {\bibfield  {journal}
  {\bibinfo  {journal} {Journal of Electron Spectroscopy and Related
  Phenomena}\ }\textbf {\bibinfo {volume} {142}},\ \bibinfo {pages} {27 -- 38}
  (\bibinfo {year} {2005})}\BibitemShut {NoStop}%
\bibitem [{\citenamefont {Rotenberg}\ and\ \citenamefont
  {Bostwick}(2014)}]{Rotenberg_JSyncRad2014}%
  \BibitemOpen
  \bibfield  {author} {\bibinfo {author} {\bibfnamefont {E.}~\bibnamefont
  {Rotenberg}}\ and\ \bibinfo {author} {\bibfnamefont {A.}~\bibnamefont
  {Bostwick}},\ }\bibfield  {title} {\enquote {\bibinfo {title} {{microARPES
  and nanoARPES at diffraction-limited light sources: opportunities and
  performance gains}},}\ }\href {\doibase 10.1107/S1600577514015409} {\bibfield
   {journal} {\bibinfo  {journal} {Journal of Synchrotron Radiation}\ }\textbf
  {\bibinfo {volume} {21}},\ \bibinfo {pages} {1048--1056} (\bibinfo {year}
  {2014})}\BibitemShut {NoStop}%
\bibitem [{\citenamefont {Sch{\"o}nhense}\ \emph {et~al.}(2018)\citenamefont
  {Sch{\"o}nhense}, \citenamefont {Medjanik}, \citenamefont {Fedchenko},
  \citenamefont {Chernov}, \citenamefont {Ellguth}, \citenamefont {Vasilyev},
  \citenamefont {Oelsner}, \citenamefont {Viefhaus}, \citenamefont
  {Kutnyakhov}, \citenamefont {Wurth}, \citenamefont {Elmers},\ and\
  \citenamefont {Sch{\"o}nhense}}]{Schoenhense_NewJPhys2018}%
  \BibitemOpen
  \bibfield  {author} {\bibinfo {author} {\bibfnamefont {B.}~\bibnamefont
  {Sch{\"o}nhense}}, \bibinfo {author} {\bibfnamefont {K.}~\bibnamefont
  {Medjanik}}, \bibinfo {author} {\bibfnamefont {O.}~\bibnamefont {Fedchenko}},
  \bibinfo {author} {\bibfnamefont {S.}~\bibnamefont {Chernov}}, \bibinfo
  {author} {\bibfnamefont {M.}~\bibnamefont {Ellguth}}, \bibinfo {author}
  {\bibfnamefont {D.}~\bibnamefont {Vasilyev}}, \bibinfo {author}
  {\bibfnamefont {A.}~\bibnamefont {Oelsner}}, \bibinfo {author} {\bibfnamefont
  {J.}~\bibnamefont {Viefhaus}}, \bibinfo {author} {\bibfnamefont
  {D.}~\bibnamefont {Kutnyakhov}}, \bibinfo {author} {\bibfnamefont
  {W.}~\bibnamefont {Wurth}}, \bibinfo {author} {\bibfnamefont {H.~J.}\
  \bibnamefont {Elmers}}, \ and\ \bibinfo {author} {\bibfnamefont
  {G.}~\bibnamefont {Sch{\"o}nhense}},\ }\bibfield  {title} {\enquote {\bibinfo
  {title} {Multidimensional photoemission spectroscopy{\textemdash}the
  space-charge limit},}\ }\href {\doibase 10.1088/1367-2630/aaa262} {\bibfield
  {journal} {\bibinfo  {journal} {New Journal of Physics}\ }\textbf {\bibinfo
  {volume} {20}},\ \bibinfo {pages} {033004} (\bibinfo {year}
  {2018})}\BibitemShut {NoStop}%
\bibitem [{\citenamefont {Sch\"onhense}\ \emph {et~al.}(2021)\citenamefont
  {Sch\"onhense}, \citenamefont {Kutnyakhov}, \citenamefont {Pressacco},
  \citenamefont {Heber}, \citenamefont {Wind}, \citenamefont {Agustsson},
  \citenamefont {Babenkov}, \citenamefont {Vasilyev}, \citenamefont
  {Fedchenko}, \citenamefont {Chernov}, \citenamefont {Rettig}, \citenamefont
  {Sch{\"o}nhense}, \citenamefont {Wenthaus}, \citenamefont {Brenner},
  \citenamefont {Dziarzhytski}, \citenamefont {Palutke}, \citenamefont
  {Mahatha}, \citenamefont {Schirmel}, \citenamefont {Redlin}, \citenamefont
  {Manschwetus}, \citenamefont {Hartl}, \citenamefont {Matveyev}, \citenamefont
  {Gloskovskii}, \citenamefont {Schlueter}, \citenamefont {Shokeen},
  \citenamefont {Duerr}, \citenamefont {Allison}, \citenamefont {Beye},
  \citenamefont {Rossnagel}, \citenamefont {Elmers},\ and\ \citenamefont
  {Medjanik}}]{Schoenhense_RSI2021}%
  \BibitemOpen
  \bibfield  {author} {\bibinfo {author} {\bibfnamefont {G.}~\bibnamefont
  {Sch\"onhense}}, \bibinfo {author} {\bibfnamefont {D.}~\bibnamefont
  {Kutnyakhov}}, \bibinfo {author} {\bibfnamefont {F.}~\bibnamefont
  {Pressacco}}, \bibinfo {author} {\bibfnamefont {M.}~\bibnamefont {Heber}},
  \bibinfo {author} {\bibfnamefont {N.}~\bibnamefont {Wind}}, \bibinfo {author}
  {\bibfnamefont {S.~Y.}\ \bibnamefont {Agustsson}}, \bibinfo {author}
  {\bibfnamefont {S.}~\bibnamefont {Babenkov}}, \bibinfo {author}
  {\bibfnamefont {D.}~\bibnamefont {Vasilyev}}, \bibinfo {author}
  {\bibfnamefont {O.}~\bibnamefont {Fedchenko}}, \bibinfo {author}
  {\bibfnamefont {S.}~\bibnamefont {Chernov}}, \bibinfo {author} {\bibfnamefont
  {L.}~\bibnamefont {Rettig}}, \bibinfo {author} {\bibfnamefont
  {B.}~\bibnamefont {Sch{\"o}nhense}}, \bibinfo {author} {\bibfnamefont
  {L.}~\bibnamefont {Wenthaus}}, \bibinfo {author} {\bibfnamefont
  {G.}~\bibnamefont {Brenner}}, \bibinfo {author} {\bibfnamefont
  {S.}~\bibnamefont {Dziarzhytski}}, \bibinfo {author} {\bibfnamefont
  {S.}~\bibnamefont {Palutke}}, \bibinfo {author} {\bibfnamefont {S.~K.}\
  \bibnamefont {Mahatha}}, \bibinfo {author} {\bibfnamefont {N.}~\bibnamefont
  {Schirmel}}, \bibinfo {author} {\bibfnamefont {H.}~\bibnamefont {Redlin}},
  \bibinfo {author} {\bibfnamefont {B.}~\bibnamefont {Manschwetus}}, \bibinfo
  {author} {\bibfnamefont {I.}~\bibnamefont {Hartl}}, \bibinfo {author}
  {\bibfnamefont {Y.}~\bibnamefont {Matveyev}}, \bibinfo {author}
  {\bibfnamefont {A.}~\bibnamefont {Gloskovskii}}, \bibinfo {author}
  {\bibfnamefont {C.}~\bibnamefont {Schlueter}}, \bibinfo {author}
  {\bibfnamefont {V.}~\bibnamefont {Shokeen}}, \bibinfo {author} {\bibfnamefont
  {H.}~\bibnamefont {Duerr}}, \bibinfo {author} {\bibfnamefont {T.~K.}\
  \bibnamefont {Allison}}, \bibinfo {author} {\bibfnamefont {M.}~\bibnamefont
  {Beye}}, \bibinfo {author} {\bibfnamefont {K.}~\bibnamefont {Rossnagel}},
  \bibinfo {author} {\bibfnamefont {H.~J.}\ \bibnamefont {Elmers}}, \ and\
  \bibinfo {author} {\bibfnamefont {K.}~\bibnamefont {Medjanik}},\ }\bibfield
  {title} {\enquote {\bibinfo {title} {{Suppression of the vacuum space-charge
  effect in fs-photoemission by a retarding electrostatic front lens}},}\
  }\href {\doibase 10.1063/5.0046567} {\bibfield  {journal} {\bibinfo
  {journal} {Review of Scientific Instruments}\ }\textbf {\bibinfo {volume}
  {92}},\ \bibinfo {pages} {053703} (\bibinfo {year} {2021})},\ \Eprint
  {http://arxiv.org/abs/https://pubs.aip.org/aip/rsi/article-pdf/doi/10.1063/5.0046567/14129235/053703\_1\_online.pdf}
  {https://pubs.aip.org/aip/rsi/article-pdf/doi/10.1063/5.0046567/14129235/053703\_1\_online.pdf}
  \BibitemShut {NoStop}%
\bibitem [{\citenamefont {Maklar}\ \emph {et~al.}(2020)\citenamefont {Maklar},
  \citenamefont {Dong}, \citenamefont {Beaulieu}, \citenamefont {Pincelli},
  \citenamefont {Dendzik}, \citenamefont {Windsor}, \citenamefont {Xian},
  \citenamefont {Wolf}, \citenamefont {Ernstorfer},\ and\ \citenamefont
  {Rettig}}]{Maklar_RSI2020}%
  \BibitemOpen
  \bibfield  {author} {\bibinfo {author} {\bibfnamefont {J.}~\bibnamefont
  {Maklar}}, \bibinfo {author} {\bibfnamefont {S.}~\bibnamefont {Dong}},
  \bibinfo {author} {\bibfnamefont {S.}~\bibnamefont {Beaulieu}}, \bibinfo
  {author} {\bibfnamefont {T.}~\bibnamefont {Pincelli}}, \bibinfo {author}
  {\bibfnamefont {M.}~\bibnamefont {Dendzik}}, \bibinfo {author} {\bibfnamefont
  {Y.~W.}\ \bibnamefont {Windsor}}, \bibinfo {author} {\bibfnamefont {R.~P.}\
  \bibnamefont {Xian}}, \bibinfo {author} {\bibfnamefont {M.}~\bibnamefont
  {Wolf}}, \bibinfo {author} {\bibfnamefont {R.}~\bibnamefont {Ernstorfer}}, \
  and\ \bibinfo {author} {\bibfnamefont {L.}~\bibnamefont {Rettig}},\
  }\bibfield  {title} {\enquote {\bibinfo {title} {A quantitative comparison of
  time-of-flight momentum microscopes and hemispherical analyzers for time- and
  angle-resolved photoemission spectroscopy experiments},}\ }\href {\doibase
  10.1063/5.0024493} {\bibfield  {journal} {\bibinfo  {journal} {Review of
  Scientific Instruments}\ }\textbf {\bibinfo {volume} {91}},\ \bibinfo {pages}
  {123112} (\bibinfo {year} {2020})},\ \Eprint
  {http://arxiv.org/abs/https://doi.org/10.1063/5.0024493}
  {https://doi.org/10.1063/5.0024493} \BibitemShut {NoStop}%
\bibitem [{\citenamefont {Chernov}\ \emph {et~al.}(2022)\citenamefont
  {Chernov}, \citenamefont {Bakalis}, \citenamefont {Kunin}, \citenamefont
  {Withers}, \citenamefont {White}, \citenamefont {Sch\"onhense},\ and\
  \citenamefont {Allison}}]{Chernov_UP2022}%
  \BibitemOpen
  \bibfield  {author} {\bibinfo {author} {\bibfnamefont {S.}~\bibnamefont
  {Chernov}}, \bibinfo {author} {\bibfnamefont {J.}~\bibnamefont {Bakalis}},
  \bibinfo {author} {\bibfnamefont {A.}~\bibnamefont {Kunin}}, \bibinfo
  {author} {\bibfnamefont {Z.~H.}\ \bibnamefont {Withers}}, \bibinfo {author}
  {\bibfnamefont {M.~G.}\ \bibnamefont {White}}, \bibinfo {author}
  {\bibfnamefont {G.}~\bibnamefont {Sch\"onhense}}, \ and\ \bibinfo {author}
  {\bibfnamefont {T.~K.}\ \bibnamefont {Allison}},\ }\bibfield  {title}
  {\enquote {\bibinfo {title} {Optimization of time of flight momentum
  microscopy for pump-probe experiments},}\ }in\ \href@noop {} {\emph {\bibinfo
  {booktitle} {The 23rd International Conference on Ultrafast Phenomena}}}\
  (\bibinfo {year} {2022.})\ p.\ \bibinfo {pages} {paper W4A.41}\BibitemShut
  {NoStop}%
\bibitem [{\citenamefont {Stadtm\"{u}ller}\ \emph {et~al.}(2019)\citenamefont
  {Stadtm\"{u}ller}, \citenamefont {Emmerich}, \citenamefont {Jungkenn},
  \citenamefont {Haag}, \citenamefont {Rollinger}, \citenamefont {Eich},
  \citenamefont {Maniraj}, \citenamefont {Aeschlimann}, \citenamefont
  {Cinchetti},\ and\ \citenamefont {Mathias}}]{Stadtmuller_NatComm2019}%
  \BibitemOpen
  \bibfield  {author} {\bibinfo {author} {\bibfnamefont {B.}~\bibnamefont
  {Stadtm\"{u}ller}}, \bibinfo {author} {\bibfnamefont {S.}~\bibnamefont
  {Emmerich}}, \bibinfo {author} {\bibfnamefont {D.}~\bibnamefont {Jungkenn}},
  \bibinfo {author} {\bibfnamefont {N.}~\bibnamefont {Haag}}, \bibinfo {author}
  {\bibfnamefont {M.}~\bibnamefont {Rollinger}}, \bibinfo {author}
  {\bibfnamefont {S.}~\bibnamefont {Eich}}, \bibinfo {author} {\bibfnamefont
  {M.}~\bibnamefont {Maniraj}}, \bibinfo {author} {\bibfnamefont
  {M.}~\bibnamefont {Aeschlimann}}, \bibinfo {author} {\bibfnamefont
  {M.}~\bibnamefont {Cinchetti}}, \ and\ \bibinfo {author} {\bibfnamefont
  {S.}~\bibnamefont {Mathias}},\ }\bibfield  {title} {\enquote {\bibinfo
  {title} {Strong modification of the transport level alignment in organic
  materials after optical excitation},}\ }\href {\doibase
  10.1038/s41467-019-09136-7} {\bibfield  {journal} {\bibinfo  {journal}
  {Nature Communications}\ }\textbf {\bibinfo {volume} {10}},\ \bibinfo {pages}
  {1470} (\bibinfo {year} {2019})}\BibitemShut {NoStop}%
\bibitem [{\citenamefont {Bennecke}\ \emph {et~al.}(2024)\citenamefont
  {Bennecke}, \citenamefont {Windischbacher}, \citenamefont {Schmitt},
  \citenamefont {Bange}, \citenamefont {Hemm}, \citenamefont {Kern},
  \citenamefont {D'Avino}, \citenamefont {Blase}, \citenamefont {Steil},
  \citenamefont {Steil}, \citenamefont {Aeschlimann}, \citenamefont
  {Stadtm{\"u}ller}, \citenamefont {Reutzel}, \citenamefont {Puschnig},
  \citenamefont {Jansen},\ and\ \citenamefont
  {Mathias}}]{Bennecke_NatComm2024}%
  \BibitemOpen
  \bibfield  {author} {\bibinfo {author} {\bibfnamefont {W.}~\bibnamefont
  {Bennecke}}, \bibinfo {author} {\bibfnamefont {A.}~\bibnamefont
  {Windischbacher}}, \bibinfo {author} {\bibfnamefont {D.}~\bibnamefont
  {Schmitt}}, \bibinfo {author} {\bibfnamefont {J.~P.}\ \bibnamefont {Bange}},
  \bibinfo {author} {\bibfnamefont {R.}~\bibnamefont {Hemm}}, \bibinfo {author}
  {\bibfnamefont {C.~S.}\ \bibnamefont {Kern}}, \bibinfo {author}
  {\bibfnamefont {G.}~\bibnamefont {D'Avino}}, \bibinfo {author} {\bibfnamefont
  {X.}~\bibnamefont {Blase}}, \bibinfo {author} {\bibfnamefont
  {D.}~\bibnamefont {Steil}}, \bibinfo {author} {\bibfnamefont
  {S.}~\bibnamefont {Steil}}, \bibinfo {author} {\bibfnamefont
  {M.}~\bibnamefont {Aeschlimann}}, \bibinfo {author} {\bibfnamefont
  {B.}~\bibnamefont {Stadtm{\"u}ller}}, \bibinfo {author} {\bibfnamefont
  {M.}~\bibnamefont {Reutzel}}, \bibinfo {author} {\bibfnamefont
  {P.}~\bibnamefont {Puschnig}}, \bibinfo {author} {\bibfnamefont {G.~S.~M.}\
  \bibnamefont {Jansen}}, \ and\ \bibinfo {author} {\bibfnamefont
  {S.}~\bibnamefont {Mathias}},\ }\bibfield  {title} {\enquote {\bibinfo
  {title} {Disentangling the multiorbital contributions of excitons by
  photoemission exciton tomography},}\ }\href {\doibase
  10.1038/s41467-024-45973-x} {\bibfield  {journal} {\bibinfo  {journal}
  {Nature Communications}\ }\textbf {\bibinfo {volume} {15}},\ \bibinfo {pages}
  {1804} (\bibinfo {year} {2024})}\BibitemShut {NoStop}%
\bibitem [{\citenamefont {Bovensiepen}, \citenamefont {Petek},\ and\
  \citenamefont {Wolf}(2010)}]{Bovensiepen_Book2010}%
  \BibitemOpen
  \bibinfo {editor} {\bibfnamefont {U.}~\bibnamefont {Bovensiepen}}, \bibinfo
  {editor} {\bibfnamefont {H.}~\bibnamefont {Petek}}, \ and\ \bibinfo {editor}
  {\bibfnamefont {M.}~\bibnamefont {Wolf}},\ eds.,\ \href@noop {} {\emph
  {\bibinfo {title} {Dynamics at Solid State Surfaces and Interfaces}}}\
  (\bibinfo  {publisher} {WILEY-VCH Verlag},\ \bibinfo {year}
  {2010})\BibitemShut {NoStop}%
\bibitem [{\citenamefont {Hofer}\ \emph {et~al.}(1997)\citenamefont {Hofer},
  \citenamefont {Shumay}, \citenamefont {Reuss}, \citenamefont {Thomann},
  \citenamefont {Wallauer},\ and\ \citenamefont {Fauster}}]{Hofer_Science1997}%
  \BibitemOpen
  \bibfield  {author} {\bibinfo {author} {\bibfnamefont {U.}~\bibnamefont
  {Hofer}}, \bibinfo {author} {\bibfnamefont {I.~L.}\ \bibnamefont {Shumay}},
  \bibinfo {author} {\bibfnamefont {C.}~\bibnamefont {Reuss}}, \bibinfo
  {author} {\bibfnamefont {U.}~\bibnamefont {Thomann}}, \bibinfo {author}
  {\bibfnamefont {W.}~\bibnamefont {Wallauer}}, \ and\ \bibinfo {author}
  {\bibfnamefont {T.}~\bibnamefont {Fauster}},\ }\bibfield  {title} {\enquote
  {\bibinfo {title} {Time-resolved coherent photoelectron spectroscopy of
  quantized electronic states on metal surfaces},}\ }\href {\doibase
  10.1126/science.277.5331.1480} {\bibfield  {journal} {\bibinfo  {journal}
  {Science}\ }\textbf {\bibinfo {volume} {277}},\ \bibinfo {pages} {1480--1482}
  (\bibinfo {year} {1997})},\ \Eprint
  {http://arxiv.org/abs/http://www.sciencemag.org/content/277/5331/1480.full.pdf}
  {http://www.sciencemag.org/content/277/5331/1480.full.pdf} \BibitemShut
  {NoStop}%
\bibitem [{\citenamefont {Petek}\ and\ \citenamefont
  {Ogawa}(1997)}]{Petek_ProgSurfSci1997}%
  \BibitemOpen
  \bibfield  {author} {\bibinfo {author} {\bibfnamefont {H.}~\bibnamefont
  {Petek}}\ and\ \bibinfo {author} {\bibfnamefont {S.}~\bibnamefont {Ogawa}},\
  }\bibfield  {title} {\enquote {\bibinfo {title} {Femtosecond time-resolved
  two-photon photoemission studies of electron dynamics in metals},}\ }\href
  {\doibase http://dx.doi.org/10.1016/S0079-6816(98)00002-1} {\bibfield
  {journal} {\bibinfo  {journal} {Progress in Surface Science}\ }\textbf
  {\bibinfo {volume} {56}},\ \bibinfo {pages} {239 -- 310} (\bibinfo {year}
  {1997})}\BibitemShut {NoStop}%
\bibitem [{\citenamefont {Regan}\ \emph {et~al.}(2022)\citenamefont {Regan},
  \citenamefont {Wang}, \citenamefont {Paik}, \citenamefont {Zeng},
  \citenamefont {Zhang}, \citenamefont {Zhu}, \citenamefont {MacDonald},
  \citenamefont {Deng},\ and\ \citenamefont {Wang}}]{Regan_NatureReviews2022}%
  \BibitemOpen
  \bibfield  {author} {\bibinfo {author} {\bibfnamefont {E.~C.}\ \bibnamefont
  {Regan}}, \bibinfo {author} {\bibfnamefont {D.}~\bibnamefont {Wang}},
  \bibinfo {author} {\bibfnamefont {E.~Y.}\ \bibnamefont {Paik}}, \bibinfo
  {author} {\bibfnamefont {Y.}~\bibnamefont {Zeng}}, \bibinfo {author}
  {\bibfnamefont {L.}~\bibnamefont {Zhang}}, \bibinfo {author} {\bibfnamefont
  {J.}~\bibnamefont {Zhu}}, \bibinfo {author} {\bibfnamefont {A.~H.}\
  \bibnamefont {MacDonald}}, \bibinfo {author} {\bibfnamefont {H.}~\bibnamefont
  {Deng}}, \ and\ \bibinfo {author} {\bibfnamefont {F.}~\bibnamefont {Wang}},\
  }\bibfield  {title} {\enquote {\bibinfo {title} {Emerging exciton physics in
  transition metal dichalcogenide heterobilayers},}\ }\href {\doibase
  10.1038/s41578-022-00440-1} {\bibfield  {journal} {\bibinfo  {journal}
  {Nature Reviews Materials}\ }\textbf {\bibinfo {volume} {7}},\ \bibinfo
  {pages} {778--795} (\bibinfo {year} {2022})}\BibitemShut {NoStop}%
\bibitem [{\citenamefont {Petek}\ \emph {et~al.}(2000)\citenamefont {Petek},
  \citenamefont {Weida}, \citenamefont {Nagano},\ and\ \citenamefont
  {Ogawa}}]{Petek_Science2000}%
  \BibitemOpen
  \bibfield  {author} {\bibinfo {author} {\bibfnamefont {H.}~\bibnamefont
  {Petek}}, \bibinfo {author} {\bibfnamefont {M.~J.}\ \bibnamefont {Weida}},
  \bibinfo {author} {\bibfnamefont {H.}~\bibnamefont {Nagano}}, \ and\ \bibinfo
  {author} {\bibfnamefont {S.}~\bibnamefont {Ogawa}},\ }\bibfield  {title}
  {\enquote {\bibinfo {title} {Real-time observation of adsorbate atom motion
  above a metal surface},}\ }\href {\doibase 10.1126/science.288.5470.1402}
  {\bibfield  {journal} {\bibinfo  {journal} {Science}\ }\textbf {\bibinfo
  {volume} {288}},\ \bibinfo {pages} {1402--1404} (\bibinfo {year} {2000})},\
  \Eprint
  {http://arxiv.org/abs/http://www.sciencemag.org/content/288/5470/1402.full.pdf}
  {http://www.sciencemag.org/content/288/5470/1402.full.pdf} \BibitemShut
  {NoStop}%
\bibitem [{\citenamefont {Onda}\ \emph {et~al.}(2005)\citenamefont {Onda},
  \citenamefont {Li}, \citenamefont {Zhao}, \citenamefont {Jordan},
  \citenamefont {Yang},\ and\ \citenamefont {Petek}}]{Onda_Science2005}%
  \BibitemOpen
  \bibfield  {author} {\bibinfo {author} {\bibfnamefont {K.}~\bibnamefont
  {Onda}}, \bibinfo {author} {\bibfnamefont {B.}~\bibnamefont {Li}}, \bibinfo
  {author} {\bibfnamefont {J.}~\bibnamefont {Zhao}}, \bibinfo {author}
  {\bibfnamefont {K.~D.}\ \bibnamefont {Jordan}}, \bibinfo {author}
  {\bibfnamefont {J.}~\bibnamefont {Yang}}, \ and\ \bibinfo {author}
  {\bibfnamefont {H.}~\bibnamefont {Petek}},\ }\bibfield  {title} {\enquote
  {\bibinfo {title} {Wet electrons at the h2o/tio2(110) surface},}\ }\href
  {\doibase 10.1126/science.1109366} {\bibfield  {journal} {\bibinfo  {journal}
  {Science}\ }\textbf {\bibinfo {volume} {308}},\ \bibinfo {pages} {1154--1158}
  (\bibinfo {year} {2005})},\ \Eprint
  {http://arxiv.org/abs/http://www.sciencemag.org/content/308/5725/1154.full.pdf}
  {http://www.sciencemag.org/content/308/5725/1154.full.pdf} \BibitemShut
  {NoStop}%
\bibitem [{\citenamefont {Stahler}\ \emph {et~al.}(2008)\citenamefont
  {Stahler}, \citenamefont {Bovensiepen}, \citenamefont {Meyer},\ and\
  \citenamefont {Wolf}}]{Stahler_ChemSocRev2008}%
  \BibitemOpen
  \bibfield  {author} {\bibinfo {author} {\bibfnamefont {J.}~\bibnamefont
  {Stahler}}, \bibinfo {author} {\bibfnamefont {U.}~\bibnamefont
  {Bovensiepen}}, \bibinfo {author} {\bibfnamefont {M.}~\bibnamefont {Meyer}},
  \ and\ \bibinfo {author} {\bibfnamefont {M.}~\bibnamefont {Wolf}},\
  }\bibfield  {title} {\enquote {\bibinfo {title} {A surface science approach
  to ultrafast electron transfer and solvation dynamics at interfaces},}\
  }\href {\doibase 10.1039/B800257F} {\bibfield  {journal} {\bibinfo  {journal}
  {Chem. Soc. Rev.}\ }\textbf {\bibinfo {volume} {37}},\ \bibinfo {pages}
  {2180--2190} (\bibinfo {year} {2008})}\BibitemShut {NoStop}%
\bibitem [{\citenamefont {Bange}\ \emph {et~al.}(2024)\citenamefont {Bange},
  \citenamefont {Schmitt}, \citenamefont {Bennecke}, \citenamefont {Meneghini},
  \citenamefont {AlMutairi}, \citenamefont {Watanabe}, \citenamefont
  {Taniguchi}, \citenamefont {Steil}, \citenamefont {Steil}, \citenamefont
  {Weitz}, \citenamefont {Jansen}, \citenamefont {Hofmann}, \citenamefont
  {Brem}, \citenamefont {Malic}, \citenamefont {Reutzel},\ and\ \citenamefont
  {Mathias}}]{Bange_SciAdv2024}%
  \BibitemOpen
  \bibfield  {author} {\bibinfo {author} {\bibfnamefont {J.~P.}\ \bibnamefont
  {Bange}}, \bibinfo {author} {\bibfnamefont {D.}~\bibnamefont {Schmitt}},
  \bibinfo {author} {\bibfnamefont {W.}~\bibnamefont {Bennecke}}, \bibinfo
  {author} {\bibfnamefont {G.}~\bibnamefont {Meneghini}}, \bibinfo {author}
  {\bibfnamefont {A.}~\bibnamefont {AlMutairi}}, \bibinfo {author}
  {\bibfnamefont {K.}~\bibnamefont {Watanabe}}, \bibinfo {author}
  {\bibfnamefont {T.}~\bibnamefont {Taniguchi}}, \bibinfo {author}
  {\bibfnamefont {D.}~\bibnamefont {Steil}}, \bibinfo {author} {\bibfnamefont
  {S.}~\bibnamefont {Steil}}, \bibinfo {author} {\bibfnamefont {R.~T.}\
  \bibnamefont {Weitz}}, \bibinfo {author} {\bibfnamefont {G.~S.~M.}\
  \bibnamefont {Jansen}}, \bibinfo {author} {\bibfnamefont {S.}~\bibnamefont
  {Hofmann}}, \bibinfo {author} {\bibfnamefont {S.}~\bibnamefont {Brem}},
  \bibinfo {author} {\bibfnamefont {E.}~\bibnamefont {Malic}}, \bibinfo
  {author} {\bibfnamefont {M.}~\bibnamefont {Reutzel}}, \ and\ \bibinfo
  {author} {\bibfnamefont {S.}~\bibnamefont {Mathias}},\ }\bibfield  {title}
  {\enquote {\bibinfo {title} {Probing electron-hole coulomb correlations in
  the exciton landscape of a twisted semiconductor heterostructure},}\ }\href
  {\doibase 10.1126/sciadv.adi1323} {\bibfield  {journal} {\bibinfo  {journal}
  {Science Advances}\ }\textbf {\bibinfo {volume} {10}},\ \bibinfo {pages}
  {eadi1323} (\bibinfo {year} {2024})},\ \Eprint
  {http://arxiv.org/abs/https://www.science.org/doi/pdf/10.1126/sciadv.adi1323}
  {https://www.science.org/doi/pdf/10.1126/sciadv.adi1323} \BibitemShut
  {NoStop}%
\bibitem [{\citenamefont {McIver}\ \emph {et~al.}(2020)\citenamefont {McIver},
  \citenamefont {Schulte}, \citenamefont {Stein}, \citenamefont {Matsuyama},
  \citenamefont {Jotzu}, \citenamefont {Meier},\ and\ \citenamefont
  {Cavalleri}}]{McIver_NatPhys2020}%
  \BibitemOpen
  \bibfield  {author} {\bibinfo {author} {\bibfnamefont {J.~W.}\ \bibnamefont
  {McIver}}, \bibinfo {author} {\bibfnamefont {B.}~\bibnamefont {Schulte}},
  \bibinfo {author} {\bibfnamefont {F.~U.}\ \bibnamefont {Stein}}, \bibinfo
  {author} {\bibfnamefont {T.}~\bibnamefont {Matsuyama}}, \bibinfo {author}
  {\bibfnamefont {G.}~\bibnamefont {Jotzu}}, \bibinfo {author} {\bibfnamefont
  {G.}~\bibnamefont {Meier}}, \ and\ \bibinfo {author} {\bibfnamefont
  {A.}~\bibnamefont {Cavalleri}},\ }\bibfield  {title} {\enquote {\bibinfo
  {title} {Light-induced anomalous hall effect in graphene},}\ }\href {\doibase
  10.1038/s41567-019-0698-y} {\bibfield  {journal} {\bibinfo  {journal} {Nature
  Physics}\ }\textbf {\bibinfo {volume} {16}},\ \bibinfo {pages} {38--41}
  (\bibinfo {year} {2020})}\BibitemShut {NoStop}%
\bibitem [{\citenamefont {Mahmood}\ \emph {et~al.}(2016)\citenamefont
  {Mahmood}, \citenamefont {Chan}, \citenamefont {Alpichshev}, \citenamefont
  {Gardner}, \citenamefont {Lee}, \citenamefont {Lee},\ and\ \citenamefont
  {Gedik}}]{Mahmood_NatPhys2016}%
  \BibitemOpen
  \bibfield  {author} {\bibinfo {author} {\bibfnamefont {F.}~\bibnamefont
  {Mahmood}}, \bibinfo {author} {\bibfnamefont {C.-K.}\ \bibnamefont {Chan}},
  \bibinfo {author} {\bibfnamefont {Z.}~\bibnamefont {Alpichshev}}, \bibinfo
  {author} {\bibfnamefont {D.}~\bibnamefont {Gardner}}, \bibinfo {author}
  {\bibfnamefont {Y.}~\bibnamefont {Lee}}, \bibinfo {author} {\bibfnamefont
  {P.~A.}\ \bibnamefont {Lee}}, \ and\ \bibinfo {author} {\bibfnamefont
  {N.}~\bibnamefont {Gedik}},\ }\bibfield  {title} {\enquote {\bibinfo {title}
  {Selective scattering between floquet--bloch and volkov states in a
  topological insulator},}\ }\href {http://dx.doi.org/10.1038/nphys3609}
  {\bibfield  {journal} {\bibinfo  {journal} {Nature Physics}\ }\textbf
  {\bibinfo {volume} {12}},\ \bibinfo {pages} {306 EP --} (\bibinfo {year}
  {2016})}\BibitemShut {NoStop}%
\bibitem [{\citenamefont {Ulstrup}\ \emph {et~al.}(2015)\citenamefont
  {Ulstrup}, \citenamefont {Johannsen}, \citenamefont {Cilento}, \citenamefont
  {Crepaldi}, \citenamefont {Miwa}, \citenamefont {Zacchigna}, \citenamefont
  {Cacho}, \citenamefont {Chapman}, \citenamefont {Springate}, \citenamefont
  {Fromm}, \citenamefont {Raidel}, \citenamefont {Seyller}, \citenamefont
  {King}, \citenamefont {Parmigiani}, \citenamefont {Grioni},\ and\
  \citenamefont {Hofmann}}]{Ultstrup_JElecSpec2015}%
  \BibitemOpen
  \bibfield  {author} {\bibinfo {author} {\bibfnamefont {S.}~\bibnamefont
  {Ulstrup}}, \bibinfo {author} {\bibfnamefont {J.~C.}\ \bibnamefont
  {Johannsen}}, \bibinfo {author} {\bibfnamefont {F.}~\bibnamefont {Cilento}},
  \bibinfo {author} {\bibfnamefont {A.}~\bibnamefont {Crepaldi}}, \bibinfo
  {author} {\bibfnamefont {J.~A.}\ \bibnamefont {Miwa}}, \bibinfo {author}
  {\bibfnamefont {M.}~\bibnamefont {Zacchigna}}, \bibinfo {author}
  {\bibfnamefont {C.}~\bibnamefont {Cacho}}, \bibinfo {author} {\bibfnamefont
  {R.~T.}\ \bibnamefont {Chapman}}, \bibinfo {author} {\bibfnamefont
  {E.}~\bibnamefont {Springate}}, \bibinfo {author} {\bibfnamefont
  {F.}~\bibnamefont {Fromm}}, \bibinfo {author} {\bibfnamefont
  {C.}~\bibnamefont {Raidel}}, \bibinfo {author} {\bibfnamefont
  {T.}~\bibnamefont {Seyller}}, \bibinfo {author} {\bibfnamefont {P.~D.}\
  \bibnamefont {King}}, \bibinfo {author} {\bibfnamefont {F.}~\bibnamefont
  {Parmigiani}}, \bibinfo {author} {\bibfnamefont {M.}~\bibnamefont {Grioni}},
  \ and\ \bibinfo {author} {\bibfnamefont {P.}~\bibnamefont {Hofmann}},\
  }\bibfield  {title} {\enquote {\bibinfo {title} {Ramifications of optical
  pumping on the interpretation of time-resolved photoemission experiments on
  graphene},}\ }\href {\doibase https://doi.org/10.1016/j.elspec.2015.04.010}
  {\bibfield  {journal} {\bibinfo  {journal} {Journal of Electron Spectroscopy
  and Related Phenomena}\ }\textbf {\bibinfo {volume} {200}},\ \bibinfo {pages}
  {340 -- 346} (\bibinfo {year} {2015})},\ \bibinfo {note} {special Anniversary
  Issue: Volume 200}\BibitemShut {NoStop}%
\bibitem [{\citenamefont {Oloff}\ \emph {et~al.}(2016)\citenamefont {Oloff},
  \citenamefont {Hanff}, \citenamefont {Stange}, \citenamefont {Rohde},
  \citenamefont {Diekmann}, \citenamefont {Bauer},\ and\ \citenamefont
  {Rossnagel}}]{Oloff_JApplPhys2016}%
  \BibitemOpen
  \bibfield  {author} {\bibinfo {author} {\bibfnamefont {L.-P.}\ \bibnamefont
  {Oloff}}, \bibinfo {author} {\bibfnamefont {K.}~\bibnamefont {Hanff}},
  \bibinfo {author} {\bibfnamefont {A.}~\bibnamefont {Stange}}, \bibinfo
  {author} {\bibfnamefont {G.}~\bibnamefont {Rohde}}, \bibinfo {author}
  {\bibfnamefont {F.}~\bibnamefont {Diekmann}}, \bibinfo {author}
  {\bibfnamefont {M.}~\bibnamefont {Bauer}}, \ and\ \bibinfo {author}
  {\bibfnamefont {K.}~\bibnamefont {Rossnagel}},\ }\bibfield  {title} {\enquote
  {\bibinfo {title} {Pump laser-induced space-charge effects in hhg-driven
  time- and angle-resolved photoelectron spectroscopy},}\ }\href {\doibase
  10.1063/1.4953643} {\bibfield  {journal} {\bibinfo  {journal} {Journal of
  Applied Physics}\ }\textbf {\bibinfo {volume} {119}},\ \bibinfo {pages}
  {225106} (\bibinfo {year} {2016})},\ \Eprint
  {http://arxiv.org/abs/https://doi.org/10.1063/1.4953643}
  {https://doi.org/10.1063/1.4953643} \BibitemShut {NoStop}%
\bibitem [{\citenamefont {Borgwardt}(2016)}]{Borgwardt_Thesis2016}%
  \BibitemOpen
  \bibfield  {author} {\bibinfo {author} {\bibfnamefont {M.}~\bibnamefont
  {Borgwardt}},\ }\emph {\bibinfo {title} {Electronic Structure and Electron
  Transfer Dynamics at Dye-Semiconductor Interfaces Studied by Means of
  Time-Resolved XUV Photoelectron Spectroscopy}},\ \href@noop {} {Ph.D.
  thesis},\ \bibinfo  {school} {Freie Universitat Berlin} (\bibinfo {year}
  {2016})\BibitemShut {NoStop}%
\bibitem [{\citenamefont {Wahl}(2024)}]{Wahl_Thesis2024}%
  \BibitemOpen
  \bibfield  {author} {\bibinfo {author} {\bibfnamefont {M.~J.}\ \bibnamefont
  {Wahl}},\ }\emph {\bibinfo {title} {A tunable wavelength pump apparatus with
  arbitrary repetition rate for time- and angle-resolved photoemission
  spectroscopy}},\ \href@noop {} {Master's thesis},\ \bibinfo  {school} {Stony
  Brook University} (\bibinfo {year} {2024})\BibitemShut {NoStop}%
\bibitem [{\citenamefont {Chang}(2011)}]{Chang_Book2011}%
  \BibitemOpen
  \bibfield  {author} {\bibinfo {author} {\bibfnamefont {Z.}~\bibnamefont
  {Chang}},\ }\href@noop {} {\emph {\bibinfo {title} {Fundamentals of
  Attosecond Optics}}}\ (\bibinfo  {publisher} {CRC Press},\ \bibinfo {year}
  {2011})\BibitemShut {NoStop}%
\bibitem [{\citenamefont {Boyd}(2003)}]{Boyd:2003}%
  \BibitemOpen
  \bibfield  {author} {\bibinfo {author} {\bibfnamefont {R.}~\bibnamefont
  {Boyd}},\ }\href@noop {} {\emph {\bibinfo {title} {Nonlinear Optics}}},\
  \bibinfo {edition} {2nd}\ ed.\ (\bibinfo  {publisher} {Academic Press},\
  \bibinfo {address} {Amsterdam},\ \bibinfo {year} {2003})\BibitemShut
  {NoStop}%
\bibitem [{\citenamefont {Agrawal}(2012)}]{Agrawal_NonlinearFiberOpticsBook}%
  \BibitemOpen
  \bibfield  {author} {\bibinfo {author} {\bibfnamefont {G.~P.}\ \bibnamefont
  {Agrawal}},\ }\href@noop {} {\emph {\bibinfo {title} {Nonlinear Fiber
  Optics}}},\ edited by\ \bibinfo {editor} {\bibnamefont {5}}\ (\bibinfo
  {publisher} {Academic Press},\ \bibinfo {year} {2012})\BibitemShut {NoStop}%
\bibitem [{\citenamefont {Krause}, \citenamefont {Schafer},\ and\ \citenamefont
  {Kulander}(1992)}]{Krause_PRL1992}%
  \BibitemOpen
  \bibfield  {author} {\bibinfo {author} {\bibfnamefont {J.~L.}\ \bibnamefont
  {Krause}}, \bibinfo {author} {\bibfnamefont {K.~J.}\ \bibnamefont {Schafer}},
  \ and\ \bibinfo {author} {\bibfnamefont {K.~C.}\ \bibnamefont {Kulander}},\
  }\bibfield  {title} {\enquote {\bibinfo {title} {High-order harmonic
  generation from atoms and ions in the high intensity regime},}\ }\href
  {\doibase 10.1103/PhysRevLett.68.3535} {\bibfield  {journal} {\bibinfo
  {journal} {Phys. Rev. Lett.}\ }\textbf {\bibinfo {volume} {68}},\ \bibinfo
  {pages} {3535--3538} (\bibinfo {year} {1992})}\BibitemShut {NoStop}%
\bibitem [{\citenamefont {Schafer}\ \emph {et~al.}(1993)\citenamefont
  {Schafer}, \citenamefont {Yang}, \citenamefont {DiMauro},\ and\ \citenamefont
  {Kulander}}]{Schafer_PRL1993}%
  \BibitemOpen
  \bibfield  {author} {\bibinfo {author} {\bibfnamefont {K.~J.}\ \bibnamefont
  {Schafer}}, \bibinfo {author} {\bibfnamefont {B.}~\bibnamefont {Yang}},
  \bibinfo {author} {\bibfnamefont {L.~F.}\ \bibnamefont {DiMauro}}, \ and\
  \bibinfo {author} {\bibfnamefont {K.~C.}\ \bibnamefont {Kulander}},\
  }\bibfield  {title} {\enquote {\bibinfo {title} {Above threshold ionization
  beyond the high harmonic cutoff},}\ }\href {\doibase
  10.1103/PhysRevLett.70.1599} {\bibfield  {journal} {\bibinfo  {journal}
  {Phys. Rev. Lett.}\ }\textbf {\bibinfo {volume} {70}},\ \bibinfo {pages}
  {1599--1602} (\bibinfo {year} {1993})}\BibitemShut {NoStop}%
\bibitem [{\citenamefont {Corkum}(1993)}]{Corkum:1993}%
  \BibitemOpen
  \bibfield  {author} {\bibinfo {author} {\bibfnamefont {P.~B.}\ \bibnamefont
  {Corkum}},\ }\bibfield  {title} {\enquote {\bibinfo {title} {Plasma
  perspective on strong-field multiphoton ionization},}\ }\href@noop {}
  {\bibfield  {journal} {\bibinfo  {journal} {Phys. Rev. Lett.}\ }\textbf
  {\bibinfo {volume} {71}},\ \bibinfo {pages} {1994--1997} (\bibinfo {year}
  {1993})}\BibitemShut {NoStop}%
\bibitem [{\citenamefont {Lewenstein}\ \emph {et~al.}(1994)\citenamefont
  {Lewenstein}, \citenamefont {Balcou}, \citenamefont {Ivanov}, \citenamefont
  {L'Huillier},\ and\ \citenamefont {Corkum}}]{Lewenstein:1994}%
  \BibitemOpen
  \bibfield  {author} {\bibinfo {author} {\bibfnamefont {M.}~\bibnamefont
  {Lewenstein}}, \bibinfo {author} {\bibfnamefont {P.}~\bibnamefont {Balcou}},
  \bibinfo {author} {\bibfnamefont {M.~Y.}\ \bibnamefont {Ivanov}}, \bibinfo
  {author} {\bibfnamefont {A.}~\bibnamefont {L'Huillier}}, \ and\ \bibinfo
  {author} {\bibfnamefont {P.~B.}\ \bibnamefont {Corkum}},\ }\bibfield  {title}
  {\enquote {\bibinfo {title} {Theory of high order harmonic generation by
  low-frequency laser fields},}\ }\href@noop {} {\bibfield  {journal} {\bibinfo
   {journal} {Phys. Rev. A}\ }\textbf {\bibinfo {volume} {49}},\ \bibinfo
  {pages} {2117--2132} (\bibinfo {year} {1994})}\BibitemShut {NoStop}%
\bibitem [{\citenamefont {Constant}\ \emph {et~al.}(1999)\citenamefont
  {Constant}, \citenamefont {Garzella}, \citenamefont {Breger}, \citenamefont
  {Mevel}, \citenamefont {Dorrer}, \citenamefont {Blanc}, \citenamefont
  {Salin},\ and\ \citenamefont {Agostini}}]{Constant_PRL1999}%
  \BibitemOpen
  \bibfield  {author} {\bibinfo {author} {\bibfnamefont {E.}~\bibnamefont
  {Constant}}, \bibinfo {author} {\bibfnamefont {D.}~\bibnamefont {Garzella}},
  \bibinfo {author} {\bibfnamefont {P.}~\bibnamefont {Breger}}, \bibinfo
  {author} {\bibfnamefont {E.}~\bibnamefont {Mevel}}, \bibinfo {author}
  {\bibfnamefont {C.}~\bibnamefont {Dorrer}}, \bibinfo {author} {\bibfnamefont
  {C.~L.}\ \bibnamefont {Blanc}}, \bibinfo {author} {\bibfnamefont
  {F.}~\bibnamefont {Salin}}, \ and\ \bibinfo {author} {\bibfnamefont
  {P.}~\bibnamefont {Agostini}},\ }\bibfield  {title} {\enquote {\bibinfo
  {title} {Optimization of high harmonic generation in absorbing gasss model
  and experiment},}\ }\href@noop {} {\bibfield  {journal} {\bibinfo  {journal}
  {Phys. Rev. Lett.}\ }\textbf {\bibinfo {volume} {82}},\ \bibinfo {pages}
  {1668} (\bibinfo {year} {1999})}\BibitemShut {NoStop}%
\bibitem [{\citenamefont {Durfee}\ \emph {et~al.}(1999)\citenamefont {Durfee},
  \citenamefont {Rundquist}, \citenamefont {Backus}, \citenamefont {Herne},
  \citenamefont {Murnane},\ and\ \citenamefont {Kapteyn}}]{Durfee_PRL1999}%
  \BibitemOpen
  \bibfield  {author} {\bibinfo {author} {\bibfnamefont {C.~G.}\ \bibnamefont
  {Durfee}}, \bibinfo {author} {\bibfnamefont {A.~R.}\ \bibnamefont
  {Rundquist}}, \bibinfo {author} {\bibfnamefont {S.}~\bibnamefont {Backus}},
  \bibinfo {author} {\bibfnamefont {C.}~\bibnamefont {Herne}}, \bibinfo
  {author} {\bibfnamefont {M.~M.}\ \bibnamefont {Murnane}}, \ and\ \bibinfo
  {author} {\bibfnamefont {H.~C.}\ \bibnamefont {Kapteyn}},\ }\bibfield
  {title} {\enquote {\bibinfo {title} {Phase matching of high-order harmonics
  in hollow waveguides},}\ }\href {\doibase 10.1103/PhysRevLett.83.2187}
  {\bibfield  {journal} {\bibinfo  {journal} {Phys. Rev. Lett.}\ }\textbf
  {\bibinfo {volume} {83}},\ \bibinfo {pages} {2187--2190} (\bibinfo {year}
  {1999})}\BibitemShut {NoStop}%
\bibitem [{\citenamefont {Allison}(2010)}]{Allison_Thesis2010}%
  \BibitemOpen
  \bibfield  {author} {\bibinfo {author} {\bibfnamefont {T.~K.}\ \bibnamefont
  {Allison}},\ }\emph {\bibinfo {title} {Femtosecond Molecular Dynamics Studied
  with Vacuum Ultraviolet Pulse Pairs}},\ \href@noop {} {Ph.D. thesis},\
  \bibinfo  {school} {University of California at Berkeley} (\bibinfo {year}
  {2010})\BibitemShut {NoStop}%
\bibitem [{\citenamefont {Heyl}\ \emph {et~al.}(2012)\citenamefont {Heyl},
  \citenamefont {G{\"u}dde}, \citenamefont {L'Huillier},\ and\ \citenamefont
  {H{\"o}fer}}]{Heyl_JPhysB2012}%
  \BibitemOpen
  \bibfield  {author} {\bibinfo {author} {\bibfnamefont {C.~M.}\ \bibnamefont
  {Heyl}}, \bibinfo {author} {\bibfnamefont {J.}~\bibnamefont {G{\"u}dde}},
  \bibinfo {author} {\bibfnamefont {A.}~\bibnamefont {L'Huillier}}, \ and\
  \bibinfo {author} {\bibfnamefont {U.}~\bibnamefont {H{\"o}fer}},\ }\bibfield
  {title} {\enquote {\bibinfo {title} {High-order harmonic generation with
  $\mu$j laser pulses at high repetition rates},}\ }\href
  {http://stacks.iop.org/0953-4075/45/i=7/a=074020} {\bibfield  {journal}
  {\bibinfo  {journal} {Journal of Physics B: Atomic, Molecular and Optical
  Physics}\ }\textbf {\bibinfo {volume} {45}},\ \bibinfo {pages} {074020}
  (\bibinfo {year} {2012})}\BibitemShut {NoStop}%
\bibitem [{\citenamefont {Hammond}, \citenamefont {Mills},\ and\ \citenamefont
  {Jones}(2011)}]{Hammond_OptExp2011}%
  \BibitemOpen
  \bibfield  {author} {\bibinfo {author} {\bibfnamefont {T.~J.}\ \bibnamefont
  {Hammond}}, \bibinfo {author} {\bibfnamefont {A.~K.}\ \bibnamefont {Mills}},
  \ and\ \bibinfo {author} {\bibfnamefont {D.~J.}\ \bibnamefont {Jones}},\
  }\bibfield  {title} {\enquote {\bibinfo {title} {Near-threshold harmonics
  from a femtosecond enhancement cavity-based euv source: effects of multiple
  quantum pathways on spatial profile and yield},}\ }\href {\doibase
  10.1364/OE.19.024871} {\bibfield  {journal} {\bibinfo  {journal} {Opt.
  Express}\ }\textbf {\bibinfo {volume} {19}},\ \bibinfo {pages} {24871--24883}
  (\bibinfo {year} {2011})}\BibitemShut {NoStop}%
\bibitem [{\citenamefont {{DiChiara}}\ \emph {et~al.}(2012)\citenamefont
  {{DiChiara}}, \citenamefont {{Ghimire}}, \citenamefont {{Blaga}},
  \citenamefont {{Sistrunk}}, \citenamefont {{Power}}, \citenamefont {{March}},
  \citenamefont {{Miller}}, \citenamefont {{Reis}}, \citenamefont
  {{Agostini}},\ and\ \citenamefont {{DiMauro}}}]{DiChiara_IEEE2012}%
  \BibitemOpen
  \bibfield  {author} {\bibinfo {author} {\bibfnamefont {A.~D.}\ \bibnamefont
  {{DiChiara}}}, \bibinfo {author} {\bibfnamefont {S.}~\bibnamefont
  {{Ghimire}}}, \bibinfo {author} {\bibfnamefont {C.~I.}\ \bibnamefont
  {{Blaga}}}, \bibinfo {author} {\bibfnamefont {E.}~\bibnamefont {{Sistrunk}}},
  \bibinfo {author} {\bibfnamefont {E.~P.}\ \bibnamefont {{Power}}}, \bibinfo
  {author} {\bibfnamefont {A.~M.}\ \bibnamefont {{March}}}, \bibinfo {author}
  {\bibfnamefont {T.~A.}\ \bibnamefont {{Miller}}}, \bibinfo {author}
  {\bibfnamefont {D.~A.}\ \bibnamefont {{Reis}}}, \bibinfo {author}
  {\bibfnamefont {P.}~\bibnamefont {{Agostini}}}, \ and\ \bibinfo {author}
  {\bibfnamefont {L.~F.}\ \bibnamefont {{DiMauro}}},\ }\bibfield  {title}
  {\enquote {\bibinfo {title} {Scaling of high-order harmonic generation in the
  long wavelength limit of a strong laser field},}\ }\href {\doibase
  10.1109/JSTQE.2011.2158391} {\bibfield  {journal} {\bibinfo  {journal} {IEEE
  Journal of Selected Topics in Quantum Electronics}\ }\textbf {\bibinfo
  {volume} {18}},\ \bibinfo {pages} {419--433} (\bibinfo {year}
  {2012})}\BibitemShut {NoStop}%
\bibitem [{\citenamefont {Brabec}\ and\ \citenamefont
  {Krausz}(2000)}]{Brabec_RMP2000}%
  \BibitemOpen
  \bibfield  {author} {\bibinfo {author} {\bibfnamefont {T.}~\bibnamefont
  {Brabec}}\ and\ \bibinfo {author} {\bibfnamefont {F.}~\bibnamefont
  {Krausz}},\ }\bibfield  {title} {\enquote {\bibinfo {title} {Intense few
  cycle laser fields: Frontiers of nonlinear optics},}\ }\href@noop {}
  {\bibfield  {journal} {\bibinfo  {journal} {Rev. Mod. Phys.}\ }\textbf
  {\bibinfo {volume} {72}},\ \bibinfo {pages} {545--591} (\bibinfo {year}
  {2000})}\BibitemShut {NoStop}%
\bibitem [{\citenamefont {Gohle}\ \emph {et~al.}(2005)\citenamefont {Gohle},
  \citenamefont {Udem}, \citenamefont {Herrmann}, \citenamefont
  {Rauschenberger}, \citenamefont {Holzwarth}, \citenamefont {Schuessler},
  \citenamefont {Krausz},\ and\ \citenamefont {H{\"a}nsch}}]{Gohle_Nature2005}%
  \BibitemOpen
  \bibfield  {author} {\bibinfo {author} {\bibfnamefont {C.}~\bibnamefont
  {Gohle}}, \bibinfo {author} {\bibfnamefont {T.}~\bibnamefont {Udem}},
  \bibinfo {author} {\bibfnamefont {M.}~\bibnamefont {Herrmann}}, \bibinfo
  {author} {\bibfnamefont {J.}~\bibnamefont {Rauschenberger}}, \bibinfo
  {author} {\bibfnamefont {R.}~\bibnamefont {Holzwarth}}, \bibinfo {author}
  {\bibfnamefont {H.~A.}\ \bibnamefont {Schuessler}}, \bibinfo {author}
  {\bibfnamefont {F.}~\bibnamefont {Krausz}}, \ and\ \bibinfo {author}
  {\bibfnamefont {T.~W.}\ \bibnamefont {H{\"a}nsch}},\ }\bibfield  {title}
  {\enquote {\bibinfo {title} {A frequency comb in the extreme ultraviolet},}\
  }\href {\doibase 10.1038/nature03851} {\bibfield  {journal} {\bibinfo
  {journal} {Nature}\ }\textbf {\bibinfo {volume} {436}},\ \bibinfo {pages}
  {234--237} (\bibinfo {year} {2005})}\BibitemShut {NoStop}%
\bibitem [{\citenamefont {Jones}\ \emph {et~al.}(2005)\citenamefont {Jones},
  \citenamefont {Moll}, \citenamefont {Thorpe},\ and\ \citenamefont
  {Ye}}]{Jones_PRL2005}%
  \BibitemOpen
  \bibfield  {author} {\bibinfo {author} {\bibfnamefont {R.~J.}\ \bibnamefont
  {Jones}}, \bibinfo {author} {\bibfnamefont {K.~D.}\ \bibnamefont {Moll}},
  \bibinfo {author} {\bibfnamefont {M.~J.}\ \bibnamefont {Thorpe}}, \ and\
  \bibinfo {author} {\bibfnamefont {J.}~\bibnamefont {Ye}},\ }\bibfield
  {title} {\enquote {\bibinfo {title} {Phase-coherent frequency combs in the
  vacuum ultraviolet via high-harmonic generation inside a femtosecond
  enhancement cavity},}\ }\href {\doibase 10.1103/PhysRevLett.94.193201}
  {\bibfield  {journal} {\bibinfo  {journal} {Physical Review Letters}\
  }\textbf {\bibinfo {volume} {94}},\ \bibinfo {eid} {193201} (\bibinfo {year}
  {2005})}\BibitemShut {NoStop}%
\bibitem [{\citenamefont {Ruehl}\ \emph {et~al.}(2010)\citenamefont {Ruehl},
  \citenamefont {Marcinkevicius}, \citenamefont {Fermann},\ and\ \citenamefont
  {Hartl}}]{Ruehl_OptLett2010}%
  \BibitemOpen
  \bibfield  {author} {\bibinfo {author} {\bibfnamefont {A.}~\bibnamefont
  {Ruehl}}, \bibinfo {author} {\bibfnamefont {A.}~\bibnamefont
  {Marcinkevicius}}, \bibinfo {author} {\bibfnamefont {M.~E.}\ \bibnamefont
  {Fermann}}, \ and\ \bibinfo {author} {\bibfnamefont {I.}~\bibnamefont
  {Hartl}},\ }\bibfield  {title} {\enquote {\bibinfo {title} {80 w, 120 fs
  yb-fiber frequency comb},}\ }\href {\doibase 10.1364/OL.35.003015} {\bibfield
   {journal} {\bibinfo  {journal} {Opt. Lett.}\ }\textbf {\bibinfo {volume}
  {35}},\ \bibinfo {pages} {3015--3017} (\bibinfo {year} {2010})}\BibitemShut
  {NoStop}%
\bibitem [{\citenamefont {Paul}\ \emph {et~al.}(2008)\citenamefont {Paul},
  \citenamefont {Johnson}, \citenamefont {Lee},\ and\ \citenamefont
  {Jones}}]{Paul_OptLett2008}%
  \BibitemOpen
  \bibfield  {author} {\bibinfo {author} {\bibfnamefont {J.}~\bibnamefont
  {Paul}}, \bibinfo {author} {\bibfnamefont {J.}~\bibnamefont {Johnson}},
  \bibinfo {author} {\bibfnamefont {J.}~\bibnamefont {Lee}}, \ and\ \bibinfo
  {author} {\bibfnamefont {R.~J.}\ \bibnamefont {Jones}},\ }\bibfield  {title}
  {\enquote {\bibinfo {title} {Generation of high-power frequency combs from
  injection-locked femtosecond amplification cavities},}\ }\href {\doibase
  10.1364/OL.33.002482} {\bibfield  {journal} {\bibinfo  {journal} {Opt.
  Lett.}\ }\textbf {\bibinfo {volume} {33}},\ \bibinfo {pages} {2482--2484}
  (\bibinfo {year} {2008})}\BibitemShut {NoStop}%
\bibitem [{\citenamefont {Carlson}\ \emph {et~al.}(2011)\citenamefont
  {Carlson}, \citenamefont {Lee}, \citenamefont {Mongelli}, \citenamefont
  {Wright},\ and\ \citenamefont {Jones}}]{Carlson_OptLett2011}%
  \BibitemOpen
  \bibfield  {author} {\bibinfo {author} {\bibfnamefont {D.~R.}\ \bibnamefont
  {Carlson}}, \bibinfo {author} {\bibfnamefont {J.}~\bibnamefont {Lee}},
  \bibinfo {author} {\bibfnamefont {J.}~\bibnamefont {Mongelli}}, \bibinfo
  {author} {\bibfnamefont {E.~M.}\ \bibnamefont {Wright}}, \ and\ \bibinfo
  {author} {\bibfnamefont {R.~J.}\ \bibnamefont {Jones}},\ }\bibfield  {title}
  {\enquote {\bibinfo {title} {Intracavity ionization and pulse formation in
  femtosecond enhancement cavities},}\ }\href {\doibase 10.1364/OL.36.002991}
  {\bibfield  {journal} {\bibinfo  {journal} {Opt. Lett.}\ }\textbf {\bibinfo
  {volume} {36}},\ \bibinfo {pages} {2991--2993} (\bibinfo {year}
  {2011})}\BibitemShut {NoStop}%
\bibitem [{\citenamefont {Cing\"{o}z}\ \emph {et~al.}(2012)\citenamefont
  {Cing\"{o}z}, \citenamefont {Yost}, \citenamefont {Allison}, \citenamefont
  {Ruehl}, \citenamefont {Fermann}, \citenamefont {Hartl},\ and\ \citenamefont
  {Ye}}]{Cingoz_Nature2012}%
  \BibitemOpen
  \bibfield  {author} {\bibinfo {author} {\bibfnamefont {A.}~\bibnamefont
  {Cing\"{o}z}}, \bibinfo {author} {\bibfnamefont {D.~C.}\ \bibnamefont
  {Yost}}, \bibinfo {author} {\bibfnamefont {T.~K.}\ \bibnamefont {Allison}},
  \bibinfo {author} {\bibfnamefont {A.}~\bibnamefont {Ruehl}}, \bibinfo
  {author} {\bibfnamefont {M.~E.}\ \bibnamefont {Fermann}}, \bibinfo {author}
  {\bibfnamefont {I.}~\bibnamefont {Hartl}}, \ and\ \bibinfo {author}
  {\bibfnamefont {J.}~\bibnamefont {Ye}},\ }\bibfield  {title} {\enquote
  {\bibinfo {title} {Direct frequency comb spectroscopy in the extreme
  ultraviolet},}\ }\href {http://dx.doi.org/10.1038/nature10711} {\bibfield
  {journal} {\bibinfo  {journal} {Nature}\ }\textbf {\bibinfo {volume} {482}},\
  \bibinfo {pages} {68--71} (\bibinfo {year} {2012})}\BibitemShut {NoStop}%
\bibitem [{\citenamefont {Pupeza}\ \emph {et~al.}(2013)\citenamefont {Pupeza},
  \citenamefont {Holzberger}, \citenamefont {Eidam}, \citenamefont {Carstens},
  \citenamefont {Esser}, \citenamefont {Weitenberg}, \citenamefont {Ruszbuldt},
  \citenamefont {Rauschenberger}, \citenamefont {Limpert}, \citenamefont
  {Udem}, \citenamefont {T\"{u}nnermann}, \citenamefont {H\"{a}nsch},
  \citenamefont {Apolonski}, \citenamefont {Krausz},\ and\ \citenamefont
  {Fill}}]{Pupeza_NatPhot2013}%
  \BibitemOpen
  \bibfield  {author} {\bibinfo {author} {\bibfnamefont {I.}~\bibnamefont
  {Pupeza}}, \bibinfo {author} {\bibfnamefont {S.}~\bibnamefont {Holzberger}},
  \bibinfo {author} {\bibfnamefont {T.}~\bibnamefont {Eidam}}, \bibinfo
  {author} {\bibfnamefont {H.}~\bibnamefont {Carstens}}, \bibinfo {author}
  {\bibfnamefont {D.}~\bibnamefont {Esser}}, \bibinfo {author} {\bibfnamefont
  {J.}~\bibnamefont {Weitenberg}}, \bibinfo {author} {\bibfnamefont
  {P.}~\bibnamefont {Ruszbuldt}}, \bibinfo {author} {\bibfnamefont
  {J.}~\bibnamefont {Rauschenberger}}, \bibinfo {author} {\bibfnamefont
  {J.}~\bibnamefont {Limpert}}, \bibinfo {author} {\bibfnamefont
  {T.}~\bibnamefont {Udem}}, \bibinfo {author} {\bibfnamefont {A.}~\bibnamefont
  {T\"{u}nnermann}}, \bibinfo {author} {\bibfnamefont {T.~W.}\ \bibnamefont
  {H\"{a}nsch}}, \bibinfo {author} {\bibfnamefont {A.}~\bibnamefont
  {Apolonski}}, \bibinfo {author} {\bibfnamefont {F.}~\bibnamefont {Krausz}}, \
  and\ \bibinfo {author} {\bibfnamefont {E.}~\bibnamefont {Fill}},\ }\bibfield
  {title} {\enquote {\bibinfo {title} {Compact high-repetition-rate source of
  coherent 100 ev radiation},}\ }\href
  {http://dx.doi.org/10.1038/nphoton.2013.156} {\bibfield  {journal} {\bibinfo
  {journal} {Nat Photon}\ }\textbf {\bibinfo {volume} {7}},\ \bibinfo {pages}
  {608--612} (\bibinfo {year} {2013})}\BibitemShut {NoStop}%
\bibitem [{\citenamefont {Pupeza}\ \emph {et~al.}(2014)\citenamefont {Pupeza},
  \citenamefont {H\"ogner}, \citenamefont {Weitenberg}, \citenamefont
  {Holzberger}, \citenamefont {Esser}, \citenamefont {Eidam}, \citenamefont
  {Limpert}, \citenamefont {T\"unnermann}, \citenamefont {Fill},\ and\
  \citenamefont {Yakovlev}}]{Pupeza_PRL2014}%
  \BibitemOpen
  \bibfield  {author} {\bibinfo {author} {\bibfnamefont {I.}~\bibnamefont
  {Pupeza}}, \bibinfo {author} {\bibfnamefont {M.}~\bibnamefont {H\"ogner}},
  \bibinfo {author} {\bibfnamefont {J.}~\bibnamefont {Weitenberg}}, \bibinfo
  {author} {\bibfnamefont {S.}~\bibnamefont {Holzberger}}, \bibinfo {author}
  {\bibfnamefont {D.}~\bibnamefont {Esser}}, \bibinfo {author} {\bibfnamefont
  {T.}~\bibnamefont {Eidam}}, \bibinfo {author} {\bibfnamefont
  {J.}~\bibnamefont {Limpert}}, \bibinfo {author} {\bibfnamefont
  {A.}~\bibnamefont {T\"unnermann}}, \bibinfo {author} {\bibfnamefont
  {E.}~\bibnamefont {Fill}}, \ and\ \bibinfo {author} {\bibfnamefont {V.~S.}\
  \bibnamefont {Yakovlev}},\ }\bibfield  {title} {\enquote {\bibinfo {title}
  {Cavity-enhanced high-harmonic generation with spatially tailored driving
  fields},}\ }\href {\doibase 10.1103/PhysRevLett.112.103902} {\bibfield
  {journal} {\bibinfo  {journal} {Phys. Rev. Lett.}\ }\textbf {\bibinfo
  {volume} {112}},\ \bibinfo {pages} {103902} (\bibinfo {year}
  {2014})}\BibitemShut {NoStop}%
\bibitem [{\citenamefont {Carstens}\ \emph {et~al.}(2016)\citenamefont
  {Carstens}, \citenamefont {H\"{o}gner}, \citenamefont {Saule}, \citenamefont
  {Holzberger}, \citenamefont {Lilienfein}, \citenamefont {Guggenmos},
  \citenamefont {Jocher}, \citenamefont {Eidam}, \citenamefont {Esser},
  \citenamefont {Tosa}, \citenamefont {Pervak}, \citenamefont {Limpert},
  \citenamefont {T\"{u}nnermann}, \citenamefont {Kleineberg}, \citenamefont
  {Krausz},\ and\ \citenamefont {Pupeza}}]{Carstens_Optica2016}%
  \BibitemOpen
  \bibfield  {author} {\bibinfo {author} {\bibfnamefont {H.}~\bibnamefont
  {Carstens}}, \bibinfo {author} {\bibfnamefont {M.}~\bibnamefont
  {H\"{o}gner}}, \bibinfo {author} {\bibfnamefont {T.}~\bibnamefont {Saule}},
  \bibinfo {author} {\bibfnamefont {S.}~\bibnamefont {Holzberger}}, \bibinfo
  {author} {\bibfnamefont {N.}~\bibnamefont {Lilienfein}}, \bibinfo {author}
  {\bibfnamefont {A.}~\bibnamefont {Guggenmos}}, \bibinfo {author}
  {\bibfnamefont {C.}~\bibnamefont {Jocher}}, \bibinfo {author} {\bibfnamefont
  {T.}~\bibnamefont {Eidam}}, \bibinfo {author} {\bibfnamefont
  {D.}~\bibnamefont {Esser}}, \bibinfo {author} {\bibfnamefont
  {V.}~\bibnamefont {Tosa}}, \bibinfo {author} {\bibfnamefont {V.}~\bibnamefont
  {Pervak}}, \bibinfo {author} {\bibfnamefont {J.}~\bibnamefont {Limpert}},
  \bibinfo {author} {\bibfnamefont {A.}~\bibnamefont {T\"{u}nnermann}},
  \bibinfo {author} {\bibfnamefont {U.}~\bibnamefont {Kleineberg}}, \bibinfo
  {author} {\bibfnamefont {F.}~\bibnamefont {Krausz}}, \ and\ \bibinfo {author}
  {\bibfnamefont {I.}~\bibnamefont {Pupeza}},\ }\bibfield  {title} {\enquote
  {\bibinfo {title} {High-harmonic generation at 250 mhz with photon energies
  exceeding 100 ev},}\ }\href {\doibase 10.1364/OPTICA.3.000366} {\bibfield
  {journal} {\bibinfo  {journal} {Optica}\ }\textbf {\bibinfo {volume} {3}},\
  \bibinfo {pages} {366--369} (\bibinfo {year} {2016})}\BibitemShut {NoStop}%
\bibitem [{\citenamefont {H\"ogner}, \citenamefont {Tosa},\ and\ \citenamefont
  {Pupeza}(2017)}]{Hogner_NJP2017}%
  \BibitemOpen
  \bibfield  {author} {\bibinfo {author} {\bibfnamefont {M.}~\bibnamefont
  {H\"ogner}}, \bibinfo {author} {\bibfnamefont {V.}~\bibnamefont {Tosa}}, \
  and\ \bibinfo {author} {\bibfnamefont {I.}~\bibnamefont {Pupeza}},\
  }\bibfield  {title} {\enquote {\bibinfo {title} {Generation of isolated
  attosecond pulses with enhancement cavities---a theoretical study},}\ }\href
  {\doibase 10.1088/1367-2630/aa6315} {\bibfield  {journal} {\bibinfo
  {journal} {New Journal of Physics}\ }\textbf {\bibinfo {volume} {19}},\
  \bibinfo {pages} {033040} (\bibinfo {year} {2017})}\BibitemShut {NoStop}%
\bibitem [{\citenamefont {Holzberger}\ \emph {et~al.}(2015)\citenamefont
  {Holzberger}, \citenamefont {Lilienfein}, \citenamefont {Trubetskov},
  \citenamefont {Carstens}, \citenamefont {L\"{u}cking}, \citenamefont
  {Pervak}, \citenamefont {Krausz},\ and\ \citenamefont
  {Pupeza}}]{Holzberger_OptLett2015}%
  \BibitemOpen
  \bibfield  {author} {\bibinfo {author} {\bibfnamefont {S.}~\bibnamefont
  {Holzberger}}, \bibinfo {author} {\bibfnamefont {N.}~\bibnamefont
  {Lilienfein}}, \bibinfo {author} {\bibfnamefont {M.}~\bibnamefont
  {Trubetskov}}, \bibinfo {author} {\bibfnamefont {H.}~\bibnamefont
  {Carstens}}, \bibinfo {author} {\bibfnamefont {F.}~\bibnamefont
  {L\"{u}cking}}, \bibinfo {author} {\bibfnamefont {V.}~\bibnamefont {Pervak}},
  \bibinfo {author} {\bibfnamefont {F.}~\bibnamefont {Krausz}}, \ and\ \bibinfo
  {author} {\bibfnamefont {I.}~\bibnamefont {Pupeza}},\ }\bibfield  {title}
  {\enquote {\bibinfo {title} {Enhancement cavities for zero-offset-frequency
  pulse trains},}\ }\href {\doibase 10.1364/OL.40.002165} {\bibfield  {journal}
  {\bibinfo  {journal} {Opt. Lett.}\ }\textbf {\bibinfo {volume} {40}},\
  \bibinfo {pages} {2165--2168} (\bibinfo {year} {2015})}\BibitemShut {NoStop}%
\bibitem [{\citenamefont {Lilienfein}\ \emph {et~al.}(2019)\citenamefont
  {Lilienfein}, \citenamefont {Hofer}, \citenamefont {H{\"o}gner},
  \citenamefont {Saule}, \citenamefont {Trubetskov}, \citenamefont {Pervak},
  \citenamefont {Fill}, \citenamefont {Riek}, \citenamefont {Leitenstorfer},
  \citenamefont {Limpert}, \citenamefont {Krausz},\ and\ \citenamefont
  {Pupeza}}]{Lilienfein_NatPhot2019}%
  \BibitemOpen
  \bibfield  {author} {\bibinfo {author} {\bibfnamefont {N.}~\bibnamefont
  {Lilienfein}}, \bibinfo {author} {\bibfnamefont {C.}~\bibnamefont {Hofer}},
  \bibinfo {author} {\bibfnamefont {M.}~\bibnamefont {H{\"o}gner}}, \bibinfo
  {author} {\bibfnamefont {T.}~\bibnamefont {Saule}}, \bibinfo {author}
  {\bibfnamefont {M.}~\bibnamefont {Trubetskov}}, \bibinfo {author}
  {\bibfnamefont {V.}~\bibnamefont {Pervak}}, \bibinfo {author} {\bibfnamefont
  {E.}~\bibnamefont {Fill}}, \bibinfo {author} {\bibfnamefont {C.}~\bibnamefont
  {Riek}}, \bibinfo {author} {\bibfnamefont {A.}~\bibnamefont {Leitenstorfer}},
  \bibinfo {author} {\bibfnamefont {J.}~\bibnamefont {Limpert}}, \bibinfo
  {author} {\bibfnamefont {F.}~\bibnamefont {Krausz}}, \ and\ \bibinfo {author}
  {\bibfnamefont {I.}~\bibnamefont {Pupeza}},\ }\bibfield  {title} {\enquote
  {\bibinfo {title} {Temporal solitons in free-space femtosecond enhancement
  cavities},}\ }\href {\doibase 10.1038/s41566-018-0341-y} {\bibfield
  {journal} {\bibinfo  {journal} {Nature Photonics}\ }\textbf {\bibinfo
  {volume} {13}},\ \bibinfo {pages} {214--218} (\bibinfo {year}
  {2019})}\BibitemShut {NoStop}%
\bibitem [{\citenamefont {Lilienfein}\ \emph {et~al.}(2017)\citenamefont
  {Lilienfein}, \citenamefont {Hofer}, \citenamefont {Holzberger},
  \citenamefont {Matzer}, \citenamefont {Zimmermann}, \citenamefont
  {Trubetskov}, \citenamefont {Pervak},\ and\ \citenamefont
  {Pupeza}}]{Lilienfein_OptLett2017}%
  \BibitemOpen
  \bibfield  {author} {\bibinfo {author} {\bibfnamefont {N.}~\bibnamefont
  {Lilienfein}}, \bibinfo {author} {\bibfnamefont {C.}~\bibnamefont {Hofer}},
  \bibinfo {author} {\bibfnamefont {S.}~\bibnamefont {Holzberger}}, \bibinfo
  {author} {\bibfnamefont {C.}~\bibnamefont {Matzer}}, \bibinfo {author}
  {\bibfnamefont {P.}~\bibnamefont {Zimmermann}}, \bibinfo {author}
  {\bibfnamefont {M.}~\bibnamefont {Trubetskov}}, \bibinfo {author}
  {\bibfnamefont {V.}~\bibnamefont {Pervak}}, \ and\ \bibinfo {author}
  {\bibfnamefont {I.}~\bibnamefont {Pupeza}},\ }\bibfield  {title} {\enquote
  {\bibinfo {title} {Enhancement cavities for few-cycle pulses},}\ }\href
  {\doibase 10.1364/OL.42.000271} {\bibfield  {journal} {\bibinfo  {journal}
  {Opt. Lett.}\ }\textbf {\bibinfo {volume} {42}},\ \bibinfo {pages} {271--274}
  (\bibinfo {year} {2017})}\BibitemShut {NoStop}%
\bibitem [{\citenamefont {Porat}\ \emph {et~al.}(2018)\citenamefont {Porat},
  \citenamefont {Heyl}, \citenamefont {Schoun}, \citenamefont {Benko},
  \citenamefont {D{\"o}rre}, \citenamefont {Corwin},\ and\ \citenamefont
  {Ye}}]{Porat_NatPhot2018}%
  \BibitemOpen
  \bibfield  {author} {\bibinfo {author} {\bibfnamefont {G.}~\bibnamefont
  {Porat}}, \bibinfo {author} {\bibfnamefont {C.~M.}\ \bibnamefont {Heyl}},
  \bibinfo {author} {\bibfnamefont {S.~B.}\ \bibnamefont {Schoun}}, \bibinfo
  {author} {\bibfnamefont {C.}~\bibnamefont {Benko}}, \bibinfo {author}
  {\bibfnamefont {N.}~\bibnamefont {D{\"o}rre}}, \bibinfo {author}
  {\bibfnamefont {K.~L.}\ \bibnamefont {Corwin}}, \ and\ \bibinfo {author}
  {\bibfnamefont {J.}~\bibnamefont {Ye}},\ }\bibfield  {title} {\enquote
  {\bibinfo {title} {Phase-matched extreme-ultraviolet frequency-comb
  generation},}\ }\href {\doibase 10.1038/s41566-018-0199-z} {\bibfield
  {journal} {\bibinfo  {journal} {Nature Photonics}\ } (\bibinfo {year}
  {2018}),\ 10.1038/s41566-018-0199-z}\BibitemShut {NoStop}%
\bibitem [{\citenamefont {Zhang}\ \emph {et~al.}(2020)\citenamefont {Zhang},
  \citenamefont {Schoun}, \citenamefont {Heyl}, \citenamefont {Porat},
  \citenamefont {Gaarde},\ and\ \citenamefont {Ye}}]{Zhang_PRL2020}%
  \BibitemOpen
  \bibfield  {author} {\bibinfo {author} {\bibfnamefont {C.}~\bibnamefont
  {Zhang}}, \bibinfo {author} {\bibfnamefont {S.~B.}\ \bibnamefont {Schoun}},
  \bibinfo {author} {\bibfnamefont {C.~M.}\ \bibnamefont {Heyl}}, \bibinfo
  {author} {\bibfnamefont {G.}~\bibnamefont {Porat}}, \bibinfo {author}
  {\bibfnamefont {M.~B.}\ \bibnamefont {Gaarde}}, \ and\ \bibinfo {author}
  {\bibfnamefont {J.}~\bibnamefont {Ye}},\ }\bibfield  {title} {\enquote
  {\bibinfo {title} {Noncollinear enhancement cavity for record-high
  out-coupling efficiency of an extreme-uv frequency comb},}\ }\href {\doibase
  10.1103/PhysRevLett.125.093902} {\bibfield  {journal} {\bibinfo  {journal}
  {Phys. Rev. Lett.}\ }\textbf {\bibinfo {volume} {125}},\ \bibinfo {pages}
  {093902} (\bibinfo {year} {2020})}\BibitemShut {NoStop}%
\bibitem [{\citenamefont {Buades}\ \emph {et~al.}(2021)\citenamefont {Buades},
  \citenamefont {Pic\'{o}n}, \citenamefont {Berger}, \citenamefont {Le\'{o}n},
  \citenamefont {Di~Palo}, \citenamefont {Cousin}, \citenamefont {Cocchi},
  \citenamefont {Pellegrin}, \citenamefont {Martin}, \citenamefont
  {Ma{\~n}as-Valero}, \citenamefont {Coronado}, \citenamefont {Danz},
  \citenamefont {Draxl}, \citenamefont {Uemoto}, \citenamefont {Yabana},
  \citenamefont {Schultze}, \citenamefont {Wall}, \citenamefont {Z\"{u}rch},\
  and\ \citenamefont {Biegert}}]{Buades_ApplPhysRev2021}%
  \BibitemOpen
  \bibfield  {author} {\bibinfo {author} {\bibfnamefont {B.}~\bibnamefont
  {Buades}}, \bibinfo {author} {\bibfnamefont {A.}~\bibnamefont {Pic\'{o}n}},
  \bibinfo {author} {\bibfnamefont {E.}~\bibnamefont {Berger}}, \bibinfo
  {author} {\bibfnamefont {I.}~\bibnamefont {Le\'{o}n}}, \bibinfo {author}
  {\bibfnamefont {N.}~\bibnamefont {Di~Palo}}, \bibinfo {author} {\bibfnamefont
  {S.~L.}\ \bibnamefont {Cousin}}, \bibinfo {author} {\bibfnamefont
  {C.}~\bibnamefont {Cocchi}}, \bibinfo {author} {\bibfnamefont
  {E.}~\bibnamefont {Pellegrin}}, \bibinfo {author} {\bibfnamefont {J.~H.}\
  \bibnamefont {Martin}}, \bibinfo {author} {\bibfnamefont {S.}~\bibnamefont
  {Ma{\~n}as-Valero}}, \bibinfo {author} {\bibfnamefont {E.}~\bibnamefont
  {Coronado}}, \bibinfo {author} {\bibfnamefont {T.}~\bibnamefont {Danz}},
  \bibinfo {author} {\bibfnamefont {C.}~\bibnamefont {Draxl}}, \bibinfo
  {author} {\bibfnamefont {M.}~\bibnamefont {Uemoto}}, \bibinfo {author}
  {\bibfnamefont {K.}~\bibnamefont {Yabana}}, \bibinfo {author} {\bibfnamefont
  {M.}~\bibnamefont {Schultze}}, \bibinfo {author} {\bibfnamefont
  {S.}~\bibnamefont {Wall}}, \bibinfo {author} {\bibfnamefont {M.}~\bibnamefont
  {Z\"{u}rch}}, \ and\ \bibinfo {author} {\bibfnamefont {J.}~\bibnamefont
  {Biegert}},\ }\bibfield  {title} {\enquote {\bibinfo {title} {{Attosecond
  state-resolved carrier motion in quantum materials probed by soft x-ray
  XANES}},}\ }\href {\doibase 10.1063/5.0020649} {\bibfield  {journal}
  {\bibinfo  {journal} {Applied Physics Reviews}\ }\textbf {\bibinfo {volume}
  {8}},\ \bibinfo {pages} {011408} (\bibinfo {year} {2021})},\ \Eprint
  {http://arxiv.org/abs/https://pubs.aip.org/aip/apr/article-pdf/doi/10.1063/5.0020649/14580075/011408\_1\_online.pdf}
  {https://pubs.aip.org/aip/apr/article-pdf/doi/10.1063/5.0020649/14580075/011408\_1\_online.pdf}
  \BibitemShut {NoStop}%
\bibitem [{\citenamefont {Saito}\ \emph {et~al.}(2019)\citenamefont {Saito},
  \citenamefont {Sannohe}, \citenamefont {Ishii}, \citenamefont {Kanai},
  \citenamefont {Kosugi}, \citenamefont {Wu}, \citenamefont {Chew},
  \citenamefont {Han}, \citenamefont {Chang},\ and\ \citenamefont
  {Itatani}}]{Saito_Optica2019}%
  \BibitemOpen
  \bibfield  {author} {\bibinfo {author} {\bibfnamefont {N.}~\bibnamefont
  {Saito}}, \bibinfo {author} {\bibfnamefont {H.}~\bibnamefont {Sannohe}},
  \bibinfo {author} {\bibfnamefont {N.}~\bibnamefont {Ishii}}, \bibinfo
  {author} {\bibfnamefont {T.}~\bibnamefont {Kanai}}, \bibinfo {author}
  {\bibfnamefont {N.}~\bibnamefont {Kosugi}}, \bibinfo {author} {\bibfnamefont
  {Y.}~\bibnamefont {Wu}}, \bibinfo {author} {\bibfnamefont {A.}~\bibnamefont
  {Chew}}, \bibinfo {author} {\bibfnamefont {S.}~\bibnamefont {Han}}, \bibinfo
  {author} {\bibfnamefont {Z.}~\bibnamefont {Chang}}, \ and\ \bibinfo {author}
  {\bibfnamefont {J.}~\bibnamefont {Itatani}},\ }\bibfield  {title} {\enquote
  {\bibinfo {title} {Real-time observation of electronic, vibrational, and
  rotational dynamics in nitric oxide with attosecond soft x-ray pulses at 400
  ev},}\ }\href {\doibase 10.1364/OPTICA.6.001542} {\bibfield  {journal}
  {\bibinfo  {journal} {Optica}\ }\textbf {\bibinfo {volume} {6}},\ \bibinfo
  {pages} {1542--1546} (\bibinfo {year} {2019})}\BibitemShut {NoStop}%
\bibitem [{\citenamefont {Summers}\ \emph {et~al.}(2023)\citenamefont
  {Summers}, \citenamefont {Severino}, \citenamefont {Reduzzi}, \citenamefont
  {Sidiropoulos}, \citenamefont {Rivas}, \citenamefont {Palo}, \citenamefont
  {Sun}, \citenamefont {Chien}, \citenamefont {Le{\'o}n}, \citenamefont
  {Buades}, \citenamefont {Cousin}, \citenamefont {Teichmann}, \citenamefont
  {Mey}, \citenamefont {Mann}, \citenamefont {Keitel}, \citenamefont
  {Pl{\"o}njes}, \citenamefont {Efetov}, \citenamefont {Schwoerer},\ and\
  \citenamefont {Biegert}}]{Summers_UltrafastScience2023}%
  \BibitemOpen
  \bibfield  {author} {\bibinfo {author} {\bibfnamefont {A.~M.}\ \bibnamefont
  {Summers}}, \bibinfo {author} {\bibfnamefont {S.}~\bibnamefont {Severino}},
  \bibinfo {author} {\bibfnamefont {M.}~\bibnamefont {Reduzzi}}, \bibinfo
  {author} {\bibfnamefont {T.~P.~H.}\ \bibnamefont {Sidiropoulos}}, \bibinfo
  {author} {\bibfnamefont {D.~E.}\ \bibnamefont {Rivas}}, \bibinfo {author}
  {\bibfnamefont {N.~D.}\ \bibnamefont {Palo}}, \bibinfo {author}
  {\bibfnamefont {H.-W.}\ \bibnamefont {Sun}}, \bibinfo {author} {\bibfnamefont
  {Y.-H.}\ \bibnamefont {Chien}}, \bibinfo {author} {\bibfnamefont
  {I.}~\bibnamefont {Le{\'o}n}}, \bibinfo {author} {\bibfnamefont
  {B.}~\bibnamefont {Buades}}, \bibinfo {author} {\bibfnamefont {S.~L.}\
  \bibnamefont {Cousin}}, \bibinfo {author} {\bibfnamefont {S.~M.}\
  \bibnamefont {Teichmann}}, \bibinfo {author} {\bibfnamefont {T.}~\bibnamefont
  {Mey}}, \bibinfo {author} {\bibfnamefont {K.}~\bibnamefont {Mann}}, \bibinfo
  {author} {\bibfnamefont {B.}~\bibnamefont {Keitel}}, \bibinfo {author}
  {\bibfnamefont {E.}~\bibnamefont {Pl{\"o}njes}}, \bibinfo {author}
  {\bibfnamefont {D.~K.}\ \bibnamefont {Efetov}}, \bibinfo {author}
  {\bibfnamefont {H.}~\bibnamefont {Schwoerer}}, \ and\ \bibinfo {author}
  {\bibfnamefont {J.}~\bibnamefont {Biegert}},\ }\bibfield  {title} {\enquote
  {\bibinfo {title} {Realizing attosecond core-level x-ray spectroscopy for the
  investigation of condensed matter systems},}\ }\href {\doibase
  10.34133/ultrafastscience.0004} {\bibfield  {journal} {\bibinfo  {journal}
  {Ultrafast Science}\ }\textbf {\bibinfo {volume} {3}},\ \bibinfo {pages}
  {0004} (\bibinfo {year} {2023})},\ \Eprint
  {http://arxiv.org/abs/https://spj.science.org/doi/pdf/10.34133/ultrafastscience.0004}
  {https://spj.science.org/doi/pdf/10.34133/ultrafastscience.0004} \BibitemShut
  {NoStop}%
\bibitem [{\citenamefont {Teichmann}\ \emph {et~al.}(2016)\citenamefont
  {Teichmann}, \citenamefont {Silva}, \citenamefont {Cousin}, \citenamefont
  {Hemmer},\ and\ \citenamefont {Biegert}}]{Teichmann_NatComm2016}%
  \BibitemOpen
  \bibfield  {author} {\bibinfo {author} {\bibfnamefont {S.~M.}\ \bibnamefont
  {Teichmann}}, \bibinfo {author} {\bibfnamefont {F.}~\bibnamefont {Silva}},
  \bibinfo {author} {\bibfnamefont {S.~L.}\ \bibnamefont {Cousin}}, \bibinfo
  {author} {\bibfnamefont {M.}~\bibnamefont {Hemmer}}, \ and\ \bibinfo {author}
  {\bibfnamefont {J.}~\bibnamefont {Biegert}},\ }\bibfield  {title} {\enquote
  {\bibinfo {title} {0.5-kev soft x-ray attosecond continua},}\ }\href
  {http://dx.doi.org/10.1038/ncomms11493} {\bibfield  {journal} {\bibinfo
  {journal} {Nature Communications}\ }\textbf {\bibinfo {volume} {7}},\
  \bibinfo {pages} {11493 EP --} (\bibinfo {year} {2016})}\BibitemShut
  {NoStop}%
\bibitem [{\citenamefont {Bakalis}\ \emph {et~al.}(2024)\citenamefont
  {Bakalis}, \citenamefont {Chernov}, \citenamefont {Li}, \citenamefont
  {Kunin}, \citenamefont {Withers}, \citenamefont {Cheng}, \citenamefont
  {Adler}, \citenamefont {Zhao}, \citenamefont {Corder}, \citenamefont {White},
  \citenamefont {Sch{\"o}nhense}, \citenamefont {Du}, \citenamefont
  {Kawakami},\ and\ \citenamefont {Allison}}]{Bakalis_NanoLett2024}%
  \BibitemOpen
  \bibfield  {author} {\bibinfo {author} {\bibfnamefont {J.}~\bibnamefont
  {Bakalis}}, \bibinfo {author} {\bibfnamefont {S.}~\bibnamefont {Chernov}},
  \bibinfo {author} {\bibfnamefont {Z.}~\bibnamefont {Li}}, \bibinfo {author}
  {\bibfnamefont {A.}~\bibnamefont {Kunin}}, \bibinfo {author} {\bibfnamefont
  {Z.~H.}\ \bibnamefont {Withers}}, \bibinfo {author} {\bibfnamefont
  {S.}~\bibnamefont {Cheng}}, \bibinfo {author} {\bibfnamefont
  {A.}~\bibnamefont {Adler}}, \bibinfo {author} {\bibfnamefont
  {P.}~\bibnamefont {Zhao}}, \bibinfo {author} {\bibfnamefont {C.}~\bibnamefont
  {Corder}}, \bibinfo {author} {\bibfnamefont {M.~G.}\ \bibnamefont {White}},
  \bibinfo {author} {\bibfnamefont {G.}~\bibnamefont {Sch{\"o}nhense}},
  \bibinfo {author} {\bibfnamefont {X.}~\bibnamefont {Du}}, \bibinfo {author}
  {\bibfnamefont {R.~K.}\ \bibnamefont {Kawakami}}, \ and\ \bibinfo {author}
  {\bibfnamefont {T.~K.}\ \bibnamefont {Allison}},\ }\bibfield  {title}
  {\enquote {\bibinfo {title} {Momentum-space observation of optically excited
  nonthermal electrons in graphene with persistent pseudospin polarization},}\
  }\href {\doibase 10.1021/acs.nanolett.4c02378} {\bibfield  {journal}
  {\bibinfo  {journal} {Nano Letters}\ } (\bibinfo {year} {2024}),\
  10.1021/acs.nanolett.4c02378}\BibitemShut {NoStop}%
\bibitem [{\citenamefont {Na}\ \emph {et~al.}(2020)\citenamefont {Na},
  \citenamefont {Boschini}, \citenamefont {Mills}, \citenamefont {Michiardi},
  \citenamefont {Day}, \citenamefont {Zwartsenberg}, \citenamefont {Levy},
  \citenamefont {Zhdanovich}, \citenamefont {Kemper}, \citenamefont {Jones},\
  and\ \citenamefont {Damascelli}}]{Na_PRB2020}%
  \BibitemOpen
  \bibfield  {author} {\bibinfo {author} {\bibfnamefont {M.~X.}\ \bibnamefont
  {Na}}, \bibinfo {author} {\bibfnamefont {F.}~\bibnamefont {Boschini}},
  \bibinfo {author} {\bibfnamefont {A.~K.}\ \bibnamefont {Mills}}, \bibinfo
  {author} {\bibfnamefont {M.}~\bibnamefont {Michiardi}}, \bibinfo {author}
  {\bibfnamefont {R.~P.}\ \bibnamefont {Day}}, \bibinfo {author} {\bibfnamefont
  {B.}~\bibnamefont {Zwartsenberg}}, \bibinfo {author} {\bibfnamefont
  {G.}~\bibnamefont {Levy}}, \bibinfo {author} {\bibfnamefont {S.}~\bibnamefont
  {Zhdanovich}}, \bibinfo {author} {\bibfnamefont {A.~F.}\ \bibnamefont
  {Kemper}}, \bibinfo {author} {\bibfnamefont {D.~J.}\ \bibnamefont {Jones}}, \
  and\ \bibinfo {author} {\bibfnamefont {A.}~\bibnamefont {Damascelli}},\
  }\bibfield  {title} {\enquote {\bibinfo {title} {Establishing nonthermal
  regimes in pump-probe electron relaxation dynamics},}\ }\href {\doibase
  10.1103/PhysRevB.102.184307} {\bibfield  {journal} {\bibinfo  {journal}
  {Phys. Rev. B}\ }\textbf {\bibinfo {volume} {102}},\ \bibinfo {pages}
  {184307} (\bibinfo {year} {2020})}\BibitemShut {NoStop}%
\bibitem [{\citenamefont {Frassetto}\ \emph {et~al.}(2011)\citenamefont
  {Frassetto}, \citenamefont {Cacho}, \citenamefont {Froud}, \citenamefont
  {Turcu}, \citenamefont {Villoresi}, \citenamefont {Bryan}, \citenamefont
  {Springate},\ and\ \citenamefont {Poletto}}]{Frassetto_OptExp2011}%
  \BibitemOpen
  \bibfield  {author} {\bibinfo {author} {\bibfnamefont {F.}~\bibnamefont
  {Frassetto}}, \bibinfo {author} {\bibfnamefont {C.}~\bibnamefont {Cacho}},
  \bibinfo {author} {\bibfnamefont {C.~A.}\ \bibnamefont {Froud}}, \bibinfo
  {author} {\bibfnamefont {I.~E.}\ \bibnamefont {Turcu}}, \bibinfo {author}
  {\bibfnamefont {P.}~\bibnamefont {Villoresi}}, \bibinfo {author}
  {\bibfnamefont {W.~A.}\ \bibnamefont {Bryan}}, \bibinfo {author}
  {\bibfnamefont {E.}~\bibnamefont {Springate}}, \ and\ \bibinfo {author}
  {\bibfnamefont {L.}~\bibnamefont {Poletto}},\ }\bibfield  {title} {\enquote
  {\bibinfo {title} {Single-grating monochromator for extreme-ultraviolet
  ultrashort pulses},}\ }\href {\doibase 10.1364/OE.19.019169} {\bibfield
  {journal} {\bibinfo  {journal} {Opt. Express}\ }\textbf {\bibinfo {volume}
  {19}},\ \bibinfo {pages} {19169--19181} (\bibinfo {year} {2011})}\BibitemShut
  {NoStop}%
\bibitem [{\citenamefont {Yost}, \citenamefont {Schibli},\ and\ \citenamefont
  {Ye}(2008)}]{Yost_OptLett2008}%
  \BibitemOpen
  \bibfield  {author} {\bibinfo {author} {\bibfnamefont {D.~C.}\ \bibnamefont
  {Yost}}, \bibinfo {author} {\bibfnamefont {T.~R.}\ \bibnamefont {Schibli}}, \
  and\ \bibinfo {author} {\bibfnamefont {J.}~\bibnamefont {Ye}},\ }\bibfield
  {title} {\enquote {\bibinfo {title} {Efficient outpult coupling of
  intracavity high harmonic generation},}\ }\href@noop {} {\bibfield  {journal}
  {\bibinfo  {journal} {Opt. Lett.}\ }\textbf {\bibinfo {volume} {33}},\
  \bibinfo {pages} {1099--1101} (\bibinfo {year} {2008})}\BibitemShut {NoStop}%
\bibitem [{\citenamefont {Zhao}(2019)}]{Zhao_Thesis2019}%
  \BibitemOpen
  \bibfield  {author} {\bibinfo {author} {\bibfnamefont {P.}~\bibnamefont
  {Zhao}},\ }\emph {\bibinfo {title} {An advanced instrument for time- and
  angle-resolved photoemission spectroscopy}},\ \href@noop {} {Ph.D. thesis},\
  \bibinfo  {school} {Stony Brook University} (\bibinfo {year}
  {2019})\BibitemShut {NoStop}%
\bibitem [{\citenamefont {Corder}\ \emph
  {et~al.}(2018{\natexlab{b}})\citenamefont {Corder}, \citenamefont {Zhao},
  \citenamefont {Bakalis}, \citenamefont {Li}, \citenamefont {Kershis},
  \citenamefont {Muraca}, \citenamefont {White},\ and\ \citenamefont
  {Allison}}]{Corder_SPIE2018}%
  \BibitemOpen
  \bibfield  {author} {\bibinfo {author} {\bibfnamefont {C.}~\bibnamefont
  {Corder}}, \bibinfo {author} {\bibfnamefont {P.}~\bibnamefont {Zhao}},
  \bibinfo {author} {\bibfnamefont {J.}~\bibnamefont {Bakalis}}, \bibinfo
  {author} {\bibfnamefont {X.}~\bibnamefont {Li}}, \bibinfo {author}
  {\bibfnamefont {M.~D.}\ \bibnamefont {Kershis}}, \bibinfo {author}
  {\bibfnamefont {A.~R.}\ \bibnamefont {Muraca}}, \bibinfo {author}
  {\bibfnamefont {M.~G.}\ \bibnamefont {White}}, \ and\ \bibinfo {author}
  {\bibfnamefont {T.~K.}\ \bibnamefont {Allison}},\ }\bibfield  {title}
  {\enquote {\bibinfo {title} {Development of a tunable high repetition rate
  xuv source for time-resolved photoemission studies of ultrafast dynamics at
  surfaces},}\ }\href {\doibase 10.1117/12.2295232} {\bibfield  {journal}
  {\bibinfo  {journal} {Proc.SPIE}\ }\textbf {\bibinfo {volume} {10519}},\
  \bibinfo {pages} {10519 -- 10519 -- 7} (\bibinfo {year}
  {2018}{\natexlab{b}})}\BibitemShut {NoStop}%
\bibitem [{\citenamefont {Udem}\ \emph {et~al.}(1999)\citenamefont {Udem},
  \citenamefont {Reichert}, \citenamefont {Holzwarth},\ and\ \citenamefont
  {H\"{a}nsch}}]{Udem_OptLett1999}%
  \BibitemOpen
  \bibfield  {author} {\bibinfo {author} {\bibfnamefont {T.}~\bibnamefont
  {Udem}}, \bibinfo {author} {\bibfnamefont {J.}~\bibnamefont {Reichert}},
  \bibinfo {author} {\bibfnamefont {R.}~\bibnamefont {Holzwarth}}, \ and\
  \bibinfo {author} {\bibfnamefont {T.~W.}\ \bibnamefont {H\"{a}nsch}},\
  }\bibfield  {title} {\enquote {\bibinfo {title} {Accurate measurement of
  large optical frequency differences with a mode-locked laser},}\ }\href
  {\doibase 10.1364/OL.24.000881} {\bibfield  {journal} {\bibinfo  {journal}
  {Opt. Lett.}\ }\textbf {\bibinfo {volume} {24}},\ \bibinfo {pages} {881--883}
  (\bibinfo {year} {1999})}\BibitemShut {NoStop}%
\bibitem [{\citenamefont {Silfies}\ \emph {et~al.}(2020)\citenamefont
  {Silfies}, \citenamefont {Kowzan}, \citenamefont {Chen}, \citenamefont
  {Lewis}, \citenamefont {Hou}, \citenamefont {Baehre}, \citenamefont {Gross},\
  and\ \citenamefont {Allison}}]{Silfies_OptLett2020}%
  \BibitemOpen
  \bibfield  {author} {\bibinfo {author} {\bibfnamefont {M.~C.}\ \bibnamefont
  {Silfies}}, \bibinfo {author} {\bibfnamefont {G.}~\bibnamefont {Kowzan}},
  \bibinfo {author} {\bibfnamefont {Y.}~\bibnamefont {Chen}}, \bibinfo {author}
  {\bibfnamefont {N.}~\bibnamefont {Lewis}}, \bibinfo {author} {\bibfnamefont
  {R.}~\bibnamefont {Hou}}, \bibinfo {author} {\bibfnamefont {R.}~\bibnamefont
  {Baehre}}, \bibinfo {author} {\bibfnamefont {T.}~\bibnamefont {Gross}}, \
  and\ \bibinfo {author} {\bibfnamefont {T.~K.}\ \bibnamefont {Allison}},\
  }\bibfield  {title} {\enquote {\bibinfo {title} {Widely tunable
  cavity-enhanced frequency combs},}\ }\href {\doibase 10.1364/OL.389412}
  {\bibfield  {journal} {\bibinfo  {journal} {Opt. Lett.}\ }\textbf {\bibinfo
  {volume} {45}},\ \bibinfo {pages} {2123--2126} (\bibinfo {year}
  {2020})}\BibitemShut {NoStop}%
\bibitem [{\citenamefont {Ozawa}\ \emph {et~al.}(2015)\citenamefont {Ozawa},
  \citenamefont {Zhao}, \citenamefont {Kuwata-Gonokami},\ and\ \citenamefont
  {Kobayashi}}]{Ozawa_OptExp2015}%
  \BibitemOpen
  \bibfield  {author} {\bibinfo {author} {\bibfnamefont {A.}~\bibnamefont
  {Ozawa}}, \bibinfo {author} {\bibfnamefont {Z.}~\bibnamefont {Zhao}},
  \bibinfo {author} {\bibfnamefont {M.}~\bibnamefont {Kuwata-Gonokami}}, \ and\
  \bibinfo {author} {\bibfnamefont {Y.}~\bibnamefont {Kobayashi}},\ }\bibfield
  {title} {\enquote {\bibinfo {title} {High average power coherent vuv
  generation at 10 mhz repetition frequency by intracavity high harmonic
  generation},}\ }\href {\doibase 10.1364/OE.23.015107} {\bibfield  {journal}
  {\bibinfo  {journal} {Opt. Express}\ }\textbf {\bibinfo {volume} {23}},\
  \bibinfo {pages} {15107--15118} (\bibinfo {year} {2015})}\BibitemShut
  {NoStop}%
\bibitem [{\citenamefont {Ozawa}\ \emph {et~al.}(2008)\citenamefont {Ozawa},
  \citenamefont {Rauschenberger}, \citenamefont {Gohle}, \citenamefont
  {Herrmann}, \citenamefont {Walker}, \citenamefont {Pervak}, \citenamefont
  {Fernandez}, \citenamefont {Graf}, \citenamefont {Apolonski}, \citenamefont
  {Holzwarth}, \citenamefont {Krausz}, \citenamefont {H\"{a}nsch},\ and\
  \citenamefont {Udem}}]{Ozawa_PRL2008}%
  \BibitemOpen
  \bibfield  {author} {\bibinfo {author} {\bibfnamefont {A.}~\bibnamefont
  {Ozawa}}, \bibinfo {author} {\bibfnamefont {J.}~\bibnamefont
  {Rauschenberger}}, \bibinfo {author} {\bibfnamefont {C.}~\bibnamefont
  {Gohle}}, \bibinfo {author} {\bibfnamefont {M.}~\bibnamefont {Herrmann}},
  \bibinfo {author} {\bibfnamefont {D.~R.}\ \bibnamefont {Walker}}, \bibinfo
  {author} {\bibfnamefont {V.}~\bibnamefont {Pervak}}, \bibinfo {author}
  {\bibfnamefont {A.}~\bibnamefont {Fernandez}}, \bibinfo {author}
  {\bibfnamefont {R.}~\bibnamefont {Graf}}, \bibinfo {author} {\bibfnamefont
  {A.}~\bibnamefont {Apolonski}}, \bibinfo {author} {\bibfnamefont
  {R.}~\bibnamefont {Holzwarth}}, \bibinfo {author} {\bibfnamefont
  {F.}~\bibnamefont {Krausz}}, \bibinfo {author} {\bibfnamefont {T.~W.}\
  \bibnamefont {H\"{a}nsch}}, \ and\ \bibinfo {author} {\bibfnamefont
  {T.}~\bibnamefont {Udem}},\ }\bibfield  {title} {\enquote {\bibinfo {title}
  {High harmonic frequency combs for high resolution spectroscopy},}\ }\href
  {\doibase 10.1103/PhysRevLett.100.253901} {\bibfield  {journal} {\bibinfo
  {journal} {Phys. Rev. Lett.}\ }\textbf {\bibinfo {volume} {100}},\ \bibinfo
  {eid} {253901} (\bibinfo {year} {2008})}\BibitemShut {NoStop}%
\bibitem [{\citenamefont {Canella}\ \emph {et~al.}(2024)\citenamefont
  {Canella}, \citenamefont {Weitenberg}, \citenamefont {Thariq}, \citenamefont
  {Schmid}, \citenamefont {Dwivedi}, \citenamefont {Galzerano}, \citenamefont
  {H\"{a}nsch}, \citenamefont {Udem},\ and\ \citenamefont
  {Ozawa}}]{Canella_Optica2024}%
  \BibitemOpen
  \bibfield  {author} {\bibinfo {author} {\bibfnamefont {F.}~\bibnamefont
  {Canella}}, \bibinfo {author} {\bibfnamefont {J.}~\bibnamefont {Weitenberg}},
  \bibinfo {author} {\bibfnamefont {M.}~\bibnamefont {Thariq}}, \bibinfo
  {author} {\bibfnamefont {F.}~\bibnamefont {Schmid}}, \bibinfo {author}
  {\bibfnamefont {P.}~\bibnamefont {Dwivedi}}, \bibinfo {author} {\bibfnamefont
  {G.}~\bibnamefont {Galzerano}}, \bibinfo {author} {\bibfnamefont {T.~W.}\
  \bibnamefont {H\"{a}nsch}}, \bibinfo {author} {\bibfnamefont
  {T.}~\bibnamefont {Udem}}, \ and\ \bibinfo {author} {\bibfnamefont
  {A.}~\bibnamefont {Ozawa}},\ }\bibfield  {title} {\enquote {\bibinfo {title}
  {Low-repetition-rate optical frequency comb},}\ }\href {\doibase
  10.1364/OPTICA.506353} {\bibfield  {journal} {\bibinfo  {journal} {Optica}\
  }\textbf {\bibinfo {volume} {11}},\ \bibinfo {pages} {1--9} (\bibinfo {year}
  {2024})}\BibitemShut {NoStop}%
\bibitem [{\citenamefont {Saule}\ \emph {et~al.}(2018)\citenamefont {Saule},
  \citenamefont {H{\"o}gner}, \citenamefont {Lilienfein}, \citenamefont
  {de~Vries}, \citenamefont {Pl{\"o}tner}, \citenamefont {Yakovlev},
  \citenamefont {Karpowicz}, \citenamefont {Limpert},\ and\ \citenamefont
  {Pupeza}}]{Saule_APLPhot2018}%
  \BibitemOpen
  \bibfield  {author} {\bibinfo {author} {\bibfnamefont {T.}~\bibnamefont
  {Saule}}, \bibinfo {author} {\bibfnamefont {M.}~\bibnamefont {H{\"o}gner}},
  \bibinfo {author} {\bibfnamefont {N.}~\bibnamefont {Lilienfein}}, \bibinfo
  {author} {\bibfnamefont {O.}~\bibnamefont {de~Vries}}, \bibinfo {author}
  {\bibfnamefont {M.}~\bibnamefont {Pl{\"o}tner}}, \bibinfo {author}
  {\bibfnamefont {V.~S.}\ \bibnamefont {Yakovlev}}, \bibinfo {author}
  {\bibfnamefont {N.}~\bibnamefont {Karpowicz}}, \bibinfo {author}
  {\bibfnamefont {J.}~\bibnamefont {Limpert}}, \ and\ \bibinfo {author}
  {\bibfnamefont {I.}~\bibnamefont {Pupeza}},\ }\bibfield  {title} {\enquote
  {\bibinfo {title} {{Cumulative plasma effects in cavity-enhanced high-order
  harmonic generation in gases}},}\ }\href {\doibase 10.1063/1.5037196}
  {\bibfield  {journal} {\bibinfo  {journal} {APL Photonics}\ }\textbf
  {\bibinfo {volume} {3}},\ \bibinfo {pages} {101301} (\bibinfo {year}
  {2018})},\ \Eprint
  {http://arxiv.org/abs/https://pubs.aip.org/aip/app/article-pdf/doi/10.1063/1.5037196/14568348/101301\_1\_online.pdf}
  {https://pubs.aip.org/aip/app/article-pdf/doi/10.1063/1.5037196/14568348/101301\_1\_online.pdf}
  \BibitemShut {NoStop}%
\bibitem [{\citenamefont {Schliesser}, \citenamefont {Picque},\ and\
  \citenamefont {Hansch}(2012)}]{Schliesser_NatPhot2012}%
  \BibitemOpen
  \bibfield  {author} {\bibinfo {author} {\bibfnamefont {A.}~\bibnamefont
  {Schliesser}}, \bibinfo {author} {\bibfnamefont {N.}~\bibnamefont {Picque}},
  \ and\ \bibinfo {author} {\bibfnamefont {T.~W.}\ \bibnamefont {Hansch}},\
  }\bibfield  {title} {\enquote {\bibinfo {title} {Mid-infrared frequency
  combs},}\ }\href {http://dx.doi.org/10.1038/nphoton.2012.142} {\bibfield
  {journal} {\bibinfo  {journal} {Nat Photon}\ }\textbf {\bibinfo {volume}
  {6}},\ \bibinfo {pages} {440--449} (\bibinfo {year} {2012})}\BibitemShut
  {NoStop}%
\bibitem [{\citenamefont {Kobayashi}\ \emph {et~al.}(2015)\citenamefont
  {Kobayashi}, \citenamefont {Torizuka}, \citenamefont {Marandi}, \citenamefont
  {Byer}, \citenamefont {McCracken}, \citenamefont {Zhang},\ and\ \citenamefont
  {Reid}}]{Kobayashi_JOpt2015}%
  \BibitemOpen
  \bibfield  {author} {\bibinfo {author} {\bibfnamefont {Y.}~\bibnamefont
  {Kobayashi}}, \bibinfo {author} {\bibfnamefont {K.}~\bibnamefont {Torizuka}},
  \bibinfo {author} {\bibfnamefont {A.}~\bibnamefont {Marandi}}, \bibinfo
  {author} {\bibfnamefont {R.~L.}\ \bibnamefont {Byer}}, \bibinfo {author}
  {\bibfnamefont {R.~A.}\ \bibnamefont {McCracken}}, \bibinfo {author}
  {\bibfnamefont {Z.}~\bibnamefont {Zhang}}, \ and\ \bibinfo {author}
  {\bibfnamefont {D.~T.}\ \bibnamefont {Reid}},\ }\bibfield  {title} {\enquote
  {\bibinfo {title} {Femtosecond optical parametric oscillator frequency
  combs},}\ }\href {\doibase 10.1088/2040-8978/17/9/094010} {\bibfield
  {journal} {\bibinfo  {journal} {Journal of Optics}\ }\textbf {\bibinfo
  {volume} {17}},\ \bibinfo {pages} {094010} (\bibinfo {year}
  {2015})}\BibitemShut {NoStop}%
\bibitem [{\citenamefont {Timmers}\ \emph {et~al.}(2018)\citenamefont
  {Timmers}, \citenamefont {Kowligy}, \citenamefont {Lind}, \citenamefont
  {Cruz}, \citenamefont {Nader}, \citenamefont {Silfies}, \citenamefont {Ycas},
  \citenamefont {Allison}, \citenamefont {Schunemann}, \citenamefont {Papp},\
  and\ \citenamefont {Diddams}}]{Timmers_Optica2018}%
  \BibitemOpen
  \bibfield  {author} {\bibinfo {author} {\bibfnamefont {H.}~\bibnamefont
  {Timmers}}, \bibinfo {author} {\bibfnamefont {A.}~\bibnamefont {Kowligy}},
  \bibinfo {author} {\bibfnamefont {A.}~\bibnamefont {Lind}}, \bibinfo {author}
  {\bibfnamefont {F.~C.}\ \bibnamefont {Cruz}}, \bibinfo {author}
  {\bibfnamefont {N.}~\bibnamefont {Nader}}, \bibinfo {author} {\bibfnamefont
  {M.}~\bibnamefont {Silfies}}, \bibinfo {author} {\bibfnamefont
  {G.}~\bibnamefont {Ycas}}, \bibinfo {author} {\bibfnamefont {T.~K.}\
  \bibnamefont {Allison}}, \bibinfo {author} {\bibfnamefont {P.~G.}\
  \bibnamefont {Schunemann}}, \bibinfo {author} {\bibfnamefont {S.~B.}\
  \bibnamefont {Papp}}, \ and\ \bibinfo {author} {\bibfnamefont {S.~A.}\
  \bibnamefont {Diddams}},\ }\bibfield  {title} {\enquote {\bibinfo {title}
  {Molecular fingerprinting with bright, broadband infrared frequency combs},}\
  }\href {\doibase 10.1364/OPTICA.5.000727} {\bibfield  {journal} {\bibinfo
  {journal} {Optica}\ }\textbf {\bibinfo {volume} {5}},\ \bibinfo {pages}
  {727--732} (\bibinfo {year} {2018})}\BibitemShut {NoStop}%
\bibitem [{\citenamefont {Seidel}\ \emph {et~al.}(2018)\citenamefont {Seidel},
  \citenamefont {Xiao}, \citenamefont {Hussain}, \citenamefont {Arisholm},
  \citenamefont {Hartung}, \citenamefont {Zawilski}, \citenamefont
  {Schunemann}, \citenamefont {Habel}, \citenamefont {Trubetskov},
  \citenamefont {Pervak}, \citenamefont {Pronin},\ and\ \citenamefont
  {Krausz}}]{Seidel_SciAdv2018}%
  \BibitemOpen
  \bibfield  {author} {\bibinfo {author} {\bibfnamefont {M.}~\bibnamefont
  {Seidel}}, \bibinfo {author} {\bibfnamefont {X.}~\bibnamefont {Xiao}},
  \bibinfo {author} {\bibfnamefont {S.~A.}\ \bibnamefont {Hussain}}, \bibinfo
  {author} {\bibfnamefont {G.}~\bibnamefont {Arisholm}}, \bibinfo {author}
  {\bibfnamefont {A.}~\bibnamefont {Hartung}}, \bibinfo {author} {\bibfnamefont
  {K.~T.}\ \bibnamefont {Zawilski}}, \bibinfo {author} {\bibfnamefont {P.~G.}\
  \bibnamefont {Schunemann}}, \bibinfo {author} {\bibfnamefont
  {F.}~\bibnamefont {Habel}}, \bibinfo {author} {\bibfnamefont
  {M.}~\bibnamefont {Trubetskov}}, \bibinfo {author} {\bibfnamefont
  {V.}~\bibnamefont {Pervak}}, \bibinfo {author} {\bibfnamefont
  {O.}~\bibnamefont {Pronin}}, \ and\ \bibinfo {author} {\bibfnamefont
  {F.}~\bibnamefont {Krausz}},\ }\bibfield  {title} {\enquote {\bibinfo {title}
  {Multi-watt, multi-octave, mid-infrared femtosecond source},}\ }\href
  {\doibase 10.1126/sciadv.aaq1526} {\bibfield  {journal} {\bibinfo  {journal}
  {Science Advances}\ }\textbf {\bibinfo {volume} {4}} (\bibinfo {year}
  {2018}),\ 10.1126/sciadv.aaq1526},\ \Eprint
  {http://arxiv.org/abs/http://advances.sciencemag.org/content/4/4/eaaq1526.full.pdf}
  {http://advances.sciencemag.org/content/4/4/eaaq1526.full.pdf} \BibitemShut
  {NoStop}%
\bibitem [{\citenamefont {Chen}\ \emph {et~al.}(2019)\citenamefont {Chen},
  \citenamefont {Silfies}, \citenamefont {Kowzan}, \citenamefont {Bautista},\
  and\ \citenamefont {Allison}}]{Chen_ApplPhysB2019}%
  \BibitemOpen
  \bibfield  {author} {\bibinfo {author} {\bibfnamefont {Y.}~\bibnamefont
  {Chen}}, \bibinfo {author} {\bibfnamefont {M.~C.}\ \bibnamefont {Silfies}},
  \bibinfo {author} {\bibfnamefont {G.}~\bibnamefont {Kowzan}}, \bibinfo
  {author} {\bibfnamefont {J.~M.}\ \bibnamefont {Bautista}}, \ and\ \bibinfo
  {author} {\bibfnamefont {T.~K.}\ \bibnamefont {Allison}},\ }\bibfield
  {title} {\enquote {\bibinfo {title} {Tunable visible frequency combs from a
  yb-fiber-laser-pumped optical parametric oscillator},}\ }\href {\doibase
  10.1007/s00340-019-7191-2} {\bibfield  {journal} {\bibinfo  {journal}
  {Applied Physics B}\ }\textbf {\bibinfo {volume} {125}},\ \bibinfo {pages}
  {81} (\bibinfo {year} {2019})}\BibitemShut {NoStop}%
\bibitem [{\citenamefont {Lesko}\ \emph {et~al.}(2021)\citenamefont {Lesko},
  \citenamefont {Timmers}, \citenamefont {Xing}, \citenamefont {Kowligy},
  \citenamefont {Lind},\ and\ \citenamefont {Diddams}}]{Lesko_NatPhot2021}%
  \BibitemOpen
  \bibfield  {author} {\bibinfo {author} {\bibfnamefont {D.~M.~B.}\
  \bibnamefont {Lesko}}, \bibinfo {author} {\bibfnamefont {H.}~\bibnamefont
  {Timmers}}, \bibinfo {author} {\bibfnamefont {S.}~\bibnamefont {Xing}},
  \bibinfo {author} {\bibfnamefont {A.}~\bibnamefont {Kowligy}}, \bibinfo
  {author} {\bibfnamefont {A.~J.}\ \bibnamefont {Lind}}, \ and\ \bibinfo
  {author} {\bibfnamefont {S.~A.}\ \bibnamefont {Diddams}},\ }\bibfield
  {title} {\enquote {\bibinfo {title} {A six-octave optical frequency comb from
  a scalable few-cycle erbium fibre laser},}\ }\href {\doibase
  10.1038/s41566-021-00778-y} {\bibfield  {journal} {\bibinfo  {journal}
  {Nature Photonics}\ }\textbf {\bibinfo {volume} {15}},\ \bibinfo {pages}
  {281--286} (\bibinfo {year} {2021})}\BibitemShut {NoStop}%
\bibitem [{\citenamefont {Rutledge}\ \emph {et~al.}(2021)\citenamefont
  {Rutledge}, \citenamefont {Catanese}, \citenamefont {Hickstein},
  \citenamefont {Diddams}, \citenamefont {Allison},\ and\ \citenamefont
  {Kowligy}}]{Rutledge_JOSAB2021}%
  \BibitemOpen
  \bibfield  {author} {\bibinfo {author} {\bibfnamefont {J.}~\bibnamefont
  {Rutledge}}, \bibinfo {author} {\bibfnamefont {A.}~\bibnamefont {Catanese}},
  \bibinfo {author} {\bibfnamefont {D.~D.}\ \bibnamefont {Hickstein}}, \bibinfo
  {author} {\bibfnamefont {S.~A.}\ \bibnamefont {Diddams}}, \bibinfo {author}
  {\bibfnamefont {T.~K.}\ \bibnamefont {Allison}}, \ and\ \bibinfo {author}
  {\bibfnamefont {A.~S.}\ \bibnamefont {Kowligy}},\ }\bibfield  {title}
  {\enquote {\bibinfo {title} {Broadband ultraviolet-visible frequency combs
  from cascaded high-harmonic generation in quasi-phase-matched waveguides},}\
  }\href {\doibase 10.1364/JOSAB.427086} {\bibfield  {journal} {\bibinfo
  {journal} {J. Opt. Soc. Am. B}\ }\textbf {\bibinfo {volume} {38}},\ \bibinfo
  {pages} {2252--2260} (\bibinfo {year} {2021})}\BibitemShut {NoStop}%
\bibitem [{\citenamefont {Catanese}\ \emph {et~al.}(2020)\citenamefont
  {Catanese}, \citenamefont {Rutledge}, \citenamefont {Silfies}, \citenamefont
  {Li}, \citenamefont {Timmers}, \citenamefont {Kowligy}, \citenamefont {Lind},
  \citenamefont {Diddams},\ and\ \citenamefont
  {Allison}}]{Catanese_OptLett2020}%
  \BibitemOpen
  \bibfield  {author} {\bibinfo {author} {\bibfnamefont {A.}~\bibnamefont
  {Catanese}}, \bibinfo {author} {\bibfnamefont {J.}~\bibnamefont {Rutledge}},
  \bibinfo {author} {\bibfnamefont {M.~C.}\ \bibnamefont {Silfies}}, \bibinfo
  {author} {\bibfnamefont {X.}~\bibnamefont {Li}}, \bibinfo {author}
  {\bibfnamefont {H.}~\bibnamefont {Timmers}}, \bibinfo {author} {\bibfnamefont
  {A.~S.}\ \bibnamefont {Kowligy}}, \bibinfo {author} {\bibfnamefont
  {A.}~\bibnamefont {Lind}}, \bibinfo {author} {\bibfnamefont {S.~A.}\
  \bibnamefont {Diddams}}, \ and\ \bibinfo {author} {\bibfnamefont {T.~K.}\
  \bibnamefont {Allison}},\ }\bibfield  {title} {\enquote {\bibinfo {title}
  {Mid-infrared frequency comb with 6.7 w average power based on difference
  frequency generation},}\ }\href {\doibase 10.1364/OL.385294} {\bibfield
  {journal} {\bibinfo  {journal} {Opt. Lett.}\ }\textbf {\bibinfo {volume}
  {45}},\ \bibinfo {pages} {1248--1251} (\bibinfo {year} {2020})}\BibitemShut
  {NoStop}%
\bibitem [{\citenamefont {Wu}\ \emph {et~al.}(2024)\citenamefont {Wu},
  \citenamefont {Ledezma}, \citenamefont {Fredrick}, \citenamefont {Sekhar},
  \citenamefont {Sekine}, \citenamefont {Guo}, \citenamefont {Briggs},
  \citenamefont {Marandi},\ and\ \citenamefont {Diddams}}]{Wu_NatPhot2024}%
  \BibitemOpen
  \bibfield  {author} {\bibinfo {author} {\bibfnamefont {T.-H.}\ \bibnamefont
  {Wu}}, \bibinfo {author} {\bibfnamefont {L.}~\bibnamefont {Ledezma}},
  \bibinfo {author} {\bibfnamefont {C.}~\bibnamefont {Fredrick}}, \bibinfo
  {author} {\bibfnamefont {P.}~\bibnamefont {Sekhar}}, \bibinfo {author}
  {\bibfnamefont {R.}~\bibnamefont {Sekine}}, \bibinfo {author} {\bibfnamefont
  {Q.}~\bibnamefont {Guo}}, \bibinfo {author} {\bibfnamefont {R.~M.}\
  \bibnamefont {Briggs}}, \bibinfo {author} {\bibfnamefont {A.}~\bibnamefont
  {Marandi}}, \ and\ \bibinfo {author} {\bibfnamefont {S.~A.}\ \bibnamefont
  {Diddams}},\ }\bibfield  {title} {\enquote {\bibinfo {title}
  {Visible-to-ultraviolet frequency comb generation in lithium niobate
  nanophotonic waveguides},}\ }\href {\doibase 10.1038/s41566-023-01364-0}
  {\bibfield  {journal} {\bibinfo  {journal} {Nature Photonics}\ }\textbf
  {\bibinfo {volume} {18}},\ \bibinfo {pages} {218--223} (\bibinfo {year}
  {2024})}\BibitemShut {NoStop}%
\bibitem [{\citenamefont {Steinle}\ \emph {et~al.}(2016)\citenamefont
  {Steinle}, \citenamefont {M\"{o}rz}, \citenamefont {Steinmann},\ and\
  \citenamefont {Giessen}}]{Steinle_OptLett2016}%
  \BibitemOpen
  \bibfield  {author} {\bibinfo {author} {\bibfnamefont {T.}~\bibnamefont
  {Steinle}}, \bibinfo {author} {\bibfnamefont {F.}~\bibnamefont {M\"{o}rz}},
  \bibinfo {author} {\bibfnamefont {A.}~\bibnamefont {Steinmann}}, \ and\
  \bibinfo {author} {\bibfnamefont {H.}~\bibnamefont {Giessen}},\ }\bibfield
  {title} {\enquote {\bibinfo {title} {Ultra-stable high average power
  femtosecond laser system tunable from 1.33 to
  20‚{\"a}{\^a}‚{\"a}{\^a}$\mu$m},}\ }\href {\doibase 10.1364/OL.41.004863}
  {\bibfield  {journal} {\bibinfo  {journal} {Opt. Lett.}\ }\textbf {\bibinfo
  {volume} {41}},\ \bibinfo {pages} {4863--4866} (\bibinfo {year}
  {2016})}\BibitemShut {NoStop}%
\bibitem [{\citenamefont {Ghotbi}, \citenamefont {Esteban-Martin},\ and\
  \citenamefont {Ebrahim-Zadeh}(2006)}]{Ghotbi_OptLett2006}%
  \BibitemOpen
  \bibfield  {author} {\bibinfo {author} {\bibfnamefont {M.}~\bibnamefont
  {Ghotbi}}, \bibinfo {author} {\bibfnamefont {A.}~\bibnamefont
  {Esteban-Martin}}, \ and\ \bibinfo {author} {\bibfnamefont {M.}~\bibnamefont
  {Ebrahim-Zadeh}},\ }\bibfield  {title} {\enquote {\bibinfo {title} {Bib3o6
  femtosecond optical parametric oscillator},}\ }\href {\doibase
  10.1364/OL.31.003128} {\bibfield  {journal} {\bibinfo  {journal} {Opt.
  Lett.}\ }\textbf {\bibinfo {volume} {31}},\ \bibinfo {pages} {3128--3130}
  (\bibinfo {year} {2006})}\BibitemShut {NoStop}%
\bibitem [{\citenamefont {Sie}\ \emph {et~al.}(2019)\citenamefont {Sie},
  \citenamefont {Rohwer}, \citenamefont {Lee},\ and\ \citenamefont
  {Gedik}}]{Sie_NatComm2019}%
  \BibitemOpen
  \bibfield  {author} {\bibinfo {author} {\bibfnamefont {E.~J.}\ \bibnamefont
  {Sie}}, \bibinfo {author} {\bibfnamefont {T.}~\bibnamefont {Rohwer}},
  \bibinfo {author} {\bibfnamefont {C.}~\bibnamefont {Lee}}, \ and\ \bibinfo
  {author} {\bibfnamefont {N.}~\bibnamefont {Gedik}},\ }\bibfield  {title}
  {\enquote {\bibinfo {title} {Time-resolved xuv arpes with tunable 24--33 ev
  laser pulses at 30 mev resolution},}\ }\href {\doibase
  10.1038/s41467-019-11492-3} {\bibfield  {journal} {\bibinfo  {journal}
  {Nature Communications}\ }\textbf {\bibinfo {volume} {10}},\ \bibinfo {pages}
  {3535} (\bibinfo {year} {2019})}\BibitemShut {NoStop}%
\bibitem [{\citenamefont {Br\"uche}(1942)}]{Bruche_KolloidZ1942}%
  \BibitemOpen
  \bibfield  {author} {\bibinfo {author} {\bibfnamefont {E.}~\bibnamefont
  {Br\"uche}},\ }\bibfield  {title} {\enquote {\bibinfo {title} {Die
  aufl\"osungsgrenze des emissions-elektronenmikroskops},}\ }\href {\doibase
  10.1007/BF01519547} {\bibfield  {journal} {\bibinfo  {journal}
  {Kolloid-Zeitschrift}\ }\textbf {\bibinfo {volume} {100}},\ \bibinfo {pages}
  {192--206} (\bibinfo {year} {1942})}\BibitemShut {NoStop}%
\bibitem [{\citenamefont {Recknagel}(1943)}]{Recknagl_ZPhys1943}%
  \BibitemOpen
  \bibfield  {author} {\bibinfo {author} {\bibfnamefont {A.}~\bibnamefont
  {Recknagel}},\ }\bibfield  {title} {\enquote {\bibinfo {title} {Das
  aufl\"osungsverm{\"o}gen des elektronenmikroskops f\"ur selbststrahler},}\
  }\href {\doibase 10.1007/BF01325849} {\bibfield  {journal} {\bibinfo
  {journal} {Zeitschrift f\"ur Physik}\ }\textbf {\bibinfo {volume} {120}},\
  \bibinfo {pages} {331--362} (\bibinfo {year} {1943})}\BibitemShut {NoStop}%
\bibitem [{\citenamefont {Medjanik}\ \emph {et~al.}(2019)\citenamefont
  {Medjanik}, \citenamefont {Babenkov}, \citenamefont {Chernov}, \citenamefont
  {Vasilyev}, \citenamefont {Sch{\"{o}}nhense}, \citenamefont {Schlueter},
  \citenamefont {Gloskovskii}, \citenamefont {Matveyev}, \citenamefont {Drube},
  \citenamefont {Elmers},\ and\ \citenamefont
  {Sch{\"{o}}nhense}}]{Medjanik_JSynchRad2019}%
  \BibitemOpen
  \bibfield  {author} {\bibinfo {author} {\bibfnamefont {K.}~\bibnamefont
  {Medjanik}}, \bibinfo {author} {\bibfnamefont {S.~V.}\ \bibnamefont
  {Babenkov}}, \bibinfo {author} {\bibfnamefont {S.}~\bibnamefont {Chernov}},
  \bibinfo {author} {\bibfnamefont {D.}~\bibnamefont {Vasilyev}}, \bibinfo
  {author} {\bibfnamefont {B.}~\bibnamefont {Sch{\"{o}}nhense}}, \bibinfo
  {author} {\bibfnamefont {C.}~\bibnamefont {Schlueter}}, \bibinfo {author}
  {\bibfnamefont {A.}~\bibnamefont {Gloskovskii}}, \bibinfo {author}
  {\bibfnamefont {Y.}~\bibnamefont {Matveyev}}, \bibinfo {author}
  {\bibfnamefont {W.}~\bibnamefont {Drube}}, \bibinfo {author} {\bibfnamefont
  {H.~J.}\ \bibnamefont {Elmers}}, \ and\ \bibinfo {author} {\bibfnamefont
  {G.}~\bibnamefont {Sch{\"{o}}nhense}},\ }\bibfield  {title} {\enquote
  {\bibinfo {title} {{Progress in HAXPES performance combining full-field {\it
  k}-imaging with time-of-flight recording}},}\ }\href {\doibase
  10.1107/S1600577519012773} {\bibfield  {journal} {\bibinfo  {journal}
  {Journal of Synchrotron Radiation}\ }\textbf {\bibinfo {volume} {26}},\
  \bibinfo {pages} {1996--2012} (\bibinfo {year} {2019})}\BibitemShut {NoStop}%
\bibitem [{\citenamefont {Elmers}\ \emph
  {et~al.}(2020{\natexlab{a}})\citenamefont {Elmers}, \citenamefont {Chernov},
  \citenamefont {D'Souza}, \citenamefont {Bommanaboyena}, \citenamefont
  {Bodnar}, \citenamefont {Medjanik}, \citenamefont {Babenkov}, \citenamefont
  {Fedchenko}, \citenamefont {Vasilyev}, \citenamefont {Agustsson},
  \citenamefont {Schlueter}, \citenamefont {Gloskovskii}, \citenamefont
  {Matveyev}, \citenamefont {Strocov}, \citenamefont {Skourski}, \citenamefont
  {{\v S}mejkal}, \citenamefont {Sinova}, \citenamefont {Min{\'a}r},
  \citenamefont {Kl\"aui}, \citenamefont {Sch\"onhense},\ and\ \citenamefont
  {Jourdan}}]{Elmers_ACSNano2020}%
  \BibitemOpen
  \bibfield  {author} {\bibinfo {author} {\bibfnamefont {H.~J.}\ \bibnamefont
  {Elmers}}, \bibinfo {author} {\bibfnamefont {S.~V.}\ \bibnamefont {Chernov}},
  \bibinfo {author} {\bibfnamefont {S.~W.}\ \bibnamefont {D'Souza}}, \bibinfo
  {author} {\bibfnamefont {S.~P.}\ \bibnamefont {Bommanaboyena}}, \bibinfo
  {author} {\bibfnamefont {S.~Y.}\ \bibnamefont {Bodnar}}, \bibinfo {author}
  {\bibfnamefont {K.}~\bibnamefont {Medjanik}}, \bibinfo {author}
  {\bibfnamefont {S.}~\bibnamefont {Babenkov}}, \bibinfo {author}
  {\bibfnamefont {O.}~\bibnamefont {Fedchenko}}, \bibinfo {author}
  {\bibfnamefont {D.}~\bibnamefont {Vasilyev}}, \bibinfo {author}
  {\bibfnamefont {S.~Y.}\ \bibnamefont {Agustsson}}, \bibinfo {author}
  {\bibfnamefont {C.}~\bibnamefont {Schlueter}}, \bibinfo {author}
  {\bibfnamefont {A.}~\bibnamefont {Gloskovskii}}, \bibinfo {author}
  {\bibfnamefont {Y.}~\bibnamefont {Matveyev}}, \bibinfo {author}
  {\bibfnamefont {V.~N.}\ \bibnamefont {Strocov}}, \bibinfo {author}
  {\bibfnamefont {Y.}~\bibnamefont {Skourski}}, \bibinfo {author}
  {\bibfnamefont {L.}~\bibnamefont {{\v S}mejkal}}, \bibinfo {author}
  {\bibfnamefont {J.}~\bibnamefont {Sinova}}, \bibinfo {author} {\bibfnamefont
  {J.}~\bibnamefont {Min{\'a}r}}, \bibinfo {author} {\bibfnamefont
  {M.}~\bibnamefont {Kl\"aui}}, \bibinfo {author} {\bibfnamefont
  {G.}~\bibnamefont {Sch\"onhense}}, \ and\ \bibinfo {author} {\bibfnamefont
  {M.}~\bibnamefont {Jourdan}},\ }\bibfield  {title} {\enquote {\bibinfo
  {title} {N\'eel vector induced manipulation of valence states in the
  collinear antiferromagnet mn$_2$au},}\ }\href {\doibase
  10.1021/acsnano.0c08215} {\bibfield  {journal} {\bibinfo  {journal} {ACS
  Nano}\ }\textbf {\bibinfo {volume} {14}},\ \bibinfo {pages} {17554--17564}
  (\bibinfo {year} {2020}{\natexlab{a}})}\BibitemShut {NoStop}%
\bibitem [{\citenamefont {Sch\"onhense}\ and\ \citenamefont
  {Elmers}(2022)}]{Schoenhense_JVST2022}%
  \BibitemOpen
  \bibfield  {author} {\bibinfo {author} {\bibfnamefont {G.}~\bibnamefont
  {Sch\"onhense}}\ and\ \bibinfo {author} {\bibfnamefont {H.-J.}\ \bibnamefont
  {Elmers}},\ }\bibfield  {title} {\enquote {\bibinfo {title} {Spin- and
  time-resolved photoelectron spectroscopy and diffraction studies using
  time-of-flight momentum microscopes},}\ }\bibfield  {booktitle} {\emph
  {\bibinfo {booktitle} {Journal of Vacuum Science and Technology A}},\ }\href
  {https://doi.org/10.1116/6.0001500} {\bibfield  {journal} {\bibinfo
  {journal} {Journal of Vacuum Science and Technology A}\ }\textbf {\bibinfo
  {volume} {40}},\ \bibinfo {pages} {020802} (\bibinfo {year}
  {2022})}\BibitemShut {NoStop}%
\bibitem [{\citenamefont {Bakalis}(2024)}]{Bakalis_Thesis2024}%
  \BibitemOpen
  \bibfield  {author} {\bibinfo {author} {\bibfnamefont {J.}~\bibnamefont
  {Bakalis}},\ }\emph {\bibinfo {title} {Imaging Electron Dynamics in Graphene
  with Momentum Microscopy}},\ \href@noop {} {Ph.D. thesis},\ \bibinfo
  {school} {Stony Brook University} (\bibinfo {year} {2024})\BibitemShut
  {NoStop}%
\bibitem [{\citenamefont {Tkach}\ \emph {et~al.}(2024)\citenamefont {Tkach},
  \citenamefont {Chernov}, \citenamefont {Babenkov}, \citenamefont
  {Lytvynenko}, \citenamefont {Fedchenko}, \citenamefont {Medjanik},
  \citenamefont {Vasilyev}, \citenamefont {Gloskowskii}, \citenamefont
  {Schlueter}, \citenamefont {Elmers},\ and\ \citenamefont
  {Sch{\"{o}}nhense}}]{Tkach_JSynchRad2024}%
  \BibitemOpen
  \bibfield  {author} {\bibinfo {author} {\bibfnamefont {O.}~\bibnamefont
  {Tkach}}, \bibinfo {author} {\bibfnamefont {S.}~\bibnamefont {Chernov}},
  \bibinfo {author} {\bibfnamefont {S.}~\bibnamefont {Babenkov}}, \bibinfo
  {author} {\bibfnamefont {Y.}~\bibnamefont {Lytvynenko}}, \bibinfo {author}
  {\bibfnamefont {O.}~\bibnamefont {Fedchenko}}, \bibinfo {author}
  {\bibfnamefont {K.}~\bibnamefont {Medjanik}}, \bibinfo {author}
  {\bibfnamefont {D.}~\bibnamefont {Vasilyev}}, \bibinfo {author}
  {\bibfnamefont {A.}~\bibnamefont {Gloskowskii}}, \bibinfo {author}
  {\bibfnamefont {C.}~\bibnamefont {Schlueter}}, \bibinfo {author}
  {\bibfnamefont {H.-J.}\ \bibnamefont {Elmers}}, \ and\ \bibinfo {author}
  {\bibfnamefont {G.}~\bibnamefont {Sch{\"{o}}nhense}},\ }\bibfield  {title}
  {\enquote {\bibinfo {title} {{Asymmetric electrostatic dodecapole: compact
  bandpass filter with low aberrations for momentum microscopy}},}\ }\href
  {\doibase 10.1107/S1600577524003540} {\bibfield  {journal} {\bibinfo
  {journal} {Journal of Synchrotron Radiation}\ }\textbf {\bibinfo {volume}
  {31}},\ \bibinfo {pages} {829--840} (\bibinfo {year} {2024})}\BibitemShut
  {NoStop}%
\bibitem [{\citenamefont {Sch{\"{o}}nhense}\ \emph {et~al.}(2021)\citenamefont
  {Sch{\"{o}}nhense}, \citenamefont {Medjanik}, \citenamefont {Fedchenko},
  \citenamefont {Zymakov{\'{a}}}, \citenamefont {Chernov}, \citenamefont
  {Kutnyakhov}, \citenamefont {Vasilyev}, \citenamefont {Babenkov},
  \citenamefont {Elmers}, \citenamefont {Baumg{\"{a}}rtel}, \citenamefont
  {Goslawski}, \citenamefont {{\"{O}}hrwall}, \citenamefont {Grunske},
  \citenamefont {Kauerhof}, \citenamefont {von Volkmann}, \citenamefont
  {Kallmayer}, \citenamefont {Ellguth},\ and\ \citenamefont
  {Oelsner}}]{Schoenhense_JSynchRad2021}%
  \BibitemOpen
  \bibfield  {author} {\bibinfo {author} {\bibfnamefont {G.}~\bibnamefont
  {Sch{\"{o}}nhense}}, \bibinfo {author} {\bibfnamefont {K.}~\bibnamefont
  {Medjanik}}, \bibinfo {author} {\bibfnamefont {O.}~\bibnamefont {Fedchenko}},
  \bibinfo {author} {\bibfnamefont {A.}~\bibnamefont {Zymakov{\'{a}}}},
  \bibinfo {author} {\bibfnamefont {S.}~\bibnamefont {Chernov}}, \bibinfo
  {author} {\bibfnamefont {D.}~\bibnamefont {Kutnyakhov}}, \bibinfo {author}
  {\bibfnamefont {D.}~\bibnamefont {Vasilyev}}, \bibinfo {author}
  {\bibfnamefont {S.}~\bibnamefont {Babenkov}}, \bibinfo {author}
  {\bibfnamefont {H.~J.}\ \bibnamefont {Elmers}}, \bibinfo {author}
  {\bibfnamefont {P.}~\bibnamefont {Baumg{\"{a}}rtel}}, \bibinfo {author}
  {\bibfnamefont {P.}~\bibnamefont {Goslawski}}, \bibinfo {author}
  {\bibfnamefont {G.}~\bibnamefont {{\"{O}}hrwall}}, \bibinfo {author}
  {\bibfnamefont {T.}~\bibnamefont {Grunske}}, \bibinfo {author} {\bibfnamefont
  {T.}~\bibnamefont {Kauerhof}}, \bibinfo {author} {\bibfnamefont
  {K.}~\bibnamefont {von Volkmann}}, \bibinfo {author} {\bibfnamefont
  {M.}~\bibnamefont {Kallmayer}}, \bibinfo {author} {\bibfnamefont
  {M.}~\bibnamefont {Ellguth}}, \ and\ \bibinfo {author} {\bibfnamefont
  {A.}~\bibnamefont {Oelsner}},\ }\bibfield  {title} {\enquote {\bibinfo
  {title} {{Time-of-flight photoelectron momentum microscopy with 80{--}500 MHz
  photon sources: electron-optical pulse picker or bandpass pre-filter}},}\
  }\href {\doibase 10.1107/S1600577521010511} {\bibfield  {journal} {\bibinfo
  {journal} {Journal of Synchrotron Radiation}\ }\textbf {\bibinfo {volume}
  {28}},\ \bibinfo {pages} {1891--1908} (\bibinfo {year} {2021})}\BibitemShut
  {NoStop}%
\bibitem [{\citenamefont {Schmitt}\ \emph {et~al.}(2024)\citenamefont
  {Schmitt}, \citenamefont {Biswas}, \citenamefont {Tkach}, \citenamefont
  {Fedchenko}, \citenamefont {Liu}, \citenamefont {Elmers}, \citenamefont
  {Sing}, \citenamefont {Claessen}, \citenamefont {Lee},\ and\ \citenamefont
  {Sch{\"o}nhense}}]{Schmitt_arXiv2024}%
  \BibitemOpen
  \bibfield  {author} {\bibinfo {author} {\bibfnamefont {M.}~\bibnamefont
  {Schmitt}}, \bibinfo {author} {\bibfnamefont {D.}~\bibnamefont {Biswas}},
  \bibinfo {author} {\bibfnamefont {O.}~\bibnamefont {Tkach}}, \bibinfo
  {author} {\bibfnamefont {O.}~\bibnamefont {Fedchenko}}, \bibinfo {author}
  {\bibfnamefont {J.}~\bibnamefont {Liu}}, \bibinfo {author} {\bibfnamefont
  {H.-J.}\ \bibnamefont {Elmers}}, \bibinfo {author} {\bibfnamefont
  {M.}~\bibnamefont {Sing}}, \bibinfo {author} {\bibfnamefont {R.}~\bibnamefont
  {Claessen}}, \bibinfo {author} {\bibfnamefont {T.-L.}\ \bibnamefont {Lee}}, \
  and\ \bibinfo {author} {\bibfnamefont {G.}~\bibnamefont {Sch{\"o}nhense}},\
  }\href {https://arxiv.org/abs/2406.00771} {\enquote {\bibinfo {title} {Hybrid
  photoelectron momentum microscope at the soft x-ray beamline i09 of the
  diamond light source},}\ } (\bibinfo {year} {2024}),\ \Eprint
  {http://arxiv.org/abs/2406.00771} {arXiv:2406.00771 [cond-mat.mtrl-sci]}
  \BibitemShut {NoStop}%
\bibitem [{\citenamefont {Cichocka}\ \emph {et~al.}(2018)\citenamefont
  {Cichocka}, \citenamefont {{\AA}ngstr{\"{o}}m}, \citenamefont {Wang},
  \citenamefont {Zou},\ and\ \citenamefont {Smeets}}]{Cichocka_JAppCryst2018}%
  \BibitemOpen
  \bibfield  {author} {\bibinfo {author} {\bibfnamefont {M.~O.}\ \bibnamefont
  {Cichocka}}, \bibinfo {author} {\bibfnamefont {J.}~\bibnamefont
  {{\AA}ngstr{\"{o}}m}}, \bibinfo {author} {\bibfnamefont {B.}~\bibnamefont
  {Wang}}, \bibinfo {author} {\bibfnamefont {X.}~\bibnamefont {Zou}}, \ and\
  \bibinfo {author} {\bibfnamefont {S.}~\bibnamefont {Smeets}},\ }\bibfield
  {title} {\enquote {\bibinfo {title} {{High-throughput continuous rotation
  electron diffraction data acquisition {\it via} software automation}},}\
  }\href {\doibase 10.1107/S1600576718015145} {\bibfield  {journal} {\bibinfo
  {journal} {Journal of Applied Crystallography}\ }\textbf {\bibinfo {volume}
  {51}},\ \bibinfo {pages} {1652--1661} (\bibinfo {year} {2018})}\BibitemShut
  {NoStop}%
\bibitem [{\citenamefont {Faruqi}\ and\ \citenamefont
  {McMullan}(2011)}]{Faruqi_QuartRevBiophys2011}%
  \BibitemOpen
  \bibfield  {author} {\bibinfo {author} {\bibfnamefont {A.~R.}\ \bibnamefont
  {Faruqi}}\ and\ \bibinfo {author} {\bibfnamefont {G.}~\bibnamefont
  {McMullan}},\ }\bibfield  {title} {\enquote {\bibinfo {title} {Electronic
  detectors for electron microscopy},}\ }\href {\doibase
  10.1017/S0033583511000035} {\bibfield  {journal} {\bibinfo  {journal}
  {Quarterly Reviews of Biophysics}\ }\textbf {\bibinfo {volume} {44}},\
  \bibinfo {pages} {357--390} (\bibinfo {year} {2011})}\BibitemShut {NoStop}%
\bibitem [{\citenamefont {Milne}\ \emph {et~al.}(2013)\citenamefont {Milne},
  \citenamefont {Borgnia}, \citenamefont {Bartesaghi}, \citenamefont {Tran},
  \citenamefont {Earl}, \citenamefont {Schauder}, \citenamefont {Lengyel},
  \citenamefont {Pierson}, \citenamefont {Patwardhan},\ and\ \citenamefont
  {Subramaniam}}]{Milne_FEBS2013}%
  \BibitemOpen
  \bibfield  {author} {\bibinfo {author} {\bibfnamefont {J.~L.~S.}\
  \bibnamefont {Milne}}, \bibinfo {author} {\bibfnamefont {M.~J.}\ \bibnamefont
  {Borgnia}}, \bibinfo {author} {\bibfnamefont {A.}~\bibnamefont {Bartesaghi}},
  \bibinfo {author} {\bibfnamefont {E.~E.~H.}\ \bibnamefont {Tran}}, \bibinfo
  {author} {\bibfnamefont {L.~A.}\ \bibnamefont {Earl}}, \bibinfo {author}
  {\bibfnamefont {D.~M.}\ \bibnamefont {Schauder}}, \bibinfo {author}
  {\bibfnamefont {J.}~\bibnamefont {Lengyel}}, \bibinfo {author} {\bibfnamefont
  {J.}~\bibnamefont {Pierson}}, \bibinfo {author} {\bibfnamefont
  {A.}~\bibnamefont {Patwardhan}}, \ and\ \bibinfo {author} {\bibfnamefont
  {S.}~\bibnamefont {Subramaniam}},\ }\bibfield  {title} {\enquote {\bibinfo
  {title} {Cryo-electron microscopy -- a primer for the non-microscopist},}\
  }\href {\doibase 10.1111/febs.12078} {\bibfield  {journal} {\bibinfo
  {journal} {The FEBS Journal}\ }\textbf {\bibinfo {volume} {280}},\ \bibinfo
  {pages} {28--45} (\bibinfo {year} {2013})},\ \Eprint
  {http://arxiv.org/abs/https://febs.onlinelibrary.wiley.com/doi/pdf/10.1111/febs.12078}
  {https://febs.onlinelibrary.wiley.com/doi/pdf/10.1111/febs.12078}
  \BibitemShut {NoStop}%
\bibitem [{\citenamefont {{Markovic}}\ \emph {et~al.}(2017)\citenamefont
  {{Markovic}}, \citenamefont {{Caragiulo}}, \citenamefont {{Dragone}},
  \citenamefont {Tamma}, \citenamefont {Osipov}, \citenamefont {Bostedt},
  \citenamefont {Kwiatkowski}, \citenamefont {Segal}, \citenamefont {Hasi},
  \citenamefont {Blaj}, \citenamefont {Kenney},\ and\ \citenamefont
  {Haller}}]{Markovic_SLACPub2017}%
  \BibitemOpen
  \bibfield  {author} {\bibinfo {author} {\bibfnamefont {B.}~\bibnamefont
  {{Markovic}}}, \bibinfo {author} {\bibfnamefont {P.}~\bibnamefont
  {{Caragiulo}}}, \bibinfo {author} {\bibfnamefont {A.}~\bibnamefont
  {{Dragone}}}, \bibinfo {author} {\bibfnamefont {C.}~\bibnamefont {Tamma}},
  \bibinfo {author} {\bibfnamefont {T.}~\bibnamefont {Osipov}}, \bibinfo
  {author} {\bibfnamefont {C.}~\bibnamefont {Bostedt}}, \bibinfo {author}
  {\bibfnamefont {M.}~\bibnamefont {Kwiatkowski}}, \bibinfo {author}
  {\bibfnamefont {J.}~\bibnamefont {Segal}}, \bibinfo {author} {\bibfnamefont
  {J.}~\bibnamefont {Hasi}}, \bibinfo {author} {\bibfnamefont {G.}~\bibnamefont
  {Blaj}}, \bibinfo {author} {\bibfnamefont {C.}~\bibnamefont {Kenney}}, \ and\
  \bibinfo {author} {\bibfnamefont {G.}~\bibnamefont {Haller}},\ }\href@noop {}
  {\enquote {\bibinfo {title} {Design and characterization of the tpix
  prototype: a spatial and time resolving front-end asic for electron and ion
  spectroscopy experiments at lcls},}\ }\bibinfo {type} {SLAC Pub}\ \bibinfo
  {number} {16891}\ (\bibinfo  {institution} {SLAC National Laboratory},\
  \bibinfo {year} {2017})\BibitemShut {NoStop}%
\bibitem [{\citenamefont {Giacomini}\ \emph {et~al.}(2019)\citenamefont
  {Giacomini}, \citenamefont {Chen}, \citenamefont {Lanni},\ and\ \citenamefont
  {Tricoli}}]{Giacomini_NIMA2019}%
  \BibitemOpen
  \bibfield  {author} {\bibinfo {author} {\bibfnamefont {G.}~\bibnamefont
  {Giacomini}}, \bibinfo {author} {\bibfnamefont {W.}~\bibnamefont {Chen}},
  \bibinfo {author} {\bibfnamefont {F.}~\bibnamefont {Lanni}}, \ and\ \bibinfo
  {author} {\bibfnamefont {A.}~\bibnamefont {Tricoli}},\ }\bibfield  {title}
  {\enquote {\bibinfo {title} {Development of a technology for the fabrication
  of low-gain avalanche diodes at bnl},}\ }\href {\doibase
  10.1016/j.nima.2019.04.073} {\bibfield  {journal} {\bibinfo  {journal}
  {Nuclear Instruments and Methods in Physics Research Section A: Accelerators,
  Spectrometers, Detectors and Associated Equipment}\ }\textbf {\bibinfo
  {volume} {934}},\ \bibinfo {pages} {52--57} (\bibinfo {year}
  {2019})}\BibitemShut {NoStop}%
\bibitem [{\citenamefont {{Markovic}}\ \emph {et~al.}(2018)\citenamefont
  {{Markovic}}, \citenamefont {{Conforti}}, \citenamefont {{de La Taille}},
  \citenamefont {{Martin-Chassard}}, \citenamefont {{Seguin-Moreau}},
  \citenamefont {{Agapopoulou}}, \citenamefont {{Makovec}}, \citenamefont
  {{Serin}}, \citenamefont {{Caragiulo}}, \citenamefont {{Dragone}},
  \citenamefont {{Koua}}, \citenamefont {{Schwartzman}}, \citenamefont {{Su}},
  \citenamefont {{Gong}}, \citenamefont {{Sun}}, \citenamefont {{Ye}},
  \citenamefont {{Zhou}},\ and\ \citenamefont
  {{Casanova}}}]{Markvoic_IEEE2018}%
  \BibitemOpen
  \bibfield  {author} {\bibinfo {author} {\bibfnamefont {B.}~\bibnamefont
  {{Markovic}}}, \bibinfo {author} {\bibfnamefont {S.}~\bibnamefont
  {{Conforti}}}, \bibinfo {author} {\bibfnamefont {C.}~\bibnamefont {{de La
  Taille}}}, \bibinfo {author} {\bibfnamefont {G.}~\bibnamefont
  {{Martin-Chassard}}}, \bibinfo {author} {\bibfnamefont {N.}~\bibnamefont
  {{Seguin-Moreau}}}, \bibinfo {author} {\bibfnamefont {C.}~\bibnamefont
  {{Agapopoulou}}}, \bibinfo {author} {\bibfnamefont {N.}~\bibnamefont
  {{Makovec}}}, \bibinfo {author} {\bibfnamefont {L.}~\bibnamefont {{Serin}}},
  \bibinfo {author} {\bibfnamefont {P.}~\bibnamefont {{Caragiulo}}}, \bibinfo
  {author} {\bibfnamefont {A.}~\bibnamefont {{Dragone}}}, \bibinfo {author}
  {\bibfnamefont {K.}~\bibnamefont {{Koua}}}, \bibinfo {author} {\bibfnamefont
  {A.~G.}\ \bibnamefont {{Schwartzman}}}, \bibinfo {author} {\bibfnamefont
  {D.}~\bibnamefont {{Su}}}, \bibinfo {author} {\bibfnamefont {D.}~\bibnamefont
  {{Gong}}}, \bibinfo {author} {\bibfnamefont {Q.}~\bibnamefont {{Sun}}},
  \bibinfo {author} {\bibfnamefont {J.}~\bibnamefont {{Ye}}}, \bibinfo {author}
  {\bibfnamefont {W.}~\bibnamefont {{Zhou}}}, \ and\ \bibinfo {author}
  {\bibfnamefont {R.}~\bibnamefont {{Casanova}}},\ }\bibfield  {title}
  {\enquote {\bibinfo {title} {Altiroc1, a 20 ps time-resolution asic prototype
  for the atlas high granularity timing detector (hgtd)},}\ }in\ \href
  {\doibase 10.1109/NSSMIC.2018.8824723} {\emph {\bibinfo {booktitle} {2018
  IEEE Nuclear Science Symposium and Medical Imaging Conference Proceedings
  (NSS/MIC)}}}\ (\bibinfo {year} {2018})\ pp.\ \bibinfo {pages}
  {1--3}\BibitemShut {NoStop}%
\bibitem [{\citenamefont {Labaye}\ \emph {et~al.}(2017)\citenamefont {Labaye},
  \citenamefont {Gaponenko}, \citenamefont {Wittwer}, \citenamefont {Diebold},
  \citenamefont {Paradis}, \citenamefont {Modsching}, \citenamefont {Merceron},
  \citenamefont {Emaury}, \citenamefont {Graumann}, \citenamefont {Phillips},
  \citenamefont {Saraceno}, \citenamefont {Kr\"{a}nkel}, \citenamefont
  {Keller},\ and\ \citenamefont {S\"{u}dmeyer}}]{Labaye_OptLett2017}%
  \BibitemOpen
  \bibfield  {author} {\bibinfo {author} {\bibfnamefont {F.}~\bibnamefont
  {Labaye}}, \bibinfo {author} {\bibfnamefont {M.}~\bibnamefont {Gaponenko}},
  \bibinfo {author} {\bibfnamefont {V.~J.}\ \bibnamefont {Wittwer}}, \bibinfo
  {author} {\bibfnamefont {A.}~\bibnamefont {Diebold}}, \bibinfo {author}
  {\bibfnamefont {C.}~\bibnamefont {Paradis}}, \bibinfo {author} {\bibfnamefont
  {N.}~\bibnamefont {Modsching}}, \bibinfo {author} {\bibfnamefont
  {L.}~\bibnamefont {Merceron}}, \bibinfo {author} {\bibfnamefont
  {F.}~\bibnamefont {Emaury}}, \bibinfo {author} {\bibfnamefont {I.~J.}\
  \bibnamefont {Graumann}}, \bibinfo {author} {\bibfnamefont {C.~R.}\
  \bibnamefont {Phillips}}, \bibinfo {author} {\bibfnamefont {C.~J.}\
  \bibnamefont {Saraceno}}, \bibinfo {author} {\bibfnamefont {C.}~\bibnamefont
  {Kr\"{a}nkel}}, \bibinfo {author} {\bibfnamefont {U.}~\bibnamefont {Keller}},
  \ and\ \bibinfo {author} {\bibfnamefont {T.}~\bibnamefont {S\"{u}dmeyer}},\
  }\bibfield  {title} {\enquote {\bibinfo {title} {Extreme ultraviolet light
  source at a megahertz repetition rate based on high-harmonic generation
  inside a mode-locked thin-disk laser oscillator},}\ }\href {\doibase
  10.1364/OL.42.005170} {\bibfield  {journal} {\bibinfo  {journal} {Opt.
  Lett.}\ }\textbf {\bibinfo {volume} {42}},\ \bibinfo {pages} {5170--5173}
  (\bibinfo {year} {2017})}\BibitemShut {NoStop}%
\bibitem [{\citenamefont {Drs}\ \emph {et~al.}(2024)\citenamefont {Drs},
  \citenamefont {Trawi}, \citenamefont {M\"{u}ller}, \citenamefont {Fischer},
  \citenamefont {Wittwer},\ and\ \citenamefont
  {S\"{u}dmeyer}}]{Drs_OptExp2024}%
  \BibitemOpen
  \bibfield  {author} {\bibinfo {author} {\bibfnamefont {J.}~\bibnamefont
  {Drs}}, \bibinfo {author} {\bibfnamefont {F.}~\bibnamefont {Trawi}}, \bibinfo
  {author} {\bibfnamefont {M.}~\bibnamefont {M\"{u}ller}}, \bibinfo {author}
  {\bibfnamefont {J.}~\bibnamefont {Fischer}}, \bibinfo {author} {\bibfnamefont
  {V.~J.}\ \bibnamefont {Wittwer}}, \ and\ \bibinfo {author} {\bibfnamefont
  {T.}~\bibnamefont {S\"{u}dmeyer}},\ }\bibfield  {title} {\enquote {\bibinfo
  {title} {Intra-oscillator high harmonic source reaching 100-ev photon
  energy},}\ }\href {\doibase 10.1364/OE.522104} {\bibfield  {journal}
  {\bibinfo  {journal} {Opt. Express}\ }\textbf {\bibinfo {volume} {32}},\
  \bibinfo {pages} {17424--17432} (\bibinfo {year} {2024})}\BibitemShut
  {NoStop}%
\bibitem [{\citenamefont {Gaida}\ \emph {et~al.}(2018)\citenamefont {Gaida},
  \citenamefont {Heuermann}, \citenamefont {Gebhardt}, \citenamefont
  {Shestaev}, \citenamefont {Butler}, \citenamefont {Gerz}, \citenamefont
  {Lilienfein}, \citenamefont {Sulzer}, \citenamefont {Fischer}, \citenamefont
  {Holzwarth}, \citenamefont {Leitenstorfer}, \citenamefont {Pupeza},\ and\
  \citenamefont {Limpert}}]{Gaida_OptLett2018_2}%
  \BibitemOpen
  \bibfield  {author} {\bibinfo {author} {\bibfnamefont {C.}~\bibnamefont
  {Gaida}}, \bibinfo {author} {\bibfnamefont {T.}~\bibnamefont {Heuermann}},
  \bibinfo {author} {\bibfnamefont {M.}~\bibnamefont {Gebhardt}}, \bibinfo
  {author} {\bibfnamefont {E.}~\bibnamefont {Shestaev}}, \bibinfo {author}
  {\bibfnamefont {T.~P.}\ \bibnamefont {Butler}}, \bibinfo {author}
  {\bibfnamefont {D.}~\bibnamefont {Gerz}}, \bibinfo {author} {\bibfnamefont
  {N.}~\bibnamefont {Lilienfein}}, \bibinfo {author} {\bibfnamefont
  {P.}~\bibnamefont {Sulzer}}, \bibinfo {author} {\bibfnamefont
  {M.}~\bibnamefont {Fischer}}, \bibinfo {author} {\bibfnamefont
  {R.}~\bibnamefont {Holzwarth}}, \bibinfo {author} {\bibfnamefont
  {A.}~\bibnamefont {Leitenstorfer}}, \bibinfo {author} {\bibfnamefont
  {I.}~\bibnamefont {Pupeza}}, \ and\ \bibinfo {author} {\bibfnamefont
  {J.}~\bibnamefont {Limpert}},\ }\bibfield  {title} {\enquote {\bibinfo
  {title} {High-power frequency comb at 2 micron wavelength emitted by a
  tm-doped fiber laser system},}\ }\href {\doibase 10.1364/OL.43.005178}
  {\bibfield  {journal} {\bibinfo  {journal} {Opt. Lett.}\ }\textbf {\bibinfo
  {volume} {43}},\ \bibinfo {pages} {5178--5181} (\bibinfo {year}
  {2018})}\BibitemShut {NoStop}%
\bibitem [{\citenamefont {Tusche}\ \emph {et~al.}(2011)\citenamefont {Tusche},
  \citenamefont {Ellguth}, \citenamefont {{\"U}nal}, \citenamefont {Chiang},
  \citenamefont {Winkelmann}, \citenamefont {Krasyuk}, \citenamefont {Hahn},
  \citenamefont {Sch{\"o}nhense},\ and\ \citenamefont
  {Kirschner}}]{Tusche_APL2011}%
  \BibitemOpen
  \bibfield  {author} {\bibinfo {author} {\bibfnamefont {C.}~\bibnamefont
  {Tusche}}, \bibinfo {author} {\bibfnamefont {M.}~\bibnamefont {Ellguth}},
  \bibinfo {author} {\bibfnamefont {A.~A.}\ \bibnamefont {{\"U}nal}}, \bibinfo
  {author} {\bibfnamefont {C.-T.}\ \bibnamefont {Chiang}}, \bibinfo {author}
  {\bibfnamefont {A.}~\bibnamefont {Winkelmann}}, \bibinfo {author}
  {\bibfnamefont {A.}~\bibnamefont {Krasyuk}}, \bibinfo {author} {\bibfnamefont
  {M.}~\bibnamefont {Hahn}}, \bibinfo {author} {\bibfnamefont {G.}~\bibnamefont
  {Sch{\"o}nhense}}, \ and\ \bibinfo {author} {\bibfnamefont {J.}~\bibnamefont
  {Kirschner}},\ }\bibfield  {title} {\enquote {\bibinfo {title} {{Spin
  resolved photoelectron microscopy using a two-dimensional spin-polarizing
  electron mirror}},}\ }\href {\doibase 10.1063/1.3611648} {\bibfield
  {journal} {\bibinfo  {journal} {Applied Physics Letters}\ }\textbf {\bibinfo
  {volume} {99}},\ \bibinfo {pages} {032505} (\bibinfo {year}
  {2011})}\BibitemShut {NoStop}%
\bibitem [{\citenamefont {Tusche}, \citenamefont {Krasyuk},\ and\ \citenamefont
  {Kirschner}(2015)}]{Tusche_Ultramicroscopy2015}%
  \BibitemOpen
  \bibfield  {author} {\bibinfo {author} {\bibfnamefont {C.}~\bibnamefont
  {Tusche}}, \bibinfo {author} {\bibfnamefont {A.}~\bibnamefont {Krasyuk}}, \
  and\ \bibinfo {author} {\bibfnamefont {J.}~\bibnamefont {Kirschner}},\
  }\bibfield  {title} {\enquote {\bibinfo {title} {Spin resolved bandstructure
  imaging with a high resolution momentum microscope},}\ }\href {\doibase
  https://doi.org/10.1016/j.ultramic.2015.03.020} {\bibfield  {journal}
  {\bibinfo  {journal} {Ultramicroscopy}\ }\textbf {\bibinfo {volume} {159}},\
  \bibinfo {pages} {520 -- 529} (\bibinfo {year} {2015})},\ \bibinfo {note}
  {special Issue: LEEM-PEEM 9}\BibitemShut {NoStop}%
\bibitem [{\citenamefont {Suga}\ and\ \citenamefont
  {Tusche}(2015)}]{Suga_JElecSpec2015}%
  \BibitemOpen
  \bibfield  {author} {\bibinfo {author} {\bibfnamefont {S.}~\bibnamefont
  {Suga}}\ and\ \bibinfo {author} {\bibfnamefont {C.}~\bibnamefont {Tusche}},\
  }\bibfield  {title} {\enquote {\bibinfo {title} {Photoelectron spectroscopy
  in a wide h$\nu$ region from 6 ev to 8 kev with full momentum and spin
  resolution},}\ }\href {\doibase https://doi.org/10.1016/j.elspec.2015.04.019}
  {\bibfield  {journal} {\bibinfo  {journal} {Journal of Electron Spectroscopy
  and Related Phenomena}\ }\textbf {\bibinfo {volume} {200}},\ \bibinfo {pages}
  {119--142} (\bibinfo {year} {2015})},\ \bibinfo {note} {special Anniversary
  Issue: Volume 200}\BibitemShut {NoStop}%
\bibitem [{\citenamefont {Chernov}\ \emph {et~al.}(2021)\citenamefont
  {Chernov}, \citenamefont {Lidig}, \citenamefont {Fedchenko}, \citenamefont
  {Medjanik}, \citenamefont {Babenkov}, \citenamefont {Vasilyev}, \citenamefont
  {Jourdan}, \citenamefont {Sch\"onhense},\ and\ \citenamefont
  {Elmers}}]{Chernov_PRB2021}%
  \BibitemOpen
  \bibfield  {author} {\bibinfo {author} {\bibfnamefont {S.}~\bibnamefont
  {Chernov}}, \bibinfo {author} {\bibfnamefont {C.}~\bibnamefont {Lidig}},
  \bibinfo {author} {\bibfnamefont {O.}~\bibnamefont {Fedchenko}}, \bibinfo
  {author} {\bibfnamefont {K.}~\bibnamefont {Medjanik}}, \bibinfo {author}
  {\bibfnamefont {S.}~\bibnamefont {Babenkov}}, \bibinfo {author}
  {\bibfnamefont {D.}~\bibnamefont {Vasilyev}}, \bibinfo {author}
  {\bibfnamefont {M.}~\bibnamefont {Jourdan}}, \bibinfo {author} {\bibfnamefont
  {G.}~\bibnamefont {Sch\"onhense}}, \ and\ \bibinfo {author} {\bibfnamefont
  {H.~J.}\ \bibnamefont {Elmers}},\ }\bibfield  {title} {\enquote {\bibinfo
  {title} {Band structure tuning of heusler compounds: Spin- and
  momentum-resolved electronic structure analysis of compounds with different
  band filling},}\ }\href {\doibase 10.1103/PhysRevB.103.054407} {\bibfield
  {journal} {\bibinfo  {journal} {Phys. Rev. B}\ }\textbf {\bibinfo {volume}
  {103}},\ \bibinfo {pages} {054407} (\bibinfo {year} {2021})}\BibitemShut
  {NoStop}%
\bibitem [{\citenamefont {Elmers}\ \emph {et~al.}(2016)\citenamefont {Elmers},
  \citenamefont {Wallauer}, \citenamefont {Liebmann}, \citenamefont {Kellner},
  \citenamefont {Morgenstern}, \citenamefont {Wang}, \citenamefont {Boschker},
  \citenamefont {Calarco}, \citenamefont {S\'anchez-Barriga}, \citenamefont
  {Rader}, \citenamefont {Kutnyakhov}, \citenamefont {Chernov}, \citenamefont
  {Medjanik}, \citenamefont {Tusche}, \citenamefont {Ellguth}, \citenamefont
  {Volfova}, \citenamefont {Borek}, \citenamefont {Braun}, \citenamefont
  {Min\'ar}, \citenamefont {Ebert},\ and\ \citenamefont
  {Sch\"onhense}}]{Elmers_PRB2016}%
  \BibitemOpen
  \bibfield  {author} {\bibinfo {author} {\bibfnamefont {H.~J.}\ \bibnamefont
  {Elmers}}, \bibinfo {author} {\bibfnamefont {R.}~\bibnamefont {Wallauer}},
  \bibinfo {author} {\bibfnamefont {M.}~\bibnamefont {Liebmann}}, \bibinfo
  {author} {\bibfnamefont {J.}~\bibnamefont {Kellner}}, \bibinfo {author}
  {\bibfnamefont {M.}~\bibnamefont {Morgenstern}}, \bibinfo {author}
  {\bibfnamefont {R.~N.}\ \bibnamefont {Wang}}, \bibinfo {author}
  {\bibfnamefont {J.~E.}\ \bibnamefont {Boschker}}, \bibinfo {author}
  {\bibfnamefont {R.}~\bibnamefont {Calarco}}, \bibinfo {author} {\bibfnamefont
  {J.}~\bibnamefont {S\'anchez-Barriga}}, \bibinfo {author} {\bibfnamefont
  {O.}~\bibnamefont {Rader}}, \bibinfo {author} {\bibfnamefont
  {D.}~\bibnamefont {Kutnyakhov}}, \bibinfo {author} {\bibfnamefont {S.~V.}\
  \bibnamefont {Chernov}}, \bibinfo {author} {\bibfnamefont {K.}~\bibnamefont
  {Medjanik}}, \bibinfo {author} {\bibfnamefont {C.}~\bibnamefont {Tusche}},
  \bibinfo {author} {\bibfnamefont {M.}~\bibnamefont {Ellguth}}, \bibinfo
  {author} {\bibfnamefont {H.}~\bibnamefont {Volfova}}, \bibinfo {author}
  {\bibfnamefont {S.}~\bibnamefont {Borek}}, \bibinfo {author} {\bibfnamefont
  {J.}~\bibnamefont {Braun}}, \bibinfo {author} {\bibfnamefont
  {J.}~\bibnamefont {Min\'ar}}, \bibinfo {author} {\bibfnamefont
  {H.}~\bibnamefont {Ebert}}, \ and\ \bibinfo {author} {\bibfnamefont
  {G.}~\bibnamefont {Sch\"onhense}},\ }\bibfield  {title} {\enquote {\bibinfo
  {title} {Spin mapping of surface and bulk rashba states in ferroelectric
  $\ensuremath{\alpha}$-gete(111) films},}\ }\href {\doibase
  10.1103/PhysRevB.94.201403} {\bibfield  {journal} {\bibinfo  {journal} {Phys.
  Rev. B}\ }\textbf {\bibinfo {volume} {94}},\ \bibinfo {pages} {201403}
  (\bibinfo {year} {2016})}\BibitemShut {NoStop}%
\bibitem [{\citenamefont {Elmers}\ \emph
  {et~al.}(2020{\natexlab{b}})\citenamefont {Elmers}, \citenamefont {Regel},
  \citenamefont {Mashoff}, \citenamefont {Braun}, \citenamefont {Babenkov},
  \citenamefont {Chernov}, \citenamefont {Fedchenko}, \citenamefont {Medjanik},
  \citenamefont {Vasilyev}, \citenamefont {Minar}, \citenamefont {Ebert},\ and\
  \citenamefont {Sch\"onhense}}]{Elmers_PhysRevResearch2020}%
  \BibitemOpen
  \bibfield  {author} {\bibinfo {author} {\bibfnamefont {H.~J.}\ \bibnamefont
  {Elmers}}, \bibinfo {author} {\bibfnamefont {J.}~\bibnamefont {Regel}},
  \bibinfo {author} {\bibfnamefont {T.}~\bibnamefont {Mashoff}}, \bibinfo
  {author} {\bibfnamefont {J.}~\bibnamefont {Braun}}, \bibinfo {author}
  {\bibfnamefont {S.}~\bibnamefont {Babenkov}}, \bibinfo {author}
  {\bibfnamefont {S.}~\bibnamefont {Chernov}}, \bibinfo {author} {\bibfnamefont
  {O.}~\bibnamefont {Fedchenko}}, \bibinfo {author} {\bibfnamefont
  {K.}~\bibnamefont {Medjanik}}, \bibinfo {author} {\bibfnamefont
  {D.}~\bibnamefont {Vasilyev}}, \bibinfo {author} {\bibfnamefont
  {J.}~\bibnamefont {Minar}}, \bibinfo {author} {\bibfnamefont
  {H.}~\bibnamefont {Ebert}}, \ and\ \bibinfo {author} {\bibfnamefont
  {G.}~\bibnamefont {Sch\"onhense}},\ }\bibfield  {title} {\enquote {\bibinfo
  {title} {Rashba splitting of the tamm surface state on re(0001) observed by
  spin-resolved photoemission and scanning tunneling spectroscopy},}\ }\href
  {\doibase 10.1103/PhysRevResearch.2.013296} {\bibfield  {journal} {\bibinfo
  {journal} {Phys. Rev. Res.}\ }\textbf {\bibinfo {volume} {2}},\ \bibinfo
  {pages} {013296} (\bibinfo {year} {2020}{\natexlab{b}})}\BibitemShut
  {NoStop}%
\bibitem [{\citenamefont {Kessler}(1981)}]{Kessler_CommentsAMO1981}%
  \BibitemOpen
  \bibfield  {author} {\bibinfo {author} {\bibfnamefont {J.}~\bibnamefont
  {Kessler}},\ }\href@noop {} {\bibfield  {journal} {\bibinfo  {journal}
  {Comments Atom. Mol. Phys.}\ }\textbf {\bibinfo {volume} {10}},\ \bibinfo
  {pages} {47} (\bibinfo {year} {1981})}\BibitemShut {NoStop}%
\bibitem [{\citenamefont {Yamamoto}\ \emph {et~al.}(2023)\citenamefont
  {Yamamoto}, \citenamefont {Yano}, \citenamefont {Karashima}, \citenamefont
  {Uenishi}, \citenamefont {Orimo}, \citenamefont {Nishitani},\ and\
  \citenamefont {Suzuki}}]{Yamamoto_BChemSocJap2023}%
  \BibitemOpen
  \bibfield  {author} {\bibinfo {author} {\bibfnamefont {Y.-i.}\ \bibnamefont
  {Yamamoto}}, \bibinfo {author} {\bibfnamefont {H.}~\bibnamefont {Yano}},
  \bibinfo {author} {\bibfnamefont {S.}~\bibnamefont {Karashima}}, \bibinfo
  {author} {\bibfnamefont {R.}~\bibnamefont {Uenishi}}, \bibinfo {author}
  {\bibfnamefont {N.}~\bibnamefont {Orimo}}, \bibinfo {author} {\bibfnamefont
  {J.}~\bibnamefont {Nishitani}}, \ and\ \bibinfo {author} {\bibfnamefont
  {T.}~\bibnamefont {Suzuki}},\ }\bibfield  {title} {\enquote {\bibinfo {title}
  {{Extreme Ultraviolet Laser Photoelectron Spectroscopy of Flat Liquid Jet
  Generated Using Microfluidic Device}},}\ }\href {\doibase
  10.1246/bcsj.20230151} {\bibfield  {journal} {\bibinfo  {journal} {Bulletin
  of the Chemical Society of Japan}\ }\textbf {\bibinfo {volume} {96}},\
  \bibinfo {pages} {938--942} (\bibinfo {year} {2023})},\ \Eprint
  {http://arxiv.org/abs/https://academic.oup.com/bcsj/article-pdf/96/9/938/56150023/bcsj.20230151.pdf}
  {https://academic.oup.com/bcsj/article-pdf/96/9/938/56150023/bcsj.20230151.pdf}
  \BibitemShut {NoStop}%
\bibitem [{\citenamefont {Tkach}\ and\ \citenamefont
  {Schoenhense}(2024)}]{Tkach_arXiv2024}%
  \BibitemOpen
  \bibfield  {author} {\bibinfo {author} {\bibfnamefont {O.}~\bibnamefont
  {Tkach}}\ and\ \bibinfo {author} {\bibfnamefont {G.}~\bibnamefont
  {Schoenhense}},\ }\href {https://arxiv.org/abs/2408.10104} {\enquote
  {\bibinfo {title} {Multi-mode lens for momentum microscopy and xpeem:
  Theory},}\ } (\bibinfo {year} {2024}),\ \Eprint
  {http://arxiv.org/abs/2408.10104} {arXiv:2408.10104 [physics.app-ph]}
  \BibitemShut {NoStop}%
\end{thebibliography}%

\end{document}